\documentclass[3p,preprint,hyperref]{elsarticle}
\input{XEFT_aux}

\begin{document}

\begin{frontmatter}
\title{XEFT, the challenging path up the hill: $\mrdim = 6$ and $\mrdim = 8$}

\author[torino]{Giampiero Passarino}
\ead{giampiero@to.infn.it}

\address[torino]{\csumb}


\begin{abstract}
There is increasing need to assess the impact and the interpretation of $\mrdim = 6$ and $\mrdim = 8$ operators within 
the context of the Standard Model Effective Field Theory (SMEFT). The observational and mathematical consistency of a 
construct based on $\mrdim = 6$ and $\mrdim = 8$ operators is critically examined in the light of known theoretical 
results.
The discussion is based on a general $\mrdim = 4$ theory $X$ and its effective extension, XEFT; it includes 
elimination of redundant operators and their higher order compensation, 
SMEFT in comparison with ultraviolet completions incorporating a proliferation of scalar and mixings, 
canonical normalization of effective field theories, 
gauge invariance and gauge fixing, 
role of tadpoles when constructing XEFT at NLO, 
heavy{-}light contributions to the low energy limit of theories containing bosons and fermions,
one-loop matching,
EFT fits and their interpretation and
effective field theory interpretation of derivative-coupled field theories.  
\end{abstract}

\end{frontmatter}

\section{Introduction \label{intro}}
Consider a four-dimensional theory, $X$, described by a Lagrangian $\Lag^{(4)}_{\ssX}$ with a symmetry group $\mrG$:
the goal of this work is to critically discuss some of the issues related to the construction of its ``effective''
extension; the XEFT extension will be described by a Lagrangian
\bq
\Lag_{\ssX} = \Lag^{(4)}_{\ssX} + \sum_{d > 4}\,\sum_i\,\frac{\mra^d_i}{\Lambda^{4-d}}\,\Ope^{(d)}_i =
\sum_{d \ge 4}\,\Lag^{(d)}_{\ssX} \spc
\label{EFTex}
\eq
where $\Lambda$ is the cutoff of the effective theory, $\mra^d_i$ are Wilson coefficients and $\Ope^{(d)}_i$ are
$\mrG\,$-invariant operators of mass-dimension $d$ involving the $\Lag^{(4)}_{\ssX}$ fields. 
The definition of EFT extension requires a more detailed description for which we need to consider $X^{\prime}$, the
ultraviolet (UV) completion of $X$ or the next theory in a tower of theories, $X^{\prime},\,X^{\prime\prime}\,\dots$.
A related point to consider: $X$ is the leading order approximation of $X^{\prime}$ but a strictly renormalizable $X$ has 
no encoded information about the scale at which it stops being a meaningful representation of a complete theory.
There are two possibilities in going from $X$ to $X^{\prime}$: 
\bei
\item[a)] $X^{\prime}$ is based on $\mrG$ but contains heavy degrees of freedom belonging to some representation of $\mrG$ or 
\item[b)] $X^{\prime}$ is based on a larger group $\mrF$ where $\mrF$ should contain $\mrG$ and $X^{\prime}$ should reduce 
to $X$ at low energies. 
\eei
An additional assumption is that there are no ``undiscovered'' degrees of freedom in $X^{\prime}$ that are light and 
weakly coupled. More broadly, one generally excludes light particles, neutral under $\mrG$, that interact with the 
$X$ particles only through the exchange of new heavy particles.
When $X$ is the standard model (SM) and $\mrG = SU(3)\,\times\,SU(2)\,\times\,U(1)$ the most simple
examples in category a) are the SM singlet extension and the THDMs, containing two scalar doublets, see
\Bref{Yagyu:2012qp} for a review. In category b) an example is given by a non supersymmetric 
$SO(10)$~\cite{Altarelli:2013aqa} which breaks down to the SM through a chain of different intermediate groups. 
Another example is $SU(3)_c\,\times\,SU(3)_{\ssL}\,\times\,U(1)_{\ssY}$, the so-called $331$ model~\cite{Okada:2016whh}.
The Lagrangian of \eqn{EFTex} assumes that all the $X$ degrees of freedom should be incorporated; for the SM 
the EFTs are further distinguished by the presence of a Higgs doublet (or not) in the construction, see 
\Bref{Brivio:2017vri} for a review. In the SMEFT the EFT is constructed with an explicit Higgs doublet, while in the 
HEFT (an Electroweak chiral Lagrangian with a dominantly $J^{\ssP} = 0^+$ scalar) no such doublet is included. 

In the rest of this paper we discuss several issues that arise in constructing an EFT 
(up to and including $\mrdim = 8$ operators), \ie using XEFT context, we derive several results that are relevant 
to the challenges presented to present and future theorists and experimentalists working to go beyond the SM:
\begin{itemize}

\item[\dnuma] The traditional way of using an EFT is based on matching the EFT to the full theory in order to 
reproduce the correct matrix elements, \ie we choose the coefficients to be those appropriate for the full theory.
After the Higgs boson discovery we have a paradigm shift, \ie we use the EFT for fitting the data; at this point
the ``fitted'' Wilson coefficients are the pseudo{-}data and we take a specific BSM model, compute the corresponding
low{-}energy limit and choose the BSM parameters to be those appropriate for the pseudo{-}data.

\item[\dnumb] SMEFT assumes a Higgs doublet, so any mixing among scalars (in general among heavy and light degrees of 
freedom) in the high{-}energy theory brings us to the HEFT/SMEFT dichotomy. 
Although there is a wide class of BSM models that support the linear SMEFT description, this realization does not always 
provide the appropriate framework.

\item[\dnumc] When looking at fitting the data and interpreting the results, several steps should be kept in mind.
Fitting requires a basis, something that everyone agrees on using, but any XEFT description truncated at $\mrdim = 6$ 
level depends on many choices; in particular a `` $\mrdim = 6\,$-quadratic'' description can be defined in many 
ways (for instance by taking into account canonical normalization of the Lagrangian, scheme dependence in the 
renormalization procedure or double insertion of $\mrdim = 6$ operators), not only by the inclusion of $(\mrdim = 6)^2$ 
terms.
A basis is defined by eliminating redundant operators through field redefinitions (akas, equation of motion), however   
the equivalence of two operators is a property of XEFT while it is possible for the $X^{\prime}$ theory to generate 
one but not the other; so, in short, we may need to replace an operator involving a field (appearing in the
$\mrdim = 6$ XEFT basis) with an operator not involving that field (appearing in the low energy limit of $X^{\prime}$). 
Interpreting the fits requires to translate the experimental bounds on the Wilson coefficients in one
basis (the XEFT basis) into limits on coefficients in another basis, the latter being computable functions of the 
parameters of $X^{\prime}$. This requires our ability to change basis, a correspondence which is linear only at 
$\mrdim = 6 $ level.

\item[\dnumd] Details regarding the introduction of heavy{-}light contributions from $X^{\prime}$ to XEFT are also important 
and explicit results are provided that highlight why they should not be ignored. Consider the set of one-loop
diagrams in the $X^{\prime}$ theory with light external legs: some of them contain heavy and light internal lines and,
after computing the low energy limit, produce non-local terms which are in correspondence with XEFT computed at
one loop. It is crucial to note that non-local terms (\eg kinematic logarithms) arise from diagrams of the underlying 
theory showing normal thresholds in the low{-}energy region. In this way one obtains comprehensive results showing 
the downside of neglecting XEFT at one-loop, which should be seen as the leading term in the Mellin-Barnes 
expansion of $X^{\prime}$. Heavy{-}light terms describe a multi scale scenario: the light masses, the Mandelstam
invariants characterizing the process and the heavy scale $\Lambda$. 

\item[\dnume] It is now generally accepted that a convenient formulation of SMEFT requires the so-called ``canonical
normalization'' which should be understood to mean ``proper normalization of the sources entering the $\mrS\,$-matrix''.
This procedure calls into question how we perform ``gauge fixing'' and how ``gauge invariance'' is preserved. 
We can simultaneously redefine couplings, mixing angles and masses so that the part of the XEFT Lagrangian which 
is quadratic in the fields has the same functional form of the original $X$ Lagrangian but with transformed fields 
and parameters. This redefinition will change how we set the $X$ parameters input. 

\end{itemize}

The paper is organized as follows. \autoref{GenA} introduces the problem.
In \autoref{exaEoM} we discuss EoMs. In \autoref{seft} -- \autoref{SMEFT} we analyze different types of EFTs, from a
simple theory containing one, real, scalar field to SMEFT. The question of redundancy is discussed and other relevant 
issues are examined, \eg canonical normalization of the EFT, mixing among scalars in $X^{\prime}$, gauge invariance 
after canonical normalization, gauge fixing.
In \autoref{tsp} -- \autoref{eYm} we study the so-called heavy{-}light effects when going from $X^{\prime}$ to XEFT and 
also their interplay with the NLO extension~\cite{Passarino:2016pzb} of XEFT. 
A final issue concerns operators belonging to the class $\upphi^2\,\Box^n$ and the interpretation of 
derivative-coupled field theories that we discuss in \autoref{Matt}. 
Our conclusions are drawn in the final section.
\section{General aspects of XEFT  \label{GenA}}
In this section we give a summary of the main ingredients which are needed to study weakly interacting modifications 
of the Higgs sector and the electroweak gauge sector of the SM.
\paragraph{The heavy scale} \hspace{0pt} \\
The choice of $\Lambda$~\footnote{For the role of $\Lambda$ in two types of empirically equivalent EFTs,
Wilsonian EFTs and continuum EFTs see \Bref{Bain2013-BAIEFT}.}
\hyperref[EFTex]{in Eq.(\ref*{EFTex})} is crucial when going from $X^{\prime}$ to $X$; more details on this topic
can be found in  \Brefs{Freitas:2016iwx,Biekotter:2016ecg,Boggia:2016asg}. For instance, in any BSM model (\eg 
the singlet extension of the SM) the scale $\Lambda$ should not be confused with the mass of the heavy Higgs boson
(or with masses of new particles) --  it is generally a ratio of mass and powers of couplings. 
It is also relevant to observe that the low energy behavior of $X^{\prime}$ should be computed in the mass eigenbasis, 
not in the weak eigenbasis.
\paragraph{Mixing} \hspace{0pt} \\
The important point to stress is that the EFT extension of $X$, as defined in \eqn{EFTex}, requires absence of mixing
between heavy and light degrees of freedom. The argument is simple, consider the singlet extension of the SM 
(HSESM) where we have one scalar doublet and one singlet; the SM scalar field $\upPhi$ (with hypercharge $1/2$) is
\bq
\upPhi = \frac{1}{\sqrt{2}}\,\Bigl(
\begin{array}{c}
\Ph_2 + \sqrt{2}\,\mrv + i\,\upphi^0 \\
\sqrt{2}\,i\,\upphi^- 
\end{array}
\Bigr)
\label{Hdoublet}
\eq
while the singlet is $\upchi = 1/\sqrt{2}\,(\Ph_1 + \mrv_s)$. There is a mixing angle such that
\bq
\Ph = \cos\alpha\,\Ph_2 - \sin\alpha\,\Ph_1 \spc
\qquad
\PH = \sin\alpha\,\Ph_2 + \cos\alpha\,\Ph_1 \spc
\eq
are the mass eigenstates, one light Higgs ($\Ph$) and one heavy Higgs ($\PH$). 
\paragraph{Mixing and gauge invariance} \hspace{0pt} \\
Of course the Lagrangian remains gauge invariant but gauge invariance of the low energy theory is more complicated 
since we integrate the $\PH$ field, and $\Ph$ does not transform as the SM Higgs boson, see sect.~$3.5$ of 
\Bref{Boggia:2016asg}. For instance $\Lag(\Ph\,,\,\PH = 0)$ alone is not invariant; of course the 
full $\Lag(\Ph\,,\,\PH)$ is gauge invariant, but $\Lag(\Ph\,,\,0)$ is not. If the doublet is the one  
\hyperref[Hdoublet]{in Eq.(\ref*{Hdoublet})} we introduce
\bq
\sin\alpha= \frac{\Delta_s}{\Lambda} + \ord{\Lambda^{-2}} \spc
\qquad
\cos\alpha= 1 - \frac{1}{2}\,\frac{\Delta^2_s}{\Lambda^{2}} + \ord{\Lambda^{-4}} \spc
\eq
where $\Delta_s$ is a function of the parameters of the HSESM Lagrangian. At $\ord{1/\Lambda^2}$ we obtain
\bq
\upPhi_{\sPh} = \frac{1}{\sqrt{2}}\,\Bigl(
\begin{array}{c}
\Ph + \sqrt{2}\,\mrv + i\,\upphi^0 \\
\sqrt{2}\,i\,\upphi^- 
\end{array}
\Bigr)
= \upPhi - \frac{1}{\sqrt{2}}\,\frac{\Delta_s}{\Lambda}\,(\Ph_1 + \frac{1}{2}\,\frac{\Delta_s}{\Lambda}\,\Ph_2)\,
\Bigl(
\begin{array}{c}
1 \\ 0
\end{array} \Bigr) \spc
\eq
and $\upPhi_{\sPh}$ does not transform as a doublet. The infinitesimal transformations are:
\bqa
\Ph &\to& \Ph + \frac{1}{2}\,g\,\cos\alpha\,\bigl( \frac{\Gamma^{\PZ}}{\ctW}\,\upphi^0 + 
         \Gamma^-\,\upphi^+ + \Gamma^+\,\upphi^- \bigr) \spc
\nl
\PH &\to& \PH + \frac{1}{2}\,g\,\sin\alpha\,\bigl( \frac{\Gamma^{\PZ}}{\ctW}\,\upphi^0 + 
         \Gamma^-\,\upphi^+ + \Gamma^+\,\upphi^- \bigr) \spc
\nl
\upphi^0 &\to& \upphi^0 - \frac{1}{2}\,g\,\frac{\Gamma_{\PZ}}{\ctW}\,(\cos\alpha\,\Ph + \sin\alpha\,\PH +
          2\,\frac{\mw}{g}) + \frac{i}{2}\,g\,(\Gamma^-\,\upphi^+ - \Gamma^+\,\upphi^-) \spc
\nl
\upphi^- &\to& \upphi^- - \frac{1}{2}\,g\,\Gamma^-\,(\cos\alpha\,\Ph + \sin\alpha\,\PH +
         2\,\frac{\mw}{g} + i\,\upphi^0) +
         \frac{i}{2}\,g\,\Bigl[(\ctWs - \stWs)\,\frac{\Gamma_{\PZ}}{\ctW} + 2\,\stW\,\Gamma^{\PA}
         \Bigr]\,\upphi^- \spc
\label{SESMt}
\eqa
where $\Gamma^i, i=\PZ\,,\,\PA\,,\,\pm$ are the parameters of the infinitesimal transformation and $\ctW(\stW)$ is the 
cosine(sine) of the weak mixing angle. To write the expansion \hyperref[SESMt]{of Eq.(\ref*{SESMt})}
we introduce 
\bq
\upPhi^{\dag}_{\sPh} = ( \Ph,\,\upphi^0,\,\upphi^-,\,\upphi^+ ) \spc
\qquad
\upGamma^{\dag} = ( 0,\,\frac{\Gamma_{\PZ}}{\ctW},\,\Gamma^-,\,\Gamma^+ ) \spc
\qquad
\mrt_1 = \Bigl(
\begin{array}{ll}
\Gamma^- & \Gamma^+ \\
0 & 0 \\
\end{array}
\Bigr) \spc
\qquad
\mrt_2 = \Bigl(
\begin{array}{lr}
\Gamma^- & \Gamma^+ \\
i\,\Gamma^- & - i\,\Gamma^+ \\
\end{array}
\Bigr) \spp
\eq
and also
\bq
\mrT_{\mrd} = \Bigl(
\begin{array}{lr}
i\,\frac{\Gamma_{\PZ}}{\ctW}\,\tau_2 & \mrt_2 \\
- \mrt^{\dag}_2 & - i\,X\,\tau_3 \\
\end{array}
\Bigr) \spc
\qquad
\mrT_{\mrn\mrd} = \Bigl(
\begin{array}{lr}
i\,\frac{\Gamma_{\PZ}}{\ctW}\,\tau_2 \quad & \quad \mrt_1 \\
- \mrt^{\dag}_1 \quad & \quad 0 \\
\end{array}
\Bigr) \spc
\eq
where $\tau_a$ are Pauli matrices and $X = (\ctWs - \stWs)\,\Gamma_{\PZ}/\ctW + 2\,\stW\,\Gamma_{\PA}$. 
At $\ord{1/\Lambda^2}$ we obtain
\bqa
\upPhi_{\sPh} &\to& 
{\overbrace{\upPhi_{\sPh} - \mw\,\upGamma  + \frac{g}{2}\,\mrT_{\mrd}\,\upPhi_{\sPh}}^{\mbox{doublet}}} - 
\frac{g}{4}\,\frac{\Delta^2_s}{\Lambda^2}\,\mrT_{\mrn\mrd}\,\upPhi_{\sPh} -
\frac{g}{2}\,\frac{\Delta_s}{\Lambda}\,\upGamma\,\PH \spc
\nl
\PH &\to& \PH - \frac{g}{2}\,\frac{\Delta_s}{\Lambda}\,\upGamma^{\dag}\,\upPhi_{\sPh} \spp
\eqa 

Working at $\ord{1/\Lambda^2}$ we can split the total Lagrangian into
\bq
\Lag_{\sPH=0} = \Lag_{\mySM}(\Ph) + \sum_{n=0,2}\,\Lambda^{2\,n - 2}\,\delta \Lag_{6 - 2\,n} \spc
\quad
\Lag_{\sPH} \to \Lag^{\TG} + \Lag^{\LG} + \Lag^{\beta} \spp
\label{SESMcompo}
\eq
The sum over $n$ in $\Lag_{\sPH=0}$ is due to the expansion of $\sin\alpha(\cos\alpha)$ in terms of $\Lambda$.
After integrating out $\PH$ in the $\PH\,$-dependent Lagrangian we will have a tree generated effective Lagrangian, 
$\Lag^{\TG} = \Lag^{\TG}_0 + \ord{1/\Lambda^2}$, a loop generated one and the tadpole contributions; for instance, 
$\Lag_{\sPH=0}$ is not invariant under the transformation of \eqn{SESMt} with $\alpha = 0$ (no mixing) but the sum 
of $\Lag_{\sPH=0}$ and of $\Lag^{\TG}_0$ restores invariance at $\ord{1/\Lambda^2}$. The procedure can be iterated 
order-by-order, \ie the gauge transformations may be seen as generating new vertices in the theory and gauge 
invariance requires that, for any Green’s function, the sum of all diagrams containing one $\Gamma\,$-vertex cancel. 
When sources are added to the Lagrangian the field transformation generates special vertices that are used to 
prove equivalence of gauges and simply-contracted Ward-Slavnov-Taylor 
identities~\cite{Veltman:1970nh,Taylor:1971ff,Slavnov:1972fg}. 
Therefore, for any ``transformed'' Green’s function we integrate the $\PH$ field and, order-by-order in
$1/\Lambda$ (including terms due to the expansion of the mixing), terms containing one $\Gamma\,$-vertex continue 
to cancel (and WST identities to be valid). 

The crucial point is that there is a substantial difference between integrating out~\cite{Bilenky:1993bt,Chiang:2015ura} 
the $\Ph_1$ field (as often done in the literature) or the $\PH$ field. The first choice overlooks the mixing 
but $\sin\alpha$ is also a function of $\Lambda$ and the limit $\Lambda \to \infty$ (as well as the choice of 
$\Lambda$) should be taken consistently~\cite{Boggia:2016asg}. 
Furthermore, SMEFT at $\ord{1/\Lambda^2}$ reproduces the effect of scalar mixing on interactions involving one 
light Higgs scalar, but fails otherwise~\cite{Gorbahn:2015gxa,Dawson:2017vgm}; additionally, the difference between 
integrating out $\Ph_1$ or $\PH$ is at the level of $\mrdim = 8$ operators, \ie of $\ord{1/\Lambda^4}$, and should not
be neglected when including $\mrdim = 8$ terms.  

Examples of local operators appearing in this context are
$\upPhi^6_{\sPh}$ or $\upPhi^2_{\sPh}\,\Box\,\upPhi^2_{\sPh}$ where $\upPhi^2_{\sPh} = \Ph^2 + \upphi^0\,\upphi^0 +
2\,\upphi^+\,\upphi^-$ ($O(4)$ invariant) and they should not be confused with $\Ope_{\upphi}$ and $\Ope_{\upphi\,\Box}$ 
of the the basis commonly referred to as the ``Warsaw basis''~\cite{Grzadkowski:2010es},
the latter being built with the $SU(2)\,\times\,U(1)$ scalar doublet.

To summarize, if $X$ is the SM and we have in mind an $X^{\prime}$ without mixing, then the EFT extension of the SM is 
what we call SMEFT. It should be clear that a geometric formulation of the so-called 
HEFT~\cite{Gavela:2014uta,Brivio:2013pma,Alonso:2015fsp,Buchalla:2015qju,Sanz-Cillero:2017jhb} seems the most promising 
road for general mixings in the scalar sector. For the developing of SMEFT and HEFT approaches, that are consistent 
versions of EFTs, systematically improvable with higher order corrections, see \Bref{Brivio:2017vri}. 

As a side note, there are other approaches where the HSESM Lagrangian is augmented with higher dimensional terms,
therefore assuming a different underlying UV completion, and mixing effects are studied in this wider 
context~\cite{Dawson:2016ugw}. 

Examples of extensions of the SM with general new vector bosons have been discussed in 
\Bref{delAguila:2010mx} where the full SM gauge symmetry has been used to classify the extra vectors, see their
tab.~$2$. It is worth noting that in their classification mass mixing terms of SM and new vectors are forbidden by gauge
invariance; however, there could be interactions with the Higgs doublet that give rise to mass mixing of the
$\PZ$ and $\PW$ bosons with the new vectors when the electroweak symmetry is broken.
The case with general scalar, spinor and vector field content and arbitrary interactions can be found
in \Brefs{delAguila:2000rc,delAguila:2008pw,deBlas:2014mba,deBlas:2017xtg}.

More informations on extra scalar fields and mixing are discussed in \Brefs{Haber:2018ltt,Grzadkowski:2018ohf} where
the notion of (approximate) alignment is discussed. For instance, exact tree-level Higgs alignment is satisfied in 
the inert doublet model~\cite{Barbieri:2006dq} (IDM); in the IDM, the SM Higgs boson resides entirely in one of the 
two scalar doublets.

The question of gauge invariance can also be illustrated by starting with the SM Lagrangian and by integrating out 
the massive electroweak gauge bosons, the Higgs boson, and the top quark fields. The gauge group of the resulting 
low-energy effective field theory (LEFT) is QCD$\;\times\;$QED~\cite{Jenkins:2017jig,Jenkins:2017dyc} and the
photon is not the $U(1)$ field in $SU(3)\,\times\,SU(2)\,\times\,U(1)$. Stated differently, $\PW/\PZ$ are integrated out,
not the $SU(2)$ fields. 
\paragraph{Proliferation of scalars and mixing} \hspace{0pt} \\
The lack of discovery of beyond-the-SM (BSM) physics suggests that the SM is ``isolated''~\cite{Wells:2017aoy},
including small mixing between light and heavy scalars. 
The small mixing scenario raises the following question: if there are may scalars then we have to assume that 
there is at least the same small mixing for every one of them. This is no longer accidental but systematic, and
so must involve a principle, such as a symmetry or some other restriction to the theory that enforces the small mixing. 
This principle is unknown, see \Bref{Wells:2016luz} for a detailed discussion.
\paragraph{Bases, over-complete sets} \hspace{0pt} \\
We now return to the XEFT \hyperref[EFTex]{defined by Eq.(\ref*{EFTex})}, in particular SMEFT; 
the main emphasis will be on consistency of
the theory and not on phenomenological applications. Buildind any XEFT means promoting a theory with a finite number of 
terms into an effective field theory with an infinite number of terms and it is important to establish its 
consistency, both observational and mathematical consistency. Let us return to the ``infinite number of terms'': 
including every operator allowed by $\mrG$ gives over-complete sets of $d\,$-dimensional operators, the well-known 
problem of redundancy. 

Given two operators, $\Ope^{(d)}_i$ and $\Ope^{(d)}_j$ the set $\mrV^d$ is over-complete if all elements of the 
$\mrS\,$-matrix depend on one linear combination of $\mra^d_i$ and $\mra^d_j$.
\paragraph{Removal of redundant operators} \hspace{0pt} \\
There are several sources of degeneracy, operators differing for a total derivative (IPB), operators related by
Fierz identities, Bianchi identities \etc
Finally, we have the so-called EoM (equation of motion) degeneracy~\footnote{General aspects of EoM for EFTs have 
been discussed in \Brefs{Barzinji:2018xvu,Helset:2018dht}.}; usually we find statements like ``by using EoM we
can remove $\dots$'', meaning that many linear combinations of operators ``vanish by the Equations
of Motion''. The original analysis of $\mrdim = 6$ operators can be found in \Bref{Buchmuller:1985jz}; the question of
redundancy has been initiated in \Bref{Grzadkowski:2010es}, the so-called Warsaw basis, and continued in 
\Brefs{Simma:1993ky,Politzer:1980me,Arzt:1994gp,Burgess:2007pt,Goldberger:2007hy,Skiba:2010xn,Einhorn:2013kja,Dedes:2017zog,Bakshi:2018ics,Gripaios:2018zrz,Criado:2019ugp}.
Extension to $\mrdim = 8$ with the introduction of novel techniques (Hilbert series) can be found in
\Brefs{Lehman:2015via,Lehman:2015coa,Henning:2015daa,Henning:2015alf,Henning:2017fpj}.
If one is interested in the $\mrdim = 6$ basis then the necessary EoMs are going to be used at $\ord{1/\Lambda^2}$, \ie
we can derive them from $\Lag^{(4)}$ alone. This last statement, taken out of context, creates the impression that 
the $\mrdim = 8$ basis requires EoMs used naively at $\ord{1/\Lambda^4}$; more details will be given in \autoref{FTEoM}.

To summarize: we can remove redundant operators directly in the Lagrangian, instead of having the ``cancellation'' 
occur when the $\mrS\,$-matrix element is constructed. To some extent we can compare this removal to the use of
the 't Hooft{-}Feynman gauge instead of the more general $\mrR_{\xi}$ gauge; since we know that the $\mrS\,$-matrix
elements do not depend on $\xi$ the calculation is more conveniently performed by starting with $\xi = 1$. On
the other hand, keeping an arbitrary $\xi$ provides a powerful check on the final result.
It has been pointed out in \Bref{Wudka:1994ny} that, even if the $\mrS\,$-matrix elements cannot distinguish between two
equivalent operators $\Ope_1$ and $\Ope_2$, there is a large quantitative difference whether the underlying theory can
generate $\Ope_2$ or not. It is equally reasonable not to eliminate redundant operators and, eventually, exploit
redundancy to check $\mrS\,$-matrix elements. 
In any case, redundancy means that we can only ``measure'' $\mra_1 + \mra_2$, where $\mra_{1,2}$ are Wilson coefficients.
Selecting $\mra_1 = 0$ or $\mra_2 = 0$ defines different bases, \ie the equivalence relation can be used to 
partition operators into equivalence classes, from which basis operators are selected. 
\paragraph{PTG and LG} \hspace{0pt} \\
Additional selection criteria have been introduced in \Bref{Einhorn:2013kja}, particularly that a basis should be 
chosen from among Potentially-Tree-Generated (PTG) operators (as compared to LG, Loop-Generated operators). In
\Bref{Einhorn:2013kja} it is shown that the SM Warsaw basis~\cite{Grzadkowski:2018ohf} for $\mrdim = 6$ operators 
satisfies this criterion. Note that suppressing operators containing field strengths by loop factors was
shown to not be a model independent EFT statement in \Bref{Jenkins:2013fya}. 
However, it has been pointed out that the classes of operators generated at a given loop level do
not form a vector subspace, in general~\cite{Gripaios:2015qya}.
The appearance of an holomorphic structure has been discussed in \Brefs{Alonso:2014rga,Cheung:2015aba}.
 
There is more: the introduction of ``phenomenological truncation'' will introduce 
ambiguities~\cite{Gripaios:2015qya,Gripaios:2018zrz}.
For instance, given a basis of operators one could then try to truncate by retaining only operators featuring a 
limited set of fields, $\{\mrf\}$, in that basis. But such a truncation certainly does not correspond to the
class of theories with new physics in $\{\mrf\}$. If we change to a basis in which we replace an operator involving a
field $\in \{\mrf\}$ with an operator not involving that field, then the truncated space of physical operators that 
we obtain will also change.

In \autoref{exaEoM} we will review the notion of higher order compensation of a $\mrd\,$-dimensional redundant
operator. Here we need only recall that removing a redundant $\mrdim = 6$ operator with a Wilson coefficient
$\mra^6_{\ssR}$ will propagate $\mra^6_{\ssR}$ into the Wilson coefficients of $\mrdim = 8$ operators. In the bottom-up
approach it does not matter since we only ``measure'' combinations of Wilson coefficients, linear in the
$\mra^8_i$ coefficients and quadratic in $\mra^6_{\ssR}$. 
Indeed, when constructing the original EFT, one must include all possible operators consistent with the symmetries 
at every order in the $1/\Lambda$ expansion. Thus, the shift due to the field redefinition can be absorbed into the 
coefficients of operators that are already present in the theory.
However, the low energy limit of the underlying theory may contain some $\Ope^{(8)}_{\ssX}$ as well as some 
$\Ope^{(6)}_{\ssR}$ whose $\mrdim = 8$ compensations contain $\Ope^{(8)}_{\ssX}$; the Wilson coefficient $\mra^8_{\ssX}$ 
is now computable in terms of the parameters of the underlying theory but what we ``measure'' at low energy is 
not $\mra^8_{\ssX}$. 

In the previous argument and for the sake of simplicity we have neglected the mixing among 
Wilson coefficients: in dimensional regularization, the UV divergences are of the form $1/(\mrd - 4) + \ln$; after 
introducing counterterms the finite part of the coefficient is fixed by matching (NLO) while the coefficient of the
logarithm is the anomalous dimension of the operator coefficient, telling us how the coefficient runs with scale. 
Not only do operator coefficients ``run'', like couplings, they can also``mix'' into each other.
This is true for renormalizable, or not, theories; if the underlying theory is not known, the effective theory still 
has to be renormalized and infinities absorbed in the various couplings or Wilson coefficients~\cite{Ghezzi:2015vva}.
Of course, the great advantage of (strictly) renormalizable theories, is that there are a finite number of terms in 
the Lagrangian.

\[
\begin{array}{ccc}
X^{\prime}     &  \longrightarrow  &  \Ope^{(8)}_{\ssX} \\
\downarrow   &  \nearrow         &  \\
\Ope^{(6)}_{\ssR}  &  \longrightarrow  &  \Ope^{(6)}_{\ssY}\,,\,\Ope^{(6)}_{\ssZ}\,\dots \\
\end{array}
\]
What we are discussing here should not be confused with the fact that one can constrain from data in the SMEFT without 
reference to any ultraviolet (UV) completion, \ie  we can certainly treat SMEFT as a real 
field theory~\cite{Passarino:2016pzb,Passarino:2016smg}.
One practical question that needs to be asked is whether or not the results of a $\mrdim= 6\,,\,8$ fit can be 
translated into weakly interacting extensions of the SM. Here many of the theoretical issues 
(can the language of higher-dimensional Lagrangians be efficiently linked to the structure of ultraviolet completions
of the SM gauge and Higgs sectors?) cannot be separated from experimental uncertainties. 
To sum up:
\bei

\item[\dnuma] We want to study weakly interacting modifications of the Higgs sector and the
electroweak gauge sector of the SM.

\item[\dnumb] For each of these models we construct the $\mrdim = 6$ and $\mrdim = 8$ Lagrangians, compute LHC observables
and EWPD, and compare the predictions from this Lagrangian and from the full model.

\item[\dnumc] The important point is that for our classes of models we can derive the structure 
of the Wilson coefficients. To give examples we mention \Brefs{Manohar:2006gz,Manohar:2006ga} and
\Bref{Manohar:2013rga} which introduces an exactly solvable large $\mrN$ model which reduces at low energies to the 
SM plus the dimension six Higgs-gauge operators that are classified as 
$\Ope_{\upphi\,\sPB},\,\Ope_{\upphi\,\sPW},\,\Ope_{\upphi\,\sPW \sPB}$ and $\Ope_{\sPW}$ in
\Bref{Grzadkowski:2018ohf}; the corresponding Wilson coefficients are given in Eqs.~$26{-}27$ of \Bref{Manohar:2013rga}.
Can we derive the size of the Wilson coefficients from the fits?  
Stated differently~\cite{Henning:2014wua}, can we interpret precision measurements as constraints on a given UV model? 
If we can make quantitative statements then the question of higher order compensation of redundant operators becomes
relevant.

\eei
\section{Equation of Motion  \label{exaEoM}}
Although well-known in the literature (it has been explicitly emphasized in \Brefs{Scherer:1994wi,Henning:2017fpj}), 
let us summarize what is meant by ``using EoM''. Consider a Lagrangian $\Lag^{(4)} + \Lag^{(6)}$ containing one real 
scalar field,
\bq
\Lag^{(4)} = - \frac{1}{2}\,\pdmu \upphi\,\pdmu \upphi - \frac{1}{2}\,m^2 \upphi^2  
       - \frac{1}{4}\,\lambda \upphi^4 \spc
\quad
\Lag^{(6)} = - \frac{1}{2}\,\frac{\mra}{\Lambda^2}\,\upphi\, \Box^2\, \upphi = - 
\frac{1}{\Lambda^2}\,\Ope\,\Box \upphi \spp
\eq
If we perform the transformation
\bq
\upphi \to \upphi + \frac{1}{2}\,\frac{\mra}{\Lambda^2}\,\Box \upphi =
\upphi + \frac{1}{\Lambda^2}\,\Ope \spc
\eq
the Lagrangian transforms as $\Lag \to \Lag + \Delta \Lag$, with
\bqa
\Delta \Lag &=& \Bigl[ \frac{\delta \Lag^{(4)}}{\delta \upphi} - \frac{\mra}{\Lambda^2}\,\Box^2 \upphi \Bigr]\,
\frac{1}{\Lambda^2}\Ope +
\frac{1}{2}\,\frac{\delta^2 \Lag}{\delta \upphi^2}\,\frac{1}{\Lambda^4}\,\Ope^2 +
\, \dots =  \frac{\delta \Lag^{(4)}}{\delta \upphi}\,\frac{1}{\Lambda^2}\,\Ope \quad + \quad \mbox{higher orders}  
\nl
{}&=& \frac{1}{2}\,\frac{\mra}{\Lambda^2}\,\upphi\, \Box^2\, \upphi -
\frac{1}{2}\,\frac{\mra}{\Lambda^2}\,(m^2\,\upphi + \lambda\,\upphi^3)\,\Box \upphi \quad + 
\quad \mbox{higher orders} \spp 
\label{EoM}
\eqa
The term of second order in the derivatives cancels in the transformed Lagrangian and the $\mrS\,$-matrix remains
unchanged. For a discussion on the Jacobian of the transformation and the inclusion of a source term see 
\Brefs{Arzt:1993gz,Passarino:2016saj}.
Therefore, ``using EoMs'' means that the redundant operators are of the form
\bq
\frac{1}{\Lambda^n}\,\Ope\,\frac{\delta \Lag^{(4)}}{\delta \upphi} \spp
\eq
Note that we have neglected higher order terms since the goal was constructing the $\mrdim = 6$ Lagrangian.
Of course one could work at second order in $\Lambda^{-2}$, including $\mrdim = 8$ operators, \etc. The result is
\bqa
\Lag &=& \Lag^{(4)} + \Lag^{(6)} + \Lag^{(8)} \spc 
\nl
\upphi \to \upphi + \frac{1}{\Lambda^2}\,\Ope \spc
&\qquad&
\Delta \Lag = \Bigl[ \frac{\delta \Lag^{(4)}}{\delta \upphi} + \frac{\delta \Lag^{(6)}}{\delta \upphi} \Bigr]\,
\frac{1}{\Lambda^2}\,\Ope + \frac{1}{2}\,\frac{\delta^2 \Lag^{(4)}}{\delta \upphi^2}\,
\frac{1}{\Lambda^4}\,\bigl[ \Ope \bigr]^2 \spc
\nl
\upphi \to \upphi + \frac{1}{\Lambda^4}\,\Ope^{\prime} \spc
&\qquad&
\Delta \Lag = \frac{\delta \Lag^{(4)}}{\delta \upphi}\,\frac{1}{\Lambda^4}\,\Ope^{\prime} \spp
\eqa
As a side note, there is a diagrammatic proof of these relations, see sect.~$10.4$ of \Bref{'tHooft:186259}.
Normally we have $\delta \Lag^{(4)}/\delta \upphi = \Box \upphi + \mrF(\upphi)$, where $\mrF$ is a polynomial in $\upphi$
and the redundant operator is
\bq
\frac{1}{\Lambda^n}\,\Ope\,\Box \upphi \spp
\eq
Of course it would be simpler to replace everywhere $\Box \upphi$ with $ - \mrF(\upphi)$ but this is only consistent
at $\ord{1/\Lambda^n}$, \ie neglecting $\ord{1/\Lambda^{n+1}}$ terms.

When XEFT is SMEFT and the group is $\mrG = SU(3)\,\times\,SU(2)\,\times\,U(1)$ we construct $\Lag^{(6)}_{\mySM}$
according to the work of \Bref{Grzadkowski:2010es}, obtaining the so-called Warsaw basis. The extension to higher
dimensional bases is better performed using the tool known as Hilbert 
series~\cite{Henning:2015daa,Henning:2015alf,Henning:2017fpj}, where the output of the Hilbert series is the number 
of invariants at each mass dimension, and the field content of the invariants. One can find in \Bref{Hays:2018zze} 
a translation of the output of the Hilbert series for SMEFT into a format that is useful for calculating Feynman rules. 
Therefore, in \Bref{Hays:2018zze} one can find a form of $\Lag^{(8)}_{\mySM}$ where all redundant operators have been 
removed, after assuming that $\Lag^{(6)}_{\mySM}$ is taken from \Bref{Grzadkowski:2010es}. 

We conclude re-emphasizing that, in fact, the effective field theory can be rewritten by freely using the naive
classical EOM (i.e. without gauge-fixing and ghost terms) derived from $\Lag^{(4)}$, provided one considers only 
on-shell matrix elements at first order in $1/\Lambda$. In particular, the effective Lagrangian may be reduced by 
the EOM before (RG improved) short-distance corrections are evaluated.

In the rest of the paper we are going to discuss the interplay between $\mrdim = 6$ and $\mrdim = 8$ redundances
by means of examples, starting from a scalar theory and ending with the SMEFT.   
\section{Scalar theory: SEFT \label{seft}}
The construction of SEFT starts by considering a Lagrangian
\bq
\Lag^{(4)} = - \frac{1}{2}\,\pdmu \upphi\,\pdmu \upphi - \frac{1}{2}\,m^2 \upphi^2  
       - \frac{1}{4\,!}\,g \upphi^4 \spc
\label{oLag}
\eq
with a symmetry $\upphi \to - \upphi$. How to construct $\Lag^{(6)}$ and $\Lag^{(8)}$? Obviously, we will have
polynomial terms, \ie $\upphi^6$ and $\upphi^8$. As a next step we write all scalar polynomials of 
mass-dimension $6$ and $8$ containing $\upphi$ and its derivatives and reduce them by using IBP identities. 
For $\mrdim = 6$ we could have $\Ope^{(6)}_0 = \upphi^6$ and
\[
\begin{array}{lll}
\Ope^{(6)}_1 = \upphi^3\,\Box \upphi \quad & \quad 
\Ope^{(6)}_2 = \upphi^2\,\pdmu \upphi\,\pdmu \upphi \quad & \quad
\Ope^{(6)}_3 = \upphi\,\Box^2 \upphi \\
&& \\
\Ope^{(6)}_4 = \pdmu \upphi\,(\pdmu \Box \upphi ) \quad & \quad
\Ope^{(6)}_5 = (\Box \upphi )^2 \quad & \quad
\Ope^{(6)}_6 = ( \pdmu \pdnu \upphi )\,( \pdmu \pdnu \upphi) \\
\end{array}
\]
IBP identities give the following relations among $\mrdim = 6$ operators:
\bq
\Ope^{(6)}_6 = - \Ope^{(6)}_3 - 2\,\Ope^{(6)}_4 \spc
\quad
\Ope^{(6)}_5 = \Ope^{(6)}_3 \spc
\quad
\Ope^{(6)}_4 = - \Ope^{(6)}_3 \spc
\quad
\Ope^{(6)}_2 = - \frac{1}{3}\,\Ope^{(6)}_1 \spp
\eq
Note that this solution of the IBP identities is based on the choice of selecting operators with a minimal
number of derivatives but we could have chosen $\Ope^{(6)}_4$ instead of $\Ope^{(6)}_3$. The $\mrdim = 6$ part of the 
Lagrangian is
\bq
\Lag^{(6)} = \frac{1}{\Lambda^2}\,\Bigl[ g^4\,\mra^6_0\,\Ope^{(6)}_0 + g^2\,\sum_{i=1}^2\,\mra^6_i\,\Ope^{(6)}_i +
               \sum_{i=3}^6\,\mra^6_i\,\Ope^{(6)}_i \Bigr] = 
\frac{1}{\Lambda^2}\,\Bigl[ g^4\,\mrab^6_0\,\upphi^6 + \mrab^6_1\,\upphi\,\Box^2 \upphi +
                            g^2\,\mrab^6_2\,\upphi^3\,\Box \upphi \Bigr] \spc 
\label{Ldim6}
\eq
where we have introduced linear combinations $\mra^6_i$ given by
\bq
\mrab^6_0 = \mra^6_0 \spc \quad
\mrab^6_1 = \mra^6_6 + \mra^6_5 - \mra^6_4 + \mra^6_3 \spc \quad
\mrab^6_2 = \mra^6_1 - \frac{1}{3}\,\mra^6_2 \spp
\eq
Note that we have rescaled the Wilson coefficients: in front of an operator $\Ope^{(k)}_i$ of dimension $k$
containing $n$ fields we write
\bq
g^{n-2}\,\frac{\mra^k_i}{\Lambda^{k-4}} \spp
\eq
This rescaling is useful when discussing NLO EFTs and their ``renormalization'', see \Bref{Ghezzi:2015vva}.

For $\mrdim = 8$ we have the following classes of operators respecting $\upphi \to - \upphi$ invariance:
$\upphi^8$, $\upphi^6\,\partial^2$, $\upphi^4\,\partial^4$ and $\upphi^2\,\partial^6$. Applying IBP 
identities we obtain
\bq
\Lag^{(8)} = \frac{1}{\Lambda^4}\,\Bigl\{
         g^6\,\mrab^8_0\,\upphi^8 
+ g^4\,\mrab^8_1\,\upphi^5\,\Box \upphi
+ \mrab^8_2\,(\Box \upphi)\,\Box^2 \upphi
+ g^2\,\Bigl[
  \mrab^8_3\,\upphi^3\,\Box^2 \upphi
+ \mrab^8_4\,\upphi^2\,(\pdmu \pdnu \upphi)\,(\pdmu \pdnu \upphi)
+ \mrab^8_5\,\upphi^2\,(\Box \upphi)^2 \Bigr] \Bigr\} \spc  
\label{Ldim8}
\eq
where the $\mrab^8_i$ coefficients are linear combinations of the original $\mra^8_i$ coefficients.
\subsection{EoM vs. FT  \label{FTEoM}}
To discuss the role of field transformations we consider the \hyperref[Ldim6]{Lagrangian of Eq.(\ref*{Ldim6})} where 
only $\upphi^3\,\Box \upphi$ is kept. The EoM is
\bq
\Box \upphi = m^2\,\upphi + \frac{1}{6}\,g^2\,\upphi^3 - \frac{g^2}{\Lambda^2}\,\mrab^6_2\,(
6\,\upphi\,\pdmu \upphi\,\pdmu \upphi + g^2\,\upphi^5 + 6\,m^2\,\upphi^3) \spp
\eq
replacing the EoM we obtain
\bq
\Lag^{(6)} \to 
\frac{g^2}{\Lambda^2}\,\mrab^6_2\,(m^2\,\upphi^4 + \frac{1}{6}\,g^2\,\upphi^6) -
(\frac{g^2}{\Lambda^2}\,\mrab^6_2)^2\,(g^2\,\upphi^8 + 6\,\upphi^4\,\pdmu \upphi\,\pdmu \upphi + 6\,m^2\upphi^6) \spp
\label{usingEoM}
\eq
The correct way of proceeding is to start from $\Lag^{(4)} + \Lag^{(6)}$ and to perform a field transformation
\bq
\upphi \to \upphi^{\prime} + \frac{1}{\Lambda^2}\,\mrP_1(\upphi^{\prime}) + 
\frac{1}{\Lambda^4}\,\mrP_2(\upphi^{\prime}) \spc
\eq
where $\mrP_{1,2}$ are polynomials in the field. The redundant $\mrdim = 6$ operator is eliminated by the choice
\bq
\mrP_1 = - g^2\,\mra^6_2\,\upphi^3 \spp
\eq
In order to check the correctness of \eqn{usingEoM} we write
\bq
\mrP_2 = \eta\,(g^2\,\mra^6_2)^2\,\upphi^5 \spc
\eq
and obtain
\bq
\Lag^{(6)} \to 
\frac{g^2}{\Lambda^2}\,\mrab^6_2\,(m^2\,\upphi^4 + \frac{1}{6}\,g^2\,\upphi^6) +
\frac{1}{2}\,(\frac{g^2}{\Lambda^2}\,\mrab^6_2)^2\,\Bigl[
(39 - 10\,\eta)\,\upphi^4\,\pdmu \upphi\,\pdmu \upphi -
(1 + 2\,\eta)\,m^2\,\upphi^6 -
6\,g^2\,(3 + 2\,\eta)\,\upphi^8 \Bigr] \spp
\label{usingFT}
\eq
The result is as follows: \eqn{usingEoM} and \eqn{usingFT} give the same result at $\ord{1/\Lambda^2}$ but
differ at $\ord{1/\Lambda^4}$ no matter what the choice of $\eta$ is. Therefore, the correct way of eliminating
the redundant operator corresponds to transform the field at $\ord{1/\Lambda^2}$, but this will introduce
other terms (at $\ord{1/\Lambda^4}$ and higher) which are the higher order compensation of the redundant operator. 
This compensation is clearly different from what we obtain by plugging in the EoM. 

The same argument applies to a redundant operator of the form
\bq
\frac{1}{\Lambda^{n-1}}\,\mra^{n+3}_{\ssF}\,\mrF(\upphi)\Box \upphi \spc
\eq
where $\mrF$ is a polynomial of degree $n$. A transformation of the form
\bq
\upphi \to \upphi^{\prime} - \frac{1}{\Lambda^{n-1}}\,\mra^{n+3}\,\mrF(\upphi^{\prime}) \spc
\eq
eliminates the redundant operator with an higher order compensation of $\ord{\Lambda^{2 - 2\,n}}$ depending on
$\mrF$, $\delta \mrF/\delta\upphi$ and $\delta^2 \mrF/\delta \upphi^2$.

Another example is as follows: given
\bq
\Lag^{(4)} = \frac{1}{2}\,\upphi\,(\Box - m^3)\,\upphi - {\overline{\uppsi}}\,\sla{\partial}\,\uppsi \spc
\qquad 
\Lag^{(6)} = g^2\,\frac{\mra}{\Lambda^2}\,{\overline{\uppsi}}\,\uppsi\,\Box \upphi^2 \spc
\eq
we define 
\bq
\Lag_{\red} = \Lag^{(6)} + g^2\,\frac{\mra}{\Lambda^2}\,{\overline{\uppsi}}\,\uppsi\,(\Box - m^2)\,\upphi \spc
\eq
and discover that, at $\ord{1/\Lambda^4}$, $\Lag^{(6)}$ and $\Lag_{\red}$ are not on-shell equivalent.  

To avoid misunderstandings we highlight examples that show explicitly the equivalence of the field redefinitions 
procedure to sub-leading order, to a procedure that uses the EoM to higher order as part of a systematic EFT 
matching~\cite{Manohar:1997qy}.
\subsection{More on field transformation  \label{FT}}
Starting with the Lagrangians of \eqns{Ldim6}{Ldim8} we want to eliminate all the operators containing $\Box^n \upphi$.
This can be achieved by transforming $\upphi$ as follows:
\bq
\upphi \to \upphi^{\prime} +
\frac{1}{\Lambda^2}\,(\mrP^{(6)}_1 + \mrP^{(6)}_2\,\Box \upphi^{\prime}) +
\frac{1}{\Lambda^4}\,(\mrP^{(8)}_1 + \mrP^{(8)}_2\,\Box \upphi^{\prime} +
                      \mrP^{(8)}_3\,\Box^2 \upphi^{\prime} +
                      \mrP^{(8)}_4\,\pdmu \upphi^{\prime}\,\pdmu \upphi^{\prime}) \spp
\label{somet}
\eq
We look for a solution where the $\mrP\,$-polynomials are of the following form:
\[
\begin{array}{ll}
\mrP^{(6)}_1(\upphi) = \eta_1\,g^2\,\upphi^3 \qquad & \qquad
\mrP^{(6)}_2(\upphi) = \eta_2 \\
& \\
\mrP^{(8)}_1(\upphi) = g^2\,(\eta_3\,m^2\,\upphi^3 + \eta_4\,g^2\,\upphi^5) \qquad & \qquad
\mrP^{(8)}_2(\upphi) = \eta_5\,m^2 + \eta_6\,g^2\,\upphi^2 \\
& \\
\mrP^{(8)}_3(\upphi) = \eta_7 \qquad & \qquad
\mrP^{(8)}_4(\upphi) = \eta_8\,g^2\,\upphi
\end{array}
\]
It may simply verified that the solution for the $\eta$ unknowns is 
\bq
\eta_1 = - \frac{1}{2}\,(\mrab^6_2 + \frac{1}{6}\,\mrab^6_1) \spc \qquad
\eta_2 = - \mrab^6_1 \spc
\eq
\bqa
\eta_3 &=& \frac{1}{2}\,\Bigl[ \frac{1}{6}\,\eta_5 - \mrab^8_5 + 3\,\mrab^8_3 + 
                         \frac{1}{2}\,\mrab^8_2 - \mrab^6_1\,\mrab^6_2 - \frac{2}{3}\,( \mrab^6_1 )^2 \Bigr] \spc
\nl
\eta_4 &=& - \frac{1}{3}\,\Bigl[ \frac{1}{6}\,\mrab^8_5 + \frac{1}{2}\,\mrab^8_3 - 
                            \frac{1}{12}\,\mrab^8_2 + \mrab^8_1 - \frac{21}{8}\,(\mrab^6_2)^2 - 
                            6\,\mrab^6_1\,\mrab^6_0 - \frac{3}{8}\,\mrab^6_1\,\mrab^6_2 + 
                            \frac{3}{32}\,( \mrab^6_1 )^2 \Bigr] \spc
\nl
\eta_5 &=& - \mrab^8_2 + 2\,( \mrab^6_1 )^2 \spc
\nl
\eta_6 &=& - \mrab^8_5 + 3\,\mrab^8_3 + \frac{1}{2}\,\mrab^8_2 - 
             \frac{3}{2}\,\mrab^6_1\,\mrab^6_2 - \frac{3}{4}\,( \mrab^6_1 )^2 \spc
\nl
\eta_7 &=& - \mrab^8_2 + \frac{3}{2}\,( \mrab^6_1 )^2 \spc
\nl
\eta_8 &=& \mrab^8_2 + 6\,\mrab^8_3 - 9\,\mrab^6_1\,\mrab^6_2 - 2\,( \mrab^6_1 )^2 \spp
\eqa
After the transformation of \eqn{somet}, the Lagrangian becomes
\bqa
\Lag &=& - \frac{1}{2}\,\pdmu \upphi\,\pdmu \upphi\,\Bigl[ 1 + 2\,(\mrab^8_2 - 2\,(\mrab^6_1)^2)\,\frac{m^4}{\Lambda^4} + 
                 2\,\mrab^6_1\,\frac{m^2}{\Lambda^2} \Bigr]
       - \frac{1}{2}\,m^2\,\upphi^2 
\nl
   {}&-& \frac{1}{24}\,g^2\,\upphi^4\,\Bigl[ 1 - 4\,(3\,\mrab^8_5 - 9\,\mrab^8_3 - \mrab^8_2 + 3\,\mrab^6_1\,\mrab^6_2 + 
                    (\mrab^6_1)^2)\,\frac{m^4}{\Lambda^4} 
                         - 2\,(6\,\mrab^6_2 + \mrab^6_1)\,\frac{m^2}{\Lambda^2} \Bigr]
\nl          
{}&+& \frac{1}{72}\,\frac{g^4}{\Lambda^2}\,\upphi^6\,\Bigl[ (72\,\mrab^6_0 + 6\,\mrab^6_2 + \mrab^6_1) + 
               2\,(5\,\mrab^8_5 - 15\,\mrab^8_3 - 2\,\mrab^8_2 + 12\,\mrab^8_1 - 
                 36\,(\mrab^6_2)^2 - 72\,\mrab^6_1\,\mrab^6_0 - 3\,\mrab^6_1\,\mrab^6_2 + 
                 2\,(\mrab^6_1)^2)\,\frac{m^2}{\Lambda^2} \Bigr]
\nl          
{}&+& \frac{1}{864}\,\frac{g^6}{\Lambda^4}\,\upphi^8\,\Bigl[ 864\,\mrab^8_0 + 8\,\mrab^8_5 - 
                24\,\mrab^8_3 - 4\,\mrab^8_2 + 48\,\mrab^8_1 - 
                2592\,\mrab^6_2\,\mrab^6_0 - 180\,(\mrab^6_2)^2 - 720\,\mrab^6_1\,\mrab^6_0 - 
                36\,\mrab^6_1\,\mrab^6_2 + 3\,(\mrab^6_1)^2 \Bigr]
\nl
{}&-& \frac{1}{4}\,\frac{g^2}{\Lambda^2}\,\upphi^2\,\pdmu \upphi\,\pdmu \upphi\,
                \Bigl[ (2\,(3\,\mrab^8_5 + 3\,\mrab^8_3 + \mrab^8_2 
               - 15\,\mrab^6_1\,\mrab^6_2 - 3\,(\mrab^6_1)^2)\,\frac{m^2}{\Lambda^2} + (6\,\mrab^6_2 + \mrab^6_1) \Bigr]
\nl          
&-& \frac{1}{72}\,\frac{g^4}{\Lambda^4}\,\upphi^4\,\pdmu \upphi\,\pdmu \upphi\,
                        \Bigl[ 40\,\mrab^8_5 - 48\,\mrab^8_3 - 8\,\mrab^8_2 + 240\,\mrab^8_1 - 468\,(\mrab^6_2)^2 - 
                             1440\,\mrab^6_1\,\mrab^6_0 - 180\,\mrab^6_1\,\mrab^6_2 - 3\,(\mrab^6_1)^2 \Bigr]
\nl          
{}&+& 2\,\frac{g^2}{\Lambda^4}\,\upphi\,(\pdmu \pdnu \upphi)\,\pdmu \upphi\,\pdnu \upphi\,
                 \Bigl[ 6\,\mrab^8_3 + \mrab^8_2 - 9\,\mrab^6_1\,\mrab^6_2 - 2\,(\mrab^6_1)^2 \Bigr]
\nl          
{}&+& \frac{g^2}{\Lambda^4}\,\upphi^2\,(\pdmu \pdnu \upphi)^2\,
                   \Bigl[\mrab^8_4 + 6\,\mrab^8_3 + \mrab^8_2 - 9\,\mrab^6_1\,\mrab^6_2 - 2\,(\mrab^6_1)^2 \Bigr] \spc
\label{TLag}
\eqa     
containing the following operators:
\bq
\upphi^6 \spc \quad \upphi^8 \quad
\upphi^2\,\pdmu \upphi\,\pdmu \upphi \spc \quad
\upphi^4\,\pdmu \upphi\,\pdmu \upphi \spc \quad
\upphi\,(\pdmu \pdnu \upphi)\,\pdmu \upphi\,\pdnu \upphi \spc \quad
\upphi^2\,(\pdmu \pdnu \upphi)^2 \spp
\eq
\begin{remark}
In fitting the data we constrain the combinations of coefficients appearing in \eqn{TLag}: after that the Wilson 
coefficients are the pseudo{-}data. When interpreting the results we should remember that the coefficient of 
$\upphi^2\,(\pdmu \pdnu \upphi)^2$ is not $\mrab^8_4$, \etc
Therefore, caution should be used in constructing the coefficients in the $\mrdim = 8$ part of the basis if we want
to extract the parameters of the high{-}energy theory from the ``fitted'' Wilson coefficients. 
An example of results for Wilson coefficients in both the redundant basis generated by the matching calculation,
and the non-redundant Warsaw basis is given in \Bref{Jiang:2018pbd}.
\end{remark}
\subsection{EFT and canonical normalization  \label{cn}}
As a next step we start from the \hyperref[TLag]{Lagrangian of Eq.(\ref*{TLag})}, 
where we find that the kinetic terms have a non-canonical normalization. This fact does not represent a 
real problem, as long as we remember the correct treatment of sources in going from amputated Green's functions 
to $\mrS\,$-matrix elements.

For the time being, we start with a simpler example: consider a Lagrangian
\bq
\Lag = \frac{1}{2}\,\mrZ_{\upphi}^2\,\upphi\,\lpar \Box - m^2 \rpar\,\upphi + 
       g\,{\overline{\upchi}}\,\upchi\,\upphi + \mrZ_{J}\,J\,\upphi + 
       {\overline{\mrK}}\,\upchi + {\overline{\upchi}}\,\mrK \spc
\eq
where $\upphi$ is a scalar field and $\upchi$ a spinor field. Furthermore, we have added the source 
terms; $\mrZ_{\upphi}$ reproduces the effect of higher dimensional operators, \eg when the 
Higgs field is replaced by its VEV or when loop corrections are included. The (properly 
normalized) propagation function for the scalar particle is
\bq
\mrZ_{J}\,J\,\frac{1}{\mrZ^2_{\upphi}\,(p^2 + m^2)}\,\mrZ_{J}\,J \spc
\eq
fixing $\mrZ_J = \mrZ_{\upphi}$. The net effect on the $\mrS\,$-matrix is that, for each
external $\upphi$ line, we have a factor $\mrZ^{-1}_{\upphi}$. Alternatively, we can define a new
field, $\upphi= \mrZ^{-1}_{\upphi}\,\upphi^{\prime}$: the Lagrangian is now
\bq
\Lag^{\prime} =  \frac{1}{2}\,\upphi^{\prime}\,\lpar \Box - m^2 \rpar\,\upphi^{\prime} + 
         g\,\mrZ^{-1}_{\upphi}\,{\overline{\upchi}}\,\upchi\,\upphi^{\prime} + \mrZ^{\prime}_J\,J\,\upphi^{\prime} + \dots 
\label{LSZ}
\eq
so that $\mrZ^{\prime}_J = 1$. However the $\mrS\,$-matrix elements has a factor $\mrZ^{-1}_{\upphi}$ for each 
external $\upphi^{\prime}\,$-line, \eg due to the coupling $g\,\mrZ^{-1}_{\upphi}\,{\overline{\upchi}}\,\upchi\,\upphi^{\prime}$.
This simple example proves that the field redefinition is a matter of taste, the crucial point 
is in the normalization of the source.

The \hyperref[TLag]{Lagrangian in Eq.(\ref*{TLag})} is not directly what we will use to generate Feynman rules and 
canonical normalization plus parameter redefinition~\cite{Grinstein:1991cd,Passarino:2012cb,Heinemeyer:2013tqa,Passarino:2016pzb,Passarino:2016saj,Dedes:2017zog,Hays:2018zze} is introduced:
\bq
\upphi = (1 + \frac{{\overline{m}}^2}{\Lambda^2}\,\xi^6_{\upphi} + 
          \frac{{\overline{m}}^4}{\Lambda^4}\,\xi^8_{\upphi})\,{\overline{\upphi}}
\spc \quad
g = (1 + \frac{{\overline{m}}^2}{\Lambda^2}\,\xi^6_g + \frac{{\overline{m}}^4}{\Lambda^4}\,\xi^8_g)\,{\overline{g}}
\spc \quad
m^2 = (1 + \frac{{\overline{m}}^2}{\Lambda^2}\,\xi^6_m + \frac{{\overline{m}}^4}{\Lambda^4}\,\xi^8_m)\,{\overline{m}}^2
\spp
\eq
In this way, in the limit $\Lambda \to \infty$, we recover the original Lagrangian written in terms of barred fields 
and parameters. We find the following solution:
\[
\begin{array}{lll}
\xi^6_{\upphi} = - \mrab^6_1 \spc\quad & \quad
\xi^6_g = 3\,\mrab^6_1 + 6\,\mrab^6_2 \spc\quad & \quad
\xi^6_m = 2\,\mrab^6_1 \\
&& \\
\xi^8_{\upphi} = - \mrab^8_2 + \frac{3}{2}\,(\mrab^6_1)^2 \spc\quad & \quad
\xi^8_g = 6\,\mrab^8_5 - 18\,\mrab^8_3 + 54\,(\mrab^6_2)^2 + 48\,\mrab^6_1\,\mrab^6_2 + \frac{15}{2}\,(\mrab^6_1)^2
\spc\quad & \quad
\xi^8_m = 2\,\mrab^8_2 \\
\end{array}
\]
Furthermore we introduce
\[
\begin{array}{lll}
\mrat^6_0 = \mrab^6_0 - \frac{1}{12}\,\mrab^6_1 - \frac{1}{2}\,\mrab^6_2 \spc\quad & \quad
\mrat^6_1 = \mrab^6_1  \spc\quad & \quad
\mrat^6_2 = \frac{3}{2}\,\mrab^6_1 + 9\,\mrab^6_2  \\ 
&& \\
\mrat^8_0 = \mrab^8_0 + \frac{1}{216}\,(2\,\mrab^8_5 - 6\,\mrab^8_3 - \mrab^8_2 + 12\,\mrab^8_1) \spc\quad & \quad 
\mrat^8_1 = \frac{1}{36}\,(5\,\mrab^8_5 - 15\,\mrab^8_3 - 2\,\mrab^8_2 + 12\,\mrab^8_1) \spc\quad & \quad \\
&& \\
\mrat^8_2= 2\,\mrab^8_2 + 12\,\mrab^8_3 \spc\quad & \quad
\mrat^8_3= \frac{1}{4}\,(6\,\mrab^8_3 - \mrat^8_2 - 6\,\mrab^8_5) \spc\quad & \quad
\mrat^8_4= \mrab^8_4 + \frac{1}{2}\,\mrat^8_2 \\
\end{array}
\]
obtaining the Lagrangian
\bqa
{\overline{\Lag}} &=&
       - \frac{1}{2}\,\pdmu \barphi\,\pdmu \barphi 
       - \frac{1}{2}\,\barm^2\,\barphi^2
       - \frac{1}{24}\,\barg^2\,\barphi^4 
       + \frac{\barg^4}{\Lambda^2}\,\barphi^6 \, \mrat^6_0
       + \frac{\barg^2}{\Lambda^2}\,\barphi^2\,\pdmu \barphi\,\pdmu \barphi \, \mrat^6_2
\nl
{}&+& \frac{\barg^6}{\Lambda^4} \, \barphi^8\,\Bigl\{
            \mrat^8_0
          - \frac{1}{216}\,\Bigl[ 72\,\mrat^6_1\,\mrat^6_0 - (\mrat^6_1)^2 - 432\,\mrat^6_2\,\mrat^6_0 + 
                         8\,\mrat^6_2\,\mrat^6_1 - 4\,(\mrat^6_2)^2 \Bigr] \Bigr\}
\nl
{}&+& \frac{\barm^2}{\Lambda^4}\,\barg^4 \, \barphi^6\,\Bigl\{
          \mrat^8_1
          + \frac{1}{72}\,\Bigl[ 3\,(\mrat^6_1)^2 - 1152\,\mrat^6_2\,\mrat^6_0 - 
                                 20\,\mrat^6_2\,\mrat^6_1 - 32\,(\mrat^6_2)^2 \Bigr] \Bigr\}
\nl
{}&+& \frac{\barg^2}{\Lambda^4}\,\barphi\,(\pdmu \barphi)\,(\pdmu \pdnu \barphi)\,\pdnu \barphi \, \Bigl[
            \mrat^8_2
          - (\mrat^6_1 - 12\,\mrat^6_2)\,\mrat^6_1 \Bigr]
\nl
{}&+& \frac{\barm^2}{\Lambda^4}\,\barg^2\,\barphi^2\,\pdmu \barphi\,\pdmu \barphi \, \Bigl\{
            \mrat^8_3
          + \frac{1}{4}\,\Bigl[ (\mrat^6_1)^2 - 20\,\mrat^6_2\,\mrat^6_1 - 32\,(\mrat^6_2)^2 \Bigr] \Bigr\}
\nl
{}&+& \frac{\barg^2}{\Lambda^4}\,\barphi^2\,(\pdmu \pdnu \barphi)\,(\pdmu \pdnu \barphi) \, \Bigl[
            \mrat^8_4
          - \frac{1}{2}\,(\mrat^6_1 - 12\,\mrat^6_2)\,\mrat^6_1 \Bigr]
\nl
{}&+& \frac{\barg^4}{\Lambda^4}\,\barphi^4\,\pdmu \barphi\,\pdmu \barphi \, \Bigl\{
            \mrat^8_5 
          + \frac{1}{36}\,\Bigl[ 720\,\mrat^6_1\,\mrat^6_0 - 7\,(\mrat^6_1)^2 + 
                                 32\,\mrat^6_2\,\mrat^6_1 + 104\,(\mrat^6_2)^2 \Bigr] \Bigr\} \spc
\label{fLag}
\eqa
showing the presence of a $\ord{1/\Lambda^4}$ compensation of the $\mrdim = 6$ redundant operators. 
The Lagrangian can be written in terms of $\{\mrc\}$ coefficients,
\bqa
{\overline{\Lag}} &=&
       - \frac{1}{2}\,\pdmu \barphi\,\pdmu \barphi 
       - \frac{1}{2}\,\barm^2\,\barphi^2
       - \frac{1}{24}\,\barg^2\,\barphi^4 
       + \frac{\barg^4}{\Lambda^2}\,\mrc^6_1\,\barphi^6 
       + \frac{\barg^2}{\Lambda^2}\,\mrc^6_2\,\barphi^2\,\pdmu \barphi\,\pdmu \barphi 
\nl
{}&+& \frac{\barg^6}{\Lambda^4}\,\mrc^8_1\,\barphi^8
   + \frac{\barm^2}{\Lambda^4}\,\barg^4\,\mrc^8_2\, \barphi^6
   + \frac{\barg^2}{\Lambda^4}\,\mrc^8_3\,\barphi\,(\pdmu \barphi)\,(\pdmu \pdnu \barphi)\,\pdnu \barphi 
\nl
{}&+& \frac{\barm^2}{\Lambda^4}\,\barg^2\,\mrc^8_4\,\barphi^2\,\pdmu \barphi\,\pdmu \barphi
   + \frac{\barg^2}{\Lambda^4}\,\mrc^8_5\,\barphi^2\,(\pdmu \pdnu \barphi)\,(\pdmu \pdnu \barphi) 
   + \frac{\barg^4}{\Lambda^4}\,\mrc^8_6\,\barphi^4\,\pdmu \barphi\,\pdmu \barphi \spc
\label{fLagc}
\eqa
and the corresponding $\mrS\,$-matrix depends only on the $\mrc\,$-set of coefficients.
\begin{remark}
Suppose that we use \eqn{fLagc} to fit the data and that $\mrc^8_6$ results to be compatible with zero.
Next, we consider an extension 
\hyperref[oLag]{of the original Lagrangian of Eq.(\ref*{oLag})} depending on a set of parameters $\{\mrp\}$; imagine
that, after computing the low energy limit, we obtain
\bei
\item[--] a set of $\mrdim = 8$ operators, including $\barphi^4\,\pdmu \barphi\,\pdmu \barphi$ with coefficient
$\mrd(\{\mrp\})$;
\item[--] a set of $\mrdim = 6$ operators, some of them redundant if we follow the classification giving \eqn{fLagc}.
\eei
Different extensions turn on different bases, to compare we need to change basis, including the higher order 
compensations; therefore, we cannot conclude that $\mrd(\{p\}) = 0$.  

An alternative solution, though with high overheads, is the following: the low{-}energy Lagrangian is expanded  
in a redundant basis and not in the non-redundant EFT basis but the $\mrS\,$-matrix elements are the same.
Taken as a whole, matching is not performed directly at the Lagrangian level; instead we have to recompute 
the $\mrS\,$-matrix elements process by process. 

Another lesson to be learned is the following: the shift due to the field redefinition at $\mrdim = 6$ level can be 
absorbed into the coefficients of $\mrdim = 8$ operators; however, the construction of the $\mrdim = 8$ basis depends 
on the way we have defined/eliminated the redundant $\mrdim = 6$ operators, \ie the corresponding higher order 
compensations must be part of the $\mrdim = 8$ basis. Furthermore, in NLO EFT (\ie when higher-dimensional operators are
used in loops) new effects will arise. Because of operator mixing, one may encounter UV divergences in the
coefficients of some of the redundant operators, \ie, renormalization fails if we do not include counterterms for 
operators that have been eliminated~\cite{Arzt:1993gz,Einhorn:2001kj}.
\end{remark}
\section{The Abelian Higgs model: AHEFT  \label{AHEFT}}
This section examines the so-called Abelian Higgs model where the Lagrangian is
\bq
\Lag^{(4)} = - (\mrD_{\mu} \upphi)^*\,\mrD_{\mu} \upphi - m^2\,\upphi^*\,\upphi 
             - \frac{1}{2}\,\lambda\,(\upphi^*\,\upphi)^2 - \frac{1}{4}\,\mrF_{\mu\nu}\,\mrF_{\mu\nu} \spc
\label{AHLag}
\eq
with $\mrF_{\mu\nu} = \pdmu \mrA_{\nu} - \pdnu \mrA_{\mu}$ and where the covariant derivative is 
$\mrD_{\mu} = \pdmu - i\,g\,\mrA_{\mu}$. Furthermore, the field $\upphi$ develops a vev,
\bq
\upphi = \frac{1}{\sqrt{2}}\,(\PH + \mrv + i\,\upchi) \spc
\eq
\bq
\mrv = \frac{M}{g} \spc \quad
m^2 = \beta_h - \frac{1}{2}\,\lambda\,\mrv^2 \spc \quad
\lambda = \frac{\mhs}{\mrv^2} = \frac{\mhs}{M^2}\,g^2 \spp 
\label{bhscheme}
\eq
The parameter $\beta_h$ will be used to cancel tadpoles, order-by-order in perturbation theory.
The Lagrangian of \eqn{AHLag} becomes
\bqa
\Lag^{(4)} &=& - \frac{1}{4}\,\mrF_{\mu\nu}\,\mrF_{\mu\nu} 
- \frac{1}{2}\,M^2\,\mrA_{\mu}\,\mrA_{\mu}
- \frac{1}{2}\,\pdmu \PH\,\pdmu \PH
- \frac{1}{2}\,\pdmu \upchi\,\pdmu \upchi
- \frac{1}{2}\,\mhs\,\PH^2
- \frac{M}{g}\,\beta_h\,\PH
\nl
{}&-& \frac{1}{2}\,\beta_h\,(\PH^2 + \upchi^2)
+ M\,\mrA_{\mu}\,\pdmu \upchi 
+ \ord{\mbox{field}^3} \spp
\eqa
Gauge fixing is given by adding a term $- 1/2\,\mrC^2$ to the Lagrangian, with
\bq
\mrC = \frac{1}{\xi}\,\pdmu \mrA_{\mu} - \xi\,M\,\upchi \spp
\label{gft}
\eq
The Warsaw-like basis for $\mrdim = 6$ operators contains
\[
\begin{array}{lll}
\Ope_{\upphi} = (\upphi^*\,\upphi)^3 \spc \quad & \quad
\Ope_{\upphi\,\Box} = (\upphi^*\,\upphi)\,\Box\,(\upphi^*\,\upphi) \spc \quad & \quad
\Ope_{\upphi\,\mrD} = (\upphi^*\,\mrD_{\mu}\,\upphi)^*\,(\upphi^*\,\mrD_{\mu}\,\upphi) \spc \\
&& \\
\Ope_{\upphi\,\mrA} = \upphi^*\,\upphi\,\mrF_{\mu\nu}\,\mrF_{\mu\nu} \spp && \\
\end{array}
\]
To give an example, consider a redundant operator of the following form:
\bq
\Ope_{\ssR} = (\upphi^*\,\upphi)\,(\mrD_{\mu}\,\upphi)^*\,(\mrD_{\mu}\,\upphi) \spp
\eq
How to eliminate it? We can use the relations
\bqa
(\mrD_{\mu}\,\upphi)^*\,\mrD_{\mu}\,\upphi &=& 
- \frac{1}{2}\,\upphi^*\,(\mrD_{\mu} \mrD_{\mu}\,\upphi) -
\frac{1}{2}\,(\mrD_{\mu} \mrD_{\mu}\,\upphi)^*\,\upphi \spc
\nl
(\upphi^*\,\upphi)\,(\mrD_{\mu} \upphi)^*\,(\mrD_{\mu}\,\upphi) &= &
- \frac{1}{2}\,(\upphi^*\,\upphi)\,\Bigl[
\upphi^*\,(\mrD_{\mu} \mrD_{\mu}\,\upphi) + (\mrD_{\mu} \mrD_{\mu}\,\upphi)^*\,\upphi \Bigr] +
\frac{1}{2}\,\Ope_{\upphi\,\Box} \spc
\label{reshape}
\eqa
and given
\bq
\Lag = \Lag^{(4)} + \Lag^{(6)} + g^2\,\frac{\mra_{\ssR}}{\Lambda^2}\,\Ope_{\ssR} \spc
\label{redO}
\eq
we find the solution by performing the transformation
\bq
\upphi \to \upphi - g^2\,\frac{\mra_{\ssR}}{\Lambda^2}\,(\upphi^*\,\upphi)\,\upphi
\spc \qquad
\upphi^* \to \upphi^* - g^2\,\frac{\mra_{\ssR}}{\Lambda^2}\,(\upphi^*\,\upphi)\,\upphi^* \spc
\label{redT}
\eq
generating $(\upphi^*\,\upphi)^2\,\mrF_{\mu\nu}\,\mrF_{\mu\nu}$ etc
\subsection{Fixing the gauge  \label{gf}}
Let us consider again the question of ``choosing the gauge''; general aspects of the problem have been discussed in
\Bref{Helset:2018fgq} while a formal definition can be found in Chapter~$6.4$ of 
\Bref{Costello2011}. Here we want to analyze the interplay between gauge fixing and the
removal of redundant operators. In order to do that we recall a basic aspect of ``continuum EFT'', see
\Bref{Bain2013-BAIEFT}. Given a scale $\Lambda$ the effective theory ($E \muchless \Lambda$) is described by a 
Lagrangian not containing heavy fields:  
\bq
\Lag^{(4)} + \delta\,\Lag \spc
\eq
where $\Delta\,\Lag$ encodes a ``matching correction'' that includes any new nonrenormalizable interactions that may 
be required. The matching correction is made so that the physics of the light fields is the same in
the two theories at the boundary. The aim of a matching calculation is to fix the values of the effective coefficients in
the XEFT Lagrangian so that they reproduce the predictions of the full theory to fixed accuracy,
\bq
\Lag_{\mathrm{\scriptscriptstyle{X^{\prime}}}}(\upphi_{\mathrm{\scriptscriptstyle{X}}}\,,\,\upPhi_{\heavy}) \mapsto 
\Lag_{\XEFT}(\upphi_{\mathrm{\scriptscriptstyle{X}}}) =
\Lag_{\mathrm{\scriptscriptstyle{X}}} + \sum_i\,\mra_i\,\Ope_i(\upphi_{\mathrm{\scriptscriptstyle{X}}}) \spp
\eq
To explicitly calculate $\delta\,\Lag$, one expands it in a complete set of local operators in the same manner that 
the expansion for Wilsonian EFTs is performed.

In Wilsonian EFT, the heavy fields are first integrated out of the underlying high-energy theory and the resulting Wilsonian
effective action is then expanded in a series of local operator terms: in this case the gauge is fixed at the level
of the underlying UV completion.

In the construction of a continuum EFT, the heavy fields are initially left alone in the underlying high-energy theory, 
which is first evolved down to the appropriate energy scale. The continuum EFT is then constructed by completely 
removing the heavy fields from the high-energy theory, as opposed to integrating them out; 
stated differently a continuum quantum field theory is a sequence of low-energy effective actions
$\mrS_{\eff}(\Lambda)$, for all $\Lambda < \infty$, as highlighted in \Bref{Costello2011}.

Having in mind our final example, SM EFT (SMEFT), we can say that any field theory valid above the scale $\Lambda$
(next-SM or NSM) should be based on a gauge group which contains $SU(3)\,\times\,SU(2)\,\times\,U(1)$ and all the SM 
degrees of freedom should be incorporated; furthermore, at low-energies, NSM should reduce to the SM, provided no 
undiscovered but weakly coupled light particles exist. In the top{-}down approach (heavy fields integrated out) the 
gauge is fixed in the NSM Lagrangian.

In our case we start with
\bq
\Lag= \Lag^{(4)}_{\inv} + \Lag^{(6)}_{\inv} + \dots
\eq
\ie with a Lagrangian invariant under transformations belonging to some group $\mrG$; in our case the (infinitesimal)
transformation is
\bq
\upphi \to (1 + i\,g\,\alpha)\,\upchi \spc \qquad
\mrA_{\mu} \to \mrA_{\mu} + \pdmu \alpha \spp
\label{U1inv}
\eq
At this level $\Lag^{(6)}_{\inv}$ may contain redundant operators. We should not apply field transformations, invoking 
the Equivalence Theorem~\cite{Kallosh:1972ap,Arzt:1993gz,Passarino:2016saj} (a statement on the invariance of 
the $\mrS\,$-matrix) since the Lagrangian is singular, \ie there is no $\mrS\,$-matrix. Therefore, we fix the gauge
(for general details see \Bref{Helset:2018fgq} and also \Bref{Misiak:2018gvl}) 
\bq
\Lag= \Lag^{(4)}_{\inv} + \Lag^{(6)}_{\inv} + \Lag_{g.f.} + \Lag_{\ssF\ssP} \spc
\eq
where $\Lag_{\ssF\ssP}$ is the associated ghost Lagrangian,
\bq
\Lag_{\ssF\ssP} = - {\overline{\upeta}}\,\Bigl[ \frac{1}{\xi}\,\Box + \xi\,M\,(M + g\,\PH) \Bigr]\,\upeta \spp
\eq
If we start with the \hyperref[redO]{invariant Lagrangian of Eq.(\ref*{redO})} and perform the 
\hyperref[redT]{transformation of Eq.(\ref*{redT})} then the gauge
fixing term of \eqn{gft} (and the corresponding ghost Lagrangian) will also change since
\bq
\PH \to \PH - \frac{g^2}{2}\,\frac{\mra_{\ssR}}{\Lambda^2}\,({\overline{\PH}}^2 + \upchi^2)\,{\overline{\PH}}
\spc 
\qquad
\upchi \to \upchi - \frac{g^2}{2}\,\frac{\mra_{\ssR}}{\Lambda^2}\,\Bigl[ {\overline{\PH}}^2 +
\upchi^2 \Bigr]\,\upchi \spc
\eq
where ${\overline{\PH}} = \PH + M/g$. However, the $\mrS\,$-matrix does not depend on the choice of the
gauge-fixing term $\mrC$~\cite{tHooft:1972qbu}, \ie will not depend on $\mra_{\ssR}$. 
More details will be given in \autoref{cnAHEFT}.
For another presentation of the problem, based on BRS~\cite{Becchi:1974md,Becchi:1975nq} variation, see 
\Bref{Simma:1993ky} where higher order compensation is also discussed.
\subsection{AHEFT: redundant operators and their higher order compensation  \label{redhoc}}
To summarize, we work with a Lagrangian having $3$ components, $\Lag^{(4)}_{\inv}$, $\Lag_{g.f.} + \Lag_{\ssF\ssP}$ and
\bq
\Lag^{(6)}_{\inv} = 
g^4\,\frac{\mra_{\upphi}}{\Lambda^2}\,\Ope_{\upphi} +
g^2\,\frac{\mra_{\upphi\,\Box}}{\Lambda^2}\,\Ope_{\upphi\,\Box} +
g^2\,\frac{\mra_{\upphi\,\mrD}}{\Lambda^2}\,\Ope_{\upphi\,\mrD} +
g^2\,\frac{\mra_{\ssR}}{\Lambda^2}\,\Ope_{\ssR} +
g^2\,\frac{\mra_{\upphi\,\mrA}}{\Lambda^2}\,\Ope_{\upphi\,\mrA}\spc
\eq
and perform the \hyperref[redT]{transformation of Eq.(\ref*{redT})}. \hyperref[reshape]{Using Eq.(\ref*{reshape})} 
we see that the transformation of $\upphi$ cancels the redundant term but we are still left with the transformation 
of $\mrD_{\mu} \mrD_{\mu}\,\upphi$. This part is
\bq
X = \upphi^*\,\mrD^2\,(\upphi^*\,\upphi\,\upphi) +
    \upphi\,(\mrD^2\,\upphi^*\,\upphi\,\upphi)^* \spc
\eq
and we can easily see that
\bq
X = \Ope_{\upphi\,\Box} - 2\,\Ope_{\upphi\,\mrD} + \;\; \mbox{tot. der.} \spp
\eq
After the transformation the Lagrangian becomes
\bqa
\Lag_{\inv} &=&
- (\mrD_{\mu} \upphi)^*\,\mrD_{\mu} \upphi
+ \frac{1}{2}\,\mhs\,\upphi^*\,\upphi
- g^2\,(\frac{1}{2}\,\frac{\mhs}{M^2} + \mra_{\ssR}\,\frac{\mhs}{\Lambda^2})\,(\upphi^*\,\upphi)^2
\nl
{}&+& \frac{g^4}{\Lambda^2}\,(\mra_{\upphi} + 2\,\frac{\mhs}{M^2}\,\mra_{\ssR})\,\Ope_{\upphi}
+ \frac{g^2}{\Lambda^2}\,(\mra_{\upphi\,\Box} - \frac{1}{2}\,\mra_{\ssR})\,\Ope_{\upphi\,\Box}
+ \frac{g^2}{\Lambda^2}\,(\mra_{\upphi\,\mrD} + 2\,\mra_{\ssR})\,\Ope_{\upphi\,\mrD} 
\nl
{}&-& \frac{1}{4}\,\mrF_{\mu\nu}\,\mrF_{\mu\nu} + \frac{g^2}{\Lambda^2}\,\mra_{\upphi\,\mrA}\,\Ope_{\upphi\,\mrA}
\spc
\eqa
with several terms representing higher order compensations, \eg
\bqa
(\upphi^*\,\upphi)^2 &\to& (\upphi^*\,\upphi)^2 -
4\,\frac{g^2\,\mra_{\ssR}}{\Lambda^2}\,(\upphi^*\,\upphi)^3 +
6\,(\frac{g^2\,\mra_{\ssR}}{\Lambda^2})^2\,(\upphi^*\,\upphi)^4 \spc
\nl
\cdots &\to& \cdots
\nl
(\upphi^*\,\upphi)\,\mrF_{\mu\nu}\,\mrF_{\mu\nu} &\to& (\upphi^*\,\upphi)\,\mrF_{\mu\nu}\,\mrF_{\mu\nu} -
\frac{g^2\,\mra_{\ssR}}{\Lambda^2}\,(\upphi^*\,\upphi)^2\,\mrF_{\mu\nu}\,\mrF_{\mu\nu} \spc
\label{nrLag}
\eqa
that should be an integral part of the $\mrdim = 8$ Lagrangian.
\subsection {Canonical normalization in the AHEFT  \label{cnAHEFT}} 
We discuss canonical normalization for the Abelian Higgs model in the case $\mra_{\ssR} = 0$. In order to deal
with tadpoles it is convenient to define new fields and parameters according to the following equations:
\bq
\mrf = (1 + \frac{\upeta_{\mrf}}{\Lambda^2})\,{\hat{\mrf}} \spc
\qquad
\mrp = (1 + \frac{\upeta_{\mrp}}{\Lambda^2})\,{\hat{\mrp}} \spc
\label{etaco}
\eq
where $\mrf = \PH, \upchi, \mrA_{\mu}$ and $\mrp = g, M, \mh$.
\paragraph{Tadpoles} \hspace{0pt} \\
A special role is played by the $\beta_h\,$-parameter (\hyperref[bhscheme]{defined in Eq.(\ref*{bhscheme})}),
which has to do with cancellation of tadpoles, order-by-order in perturbation theory: this is the so-called
$\beta_h\,$-scheme~\cite{Actis:2006ra}. Alternatively we could use the $\beta_{\mrt}\,$-scheme, as mentioned in
sect.~$2.3.1$ of \Bref{Actis:2006ra}; for a discussion on tadpoles and gauge invariance see 
\Bref{Fleischer:1980ub,Actis:2006ra,Denner:2016etu}. In the $\beta_h\,$-scheme we define
\bq
\beta_h = (1 + \frac{\upeta_{\beta}}{\Lambda^2})\,{\hat{\beta}_h} + \frac{\delta_{\beta}}{\Lambda^2} \spc
\eq
and fix 
\bq
\delta_{\beta} = \frac{3}{4}\,\hat{M}^4\,\mra_{\upphi} \spc
\eq
and the $\eta\,$-coefficients of \eqn{etaco} according to
\[
\begin{array}{llll}
\upeta_{\sPH} = \frac{1}{4}\,{\hat{M}}^2\,(\mra_{\upphi\,\mrD} - 4\,\mra_{\upphi\,\Box}) \spc
\quad & \quad
\upeta_{\upchi} = \frac{1}{4}\,{\hat{M}}^2\,\mra_{\upphi\,\mrD} \spc
\quad & \quad
\upeta_{\mrA} = {\hat{M}}^2\,\mra_{\upphi\,\mrA} \spc
\quad & \quad
\\
&&& \\
\upeta_{\mh} = \frac{3}{2}\,\frac{{\hat{M}}^4}{{\hat{M}}^2_{\sPH}}\,\mra_{\upphi} - \upeta_{\sPH} \spc
\quad & \quad
\upeta_M = \frac{1}{4}\,{\hat{M}}^2\,(\mra_{\upphi\,\mrD} - 4\,\mra_{\upphi\,\mrA}) \spc
\quad & \quad
\upeta_g = - {\hat{M}}^2\,\mra_{\upphi\,\mrA} \spc
\quad & \quad
\upeta_{\beta} = \upeta_g - \upeta_M - \upeta_{\sPH} \spp \\
\end{array}
\]
The part of the invariant Lagrangian containing at most two fields becomes
\bqa
{\widehat{\Lag}}_{\inv} &=&
 - \frac{1}{2}\,\pdmu \hH\,\pdmu \hH
 - \frac{1}{2}\,\pdmu \hchi\,\pdmu \hchi
 - \frac{1}{4}\,{\hat{\mrF}}_{\mu\nu}\,{\hat{\mrF}}_{\mu\nu}
 + \hM\,\hA_{\mu}\,\pdmu \hchi
\nl
{}&-& \frac{1}{2}\,\hMH^2\,\hH^2
 - \frac{1}{2}\,\hM^2\,\hA_{\mu}\,\hA_{\mu}
 - \frac{1}{2}\,{\hat{\beta}}_{hh}\,\hH^2
 - \frac{1}{2}\,{\hat{\beta}}_{h\upchi}\,\hchi^2
\nl
{}&-& \frac{\hM}{\hg}\,{\hat{\beta}}_h\,\hH + \ord{\mbox{field}^3} + \ord{\Lambda^{-4}} \spc
\label{aftercn}
\eqa
where we have introduced auxiliary quantities,
\bq
{\hat{\beta}}_{hh} = {\hat{\beta}}_h\,(1 - \frac{\hM^2}{\Lambda^2}\,\mra_{\upphi\,\Box}) \spc
\qquad
{\hat{\beta}}_{h\upchi} = {\hat{\beta}}_h\,(1 + \frac{\hM^2}{\Lambda^2}\,\mra_{\upphi\,\Box}) \spp
\label{dbeta}
\eq
Therefore, ${\hat{\beta}}_h$ is designed to cancel tadpoles order-by-order while 
${\hat{\beta}}_{hh}$ and ${\hat{\beta}}_{h\upchi}$ contribute to various Green's functions and are crucial for the
validity of the WST identities.
\paragraph{Shifted gauge invariance} \hspace{0pt} \\
The invariance of the full ${\widehat{\Lag}}_{\inv}$ is not ``broken'' but ``shifted'': at $\ord{1/\Lambda^2}$ the
canonical normalized Lagrangian is invariant under the following ``shifted'' transformations:
\bqa
\hH &\to& \hH - \hg\,\Bigl[ 1 - \frac{\hM^2}{\Lambda^2}\,
 (\mra_{\upphi\,\mrA} - \mra_{\upphi\,\Box}) \Bigr]\,\alpha\,\hchi \spc
\nl
\hchi &\to& \hchi + \hM\,(1 - \frac{\hM^2}{\Lambda^2}\,\mra_{\upphi\,\mrA})\,\alpha 
 + \hg\,\Bigl[ 1 - \frac{\hM^2}{\Lambda^2}\,(\mra_{\upphi\,\mrA} + \mra_{\upphi\,\Box}) \Bigr]\,\alpha\,\hH \spc
\nl
\hA_{\mu} &\to& \hA_{\mu} + (1 - \frac{\hM^2}{\Lambda^2}\,\mra_{\upphi\,\mrA})\,\pdmu \alpha \spp
\eqa
The argument can be repeated order-by-order to all orders in inverse powers of $\Lambda$. Consider again
the part of the invariant Lagrangian containing at most two fields which can be written as
\bqa
\Lag_{\inv} &=&
 - \frac{1}{2}\,X_{\ssH}\,\pdmu \PH\,\pdmu \PH
 - \frac{1}{2}\,X_{\upchi}\,\pdmu \upchi\,\pdmu \upchi
 - \frac{1}{4}\,X_{\ssA}\,\mrF_{\mu\nu}\,\mrF_{\mu\nu}
 + M\,X_t\,\mrA_{\mu}\,\pdmu \upchi
\nl
{}&-& \frac{1}{2}\,X_{\mh}\,\mhs\,\PH^2
 - \frac{1}{2}\,X_{\ssM}\,M^2\,\upchi^2
 - \frac{1}{2}\,X_{\ma}\,M^2\,\mrA_{\mu}\,\mrA_{\mu}
 - \frac{M}{g}\,(\beta_h + X_{\beta})\,\PH
 - \frac{1}{2}\,\beta_h\,(\PH^2 + \upchi^2) \spc
\eqa
where $X_{\ssM}$ and $X_{\beta}$ start at $\ord{1/\Lambda^2}$ while ther remaining $X_i$ start at $\ord{1}$. In full
generality we define
\[
\begin{array}{lll}
\PH = \mrZ^{1/2}_{\ssH}\,\hH \spc \quad & \quad
\upchi = \mrZ^{1/2}_{\upchi}\,\hchi \spc \quad & \quad
\mrA_{\mu} = \mrZ^{1/2}_{\ssA}\,\hA_{\mu} \spc \\      
&& \\
M = \mrZ_{\ssM}\,\hM \spc \quad & \quad
\mh = \mrZ_{\mh}\,\hMH \quad & \quad
g = \mrZ_g\,\hg
\end{array}
\]
Furthermore, we introduce $\beta_h= \mrZ_{\beta}\,{\hat{\beta}}_h + \Delta \beta$.
The general definition of canonical normalization is as follows:
$\Delta \beta = - \hM^2\,X_{\ssM}\,\mrZ^2_{\ssM}$ and
\[
\begin{array}{lll}
\mrZ_{\ssH} = X^{-2}_{\ssH} \spc \quad & \quad
\mrZ_{\upchi} = X^{-2}_{\upchi} \spc \quad & \quad
\mrZ_{\ssA} = X^{-2}_{\ssA} \spc \\ 
&& \\
\mrZ^2_{\ssM} = \frac{X_{\ssA}}{X_{\ma}} \spc \quad & \quad
\mrZ^2_{\mh} = (\frac{\hM^2}{\hMH^2}\,\frac{X_{\ssM}}{X_{\ma}}\,X_{\ssA} + X_{\ssH})\,X^{-1}_{\mh} \quad & \quad \\
\end{array}
\]
The EFT correction terms are not independent, due to the invariance of the Lagrangian, \eg
$X_t = X_{\upchi} = X_{\ma}$. Furthermore
\bq
X_{\ma}\,X_{\beta} = M^2\,X_{\ssA}\,X_{\ssM} + \ord{\Lambda^{-4}} \spp
\eq
Clearly, also the interaction part of the Lagrangian is modified by the shift; the transformation law for the fields 
can be written as:
\bqa
\hH &\to& \hH - \bigl( \frac{\mrZ_{\upchi}}{\mrZ_{\ssH}} \bigr)^{1/2}\,\mrZ_g\,\hg\,\hchi\,\alpha \spc
\nl
\hchi &\to& \hchi + \Bigl[ \bigl( \frac{\mrZ_{\ssH}}{\mrZ_{\upchi}} \bigr)^{1/2}\,\mrZ_g\,\hg\,\hH +
\frac{\mrZ_{\ssM}}{\mrZ^{1/2}_{\upchi}}\,\hH \Bigr]\,\alpha \spc
\nl
\hA_{\mu} &\to& \hA_{\mu} + \mrZ^{-1/2}_{\ssA}\,\pdmu \alpha \spp
\eqa
\paragraph{Shifted gauge-fixing} \hspace{0pt} \\
Special attention should be paid to the gauge-fixing term when we canonically normalize the Lagrangian. 
\hyperref[gft]{Instead of Eq.(\ref*{gft})} is more convenient to start with a two-parameter gauge fixing,
\bq
\Lag_{g.f.} = \frac{1}{\xi_{\mrA}}\,\pdmu \mrA_{\mu} - \xi_{\upphi}\,M\,\upchi \spc
\eq
and to define ``shifted'' gauge parameters,
\bq
\xi_{\mrA} = ( 1 + \frac{\upeta_{\xi_{\mrA}}}{\Lambda^2})\,{\hat{\xi}}_{\mrA} \spc
\qquad
\xi_{\upphi} = (1 + \frac{\upeta_{\xi_{\upphi}}}{\Lambda^2})\,{\hat{\xi}}_{\upphi} \spp
\eq
If we set
\bq
\upeta_{\xi_{\mrA}} = \hM^2\,\mra_{\upphi\,\mrA} \spc
\qquad
\upeta_{\xi_{\upphi}} = \hM^2\,(\mra_{\upphi\,\mrA} - \frac{1}{2}\,\mrA_{\upphi\,\mrD}) \spc
\eq
the gauge-fixing part of the Lagrangian becomes
\bq
\Lag_{g.f.} \to {\widehat{\Lag}}_{g.f.} = 
\frac{1}{{\hat{\xi}}_{\mrA}}\,\pdmu \hA_{\mu} - {\hat{\xi}}_{\upphi}\,\hM\,\hchi \spp
\label{gfcn}
\eq
Only at this point we are free to set the ${\hat{\xi}}$ parameters to one (the 't Hooft{-}Feynman gauge) or equal 
to each other (the $\mrR_{\xi}$ gauge). The ghost Lagrangian will change accordingly.
\subsection{AHEFT: including  \texorpdfstring{$\mrdim = 8$}{eight} operators  \label{dim8AHEFT}}
Phenomenological applications for $\mrdim = 8$ operators have been discussed in 
\Brefs{Degrande:2013kka,Senol:2018cks,Hays:2018zze}.

In the Abelian Higgs model we can introduce the set of \hyperref[AHEFTlist]{$\mrdim = 6,8$ operators 
shown in Tab.~\ref*{AHEFTlist}}.
\begin{table}
\begin{center}
\scalebox{0.8}{
\begin{tabular}{llll}
\hline
class &&& \\
\hline
$\upphi^6$ & $\Ope^{(6)}_{\upphi} = \upPhi^3$ && 
\\
&&& \\
$\upphi^4\,\mrD^2$ & 
$\Ope^{(6)}_{\upphi\,\mrD} = (\upphi^*\,\mrD_{\mu} \upphi)^*\,(\upphi^*\,\mrD_{\mu} \upphi)$ &
$\Ope^{(6)}_{\upphi\,\Box} = \upPhi\,\Box\,\upPhi$ & 
\\
&&&\\
$\upphi^2\,X^2$ &
$\Ope^{(6)}_{\upphi\,\mrA} = \upPhi\,\mrF_{\mu\nu}\,\mrF_{\mu\nu}$ &&
\\
&&& \\
\hline
&&& \\
$\upphi^4$ &
$\Ope^{(8)}_{\upphi} = \upPhi^4$ &&
\\
&&& \\
$\upphi^6\,\mrD^2$ &
$\Ope^{(8)}_{\upphi\,\mrD\,;\,1} = \upPhi\,(\upphi^*\,\mrD_{\mu} \upphi)^*\,(\upphi^*\,\mrD_{\mu} \upphi)$ &
$\Ope^{(8)}_{\upphi\,\Box} = \upPhi^2\,\Box\,\upPhi$ & 
\\
&&& \\
$\upphi^4\,\mrD^4$ &
$\Ope^{(8)}_{\upphi\,\mrD\,;\,2} = 
(\mrD_{\mu} \upphi)^*\,(\mrD_{\nu} \upphi)\,(\mrD_{\nu}\,\upphi)^*\,\mrD_{\mu} \upphi$ &
$\Ope^{(8)}_{\upphi\,\mrD\,;\,3} = 
(\mrD_{\mu} \upphi)^*\,(\mrD_{\nu} \upphi)\,(\mrD_{\mu}\,\upphi)^*\,\mrD_{\nu} \upphi$ &
$\Ope^{(8)}_{\upphi\,\mrD\,;\,4} = 
(\mrD_{\mu} \upphi)^*\,(\mrD_{\mu} \upphi)\,(\mrD_{\nu}\,\upphi)^*\,\mrD_{\nu} \upphi$ 
\\
&&& \\
$X^4$ &
$\Ope^{(8)}_{\ssA\,;\,1} = (\mrF_{\mu\nu}\,\mrF_{\mu\nu})^2$ &
$\Ope^{(8)}_{\ssA\,;\,2} = \mrF_{\mu\nu}\,\mrF_{\nu\rho}\,\mrF_{\rho\sigma}\,\mrF_{\sigma\mu}$ &
\\
&&& \\
$X^2\,\upphi^4$ &
$\Ope^{(8)}_{\upphi\,\mrA} = \upPhi^2\,\mrF_{\mu\nu}\,\mrF_{\mu\nu}$ &&
\\
&&& \\ 
$X\,\upphi^4\,\mrD^2$ &
$\Ope^{(8)}_{\upphi\,\mrD\,\mrA\,;\,1} = 
i\,(\upphi^*\,\mrD_{\mu} \upphi)^*\,(\upphi^*\,\mrD_{\nu} \upphi)\,\mrF_{\mu\nu}$ &&
\\
&&& \\
$X^2\,\upphi^2\,\mrD^2$ &
$\Ope^{(8)}_{\upphi\,\mrD\,\mrA\,;\,2} = (\mrD_{\mu} \upphi)^*\,\mrD_{\nu} \upphi\,\mrF_{\mu\rho}\,\mrF_{\nu\rho}$ &
$\Ope^{(8)}_{\upphi\,\mrD\,\mrA\,;\,3} = (\mrD_{\mu} \upphi)^*\,\mrD_{\mu} \upphi\,\mrF_{\rho\sigma}\,\mrF_{\rho\sigma}$ &
\\
&&&\\
\hline
\end{tabular}
}
\end{center}
\caption[]{List of $\mrdim = 6$ and $\mrdim = 8$ operators for the AHEFT theory. We have introduced 
$\upPhi = \upphi^*\,\upphi$.}
\label{AHEFTlist}
\end{table}
where we have introduced $\upPhi^2 = \upphi^*\,\upphi$. The corresponding Wilson coefficients mix with those coming from 
$\mrdim = 8$ compensation of $\mrdim = 6$ redundant operators, \hyperref[redhoc]{as explained in \ref*{redhoc}}.
\paragraph{Canonical normalization up to  $\mrdim = 8$} \hspace{0pt} \\
We introduce the following field and parameter transformations:
\bq
\mrf = (1 + \frac{\upeta^6_{\mrf}}{\Lambda^2} + \frac{\upeta^8_{\mrf}}{\Lambda^4})\,{\hat{\mrf}}
\spc \qquad
\mrp = (1 + \frac{\upeta^6_{\mrp}}{\Lambda^2} + \frac{\upeta^8_{\mrp}}{\Lambda^4})\,{\hat{\mrp}} \spc
\eq
\bq
\beta_h = (1 + \frac{\upeta^6_h}{\Lambda^2} + \frac{\upeta^8_h}{\Lambda^4})\,{\hat{\beta}}_h +
          \frac{\delta\beta^6}{\Lambda^2} + \frac{\delta\beta^8}{\Lambda^4} \spc
\eq
with the \hyperref[AHEFTred]{shifts defined in Tab.~\ref*{AHEFTred}}. 
\begin{table}
\begin{center}
\scalebox{0.8}{
\begin{tabular}{ll}
\hline
& \\
$   \delta\beta^6 =
      \frac{3}{4}\,\mra^6_{\upphi}\,\hM^4$ &
$   \delta\beta^8 =
      \frac{1}{4}\,(2\,\mra^8_{\upphi} - 12\,\mra^6_{\upphi}\,\mra^6_{\upphi\,\mrA} + 
           3\,\mra^6_{\upphi}\,\mra^6_{\upphi\,\mrD})\,\hM^6$ \\
& \\
$   \upeta^6_{\beta} =
      \frac{1}{2}\,(2\,\mra^6_{\upphi\,\Box} - \mra^6_{\upphi\,\mrD})\,\hM^2$ &
$   \upeta^8_{\beta} =
       - \frac{1}{4}\,\Bigl[ \mra^8_{\upphi\,\mrD\,;\,1} - 4\,\mra^8_{\upphi\,\Box} + 
         8\,\mra^6_{\upphi\,\Box}\,\mra^6_{\upphi\,\mrA} + 2\,(\mra^6_{\upphi\,\Box})^2 - 
         4\,\mra^6_{\upphi\,\mrD}\,\mra^6_{\upphi\,\mrA} - 2\,\mra^6_{\upphi\,\mrD}\,
         \mra^6_{\upphi\,\Box} + (\mra^6_{\upphi\,\mrD})^2 \Bigr]\,\hM^4$ \\
& \\
$   \upeta^6_{\mh} =
      \frac{1}{4}\,(4\,\mra^6_{\upphi\,\Box} - \mra^6_{\upphi\,\mrD})\,\hM^2 + 
      \frac{3}{2}\,\mra^6_{\upphi}\,\frac{\hM^4}{\hMH^2}$ &
$   \upeta^8_{\mh} =
       - \frac{1}{32}\,\Bigl[ 4\,\mra^8_{\upphi\,\mrD\,;\,1} - 32\,\mra^8_{\upphi\,\Box} + 
         64\,\mra^6_{\upphi\,\Box}\,\mra^6_{\upphi\,\mrA} + 16\,(\mra^6_{\upphi\,\Box})^2 - 
         16\,\mra^6_{\upphi\,\mrD}\,\mra^6_{\upphi\,\mrA} - 
         24\,\mra^6_{\upphi\,\mrD}\,\mra^6_{\upphi\,\Box} + 5\,(\mra^6_{\upphi\,\mrD})^2 \Bigr]\,\hM^4$ 
\\
 & $+ \frac{3}{8}\,(4\,\mra^8_{\upphi} - 16\,\mra^6_{\upphi}\,\mra^6_{\upphi\,\mrA} - 
         4\,\mra^6_{\upphi}\,\mra^6_{\upphi\,\Box} + 
         5\,\mra^6_{\upphi}\,\mra^6_{\upphi\,\mrD})\,\frac{\hM^6}{\hMH^2} - 
         \frac{9}{8}\,(\mra^6_{\upphi})^2\,\frac{\hM^8}{\hMH^4}$ \\
& \\
$   \upeta^6_{\ssH} =
       - \frac{1}{4}\,(4\,\mra^6_{\upphi\,\Box} - \mra^6_{\upphi\,\mrD})\,\hM^2$ &
$   \upeta^8_{\ssH} =
      \frac{1}{32}\,\Bigl[ 4\,\mra^8_{\upphi\,\mrD\,;\,1} - 32\,\mra^8_{\upphi\,\Box} + 
      64\,\mra^6_{\upphi\,\Box}\,\mra^6_{\upphi\,\mrA} + 48\,(\mra^6_{\upphi\,\Box})^2 - 
      16\,\mra^6_{\upphi\,\mrD}\,\mra^6_{\upphi\,\mrA} - 40\,
      \mra^6_{\upphi\,\mrD}\,\mra^6_{\upphi\,\Box} + 7\,(\mra^6_{\upphi\,\mrD})^2 \Bigr]\,\hM^4$ \\
& \\
$   \upeta^6_{\upchi} =
      \frac{1}{4}\,\mra^6_{\upphi\,\mrD}\,\hM^2$ &
$   \upeta^8_{\upchi} =
      \frac{1}{32}\,\Bigl[ 4\,\mra^8_{\upphi\,\mrD\,;\,1} - 16\,\mra^6_{\upphi\,\mrD}\,\mra^6_{\upphi\,\mrA} + 
      7\,(\mra^6_{\upphi\,\mrD})^2 \Bigr]\,\hM^4$ \\
& \\
$   \upeta^6_{\mrA} =
      \mra^6_{\upphi\,\mrA}\,\hM^2$ &
$   \upeta^8_{\mrA} =
      \frac{1}{2}\,\Bigl[ \mra^8_{\upphi\,\mrA} - (\mra^6_{\upphi\,\mrA})^2 + 
      \mra^6_{\upphi\,\mrD}\,\mra^6_{\upphi\,\mrA} \Bigr]\,\hM^4$ \\
& \\
$   \upeta^6_{\ssM} =
       - \frac{1}{4}\,(4\,\mra^6_{\upphi\,\mrA} - \mra^6_{\upphi\,\mrD})\,\hM^2$ &
$   \upeta^8_{\ssM} =
       - \frac{1}{32}\,\Bigl[ 16\,\mra^8_{\upphi\,\mrA} - 4\,\mra^8_{\upphi\,\mrD\,;\,1} - 
         48\,(\mra^6_{\upphi\,\mrA})^2 + 40\,\mra^6_{\upphi\,\mrD}\,\mra^6_{\upphi\,\mrA} - 
         7\,(\mra^6_{\upphi\,\mrD})^2 \Bigr]\,\hM^4$ \\
& \\
$   \upeta^6_{g} =
       - \mra^6_{\upphi\,\mrA}\,\hM^2$ &
$   \upeta^8_{g} =
       - \frac{1}{2}\,\Bigl[ \mra^8_{\upphi\,\mrA} - 3\,(\mra^6_{\upphi\,\mrA})^2 + 
         \mra^6_{\upphi\,\mrD}\,\mra^6_{\upphi\,\mrA} \Bigr]\,\hM^4$ \\
& \\
$   \upeta^6_{\xi_{\mrA}} =
      \mra^6_{\upphi\,\mrA}\,\hM^2$ &
$   \upeta^8_{\xi_{\mrA}} =
      \frac{1}{2}\,\Bigl[ \mra^8_{\upphi\,\mrA} - (\mra^6_{\upphi\,\mrA})^2 + 
      \mra^6_{\upphi\,\mrD}\,\mra^6_{\upphi\,\mrA} \Bigr]\,\hM^4$ \\
& \\
$   \upeta^6_{\xi_{\upchi}} =
      \frac{1}{2}\,(2\,\mra^6_{\upphi\,\mrA} - \mra^6_{\upphi\,\mrD})\,\hM^2$ &
$   \upeta^8_{\xi_{\upchi}} =
      \frac{1}{4}\,\Bigl[ 2\,\mra^8_{\upphi\,\mrA} - \mra^8_{\upphi\,\mrD\,;\,1} - 
      2\,(\mra^6_{\upphi\,\mrA})^2 + 4\,\mra^6_{\upphi\,\mrD}\,\mra^6_{\upphi\,\mrA} - 
      (\mra^6_{\upphi\,\mrD})^2 \Bigr]\,\hM^4$ \\
& \\
\hline
\end{tabular}
}
\end{center}
\caption[]{List of $\mrdim = 6$ and $\mrdim = 8$ field and parameter redefinitions for the AHEFT theory.}
\label{AHEFTred}
\end{table}
The quadratic part of the Lagrangian remains as in \eqn{aftercn} with the only difference that \eqn{dbeta} is
replaced by
\bqa
{\hat{\beta}}_{hh} &=& {\hat{\beta}}_h\,\Bigl\{
1 - \frac{\hM^2}{\Lambda^2}\,\mra^6_{\upphi\,\Box} +
\frac{\hM^4}{\Lambda^4}\,\Bigl[
  2\,\mra^6_{\upphi\,\Box}\,\mra^6_{\upphi\,\mrA}
- \mra^6_{\upphi\,\Box}\,\mra^6_{\upphi\,\mrD}
+ \frac{3}{2}\,(\mra^6_{\upphi\,\Box})^2
- \mra^8_{\upphi\,\Box} \Bigr] \Bigr\} \spc
\nl
{\hat{\beta}}_{h\upchi} &=& {\hat{\beta}}_h\,\Bigl\{
1 + \frac{\hM^2}{\Lambda^2}\,\mra^6_{\upphi\,\Box} -
\frac{\hM^4}{\Lambda^4}\,\Bigl[
  2\,\mra^6_{\upphi\,\Box}\,\mra^6_{\upphi\,\mrA}
- \mra^6_{\upphi\,\Box}\,\mra^6_{\upphi\,\mrD}
+ \frac{3}{2}\,(\mra^6_{\upphi\,\Box})^2
- \mra^8_{\upphi\,\Box} \Bigr] \Bigr\} \spc
\eqa
\paragraph{Gauge invariance up to  $\mrdim = 8$} \hspace{0pt} \\
After canonical normalization the Lagrangian, up to $\mrdim = 8$ terms (\ie $\ord{1/\Lambda^4}$), is invariant under 
the following set of shifted transformations:
\bqa
\hH &\to& \hH
 - \hg\,\hchi\,\alpha
 + \hg\,\frac{\hM^2}{\Lambda^2}\,(\mra^6_{\upphi\,\mrA} - \mra^6_{\upphi\,\Box})\,\hchi\,\alpha
\nl
{}&+& \frac{1}{2}\,\hg\,\frac{\hM^4}{\Lambda^4}\,\Bigl[
               \mra^8_{\upphi\,\mrA}
          - 2\,\mra^8_{\upphi\,\Box}
          - 3\,(\mra^6_{\upphi\,\mrA})^2
          + 6\,\mra^6_{\upphi\,\Box}\,\mra^6_{\upphi\,\mrA}
          + (\mra^6_{\upphi\,\Box})^2
          + \mra^6_{\upphi\,\mrD}\,\mra^6_{\upphi\,\mrA}
          - 2\,\mra^6_{\upphi\,\mrD}\,\mra^6_{\upphi\,\Box}
          \Bigr]\,\hchi\,\alpha \spc
\nl\nl
\hchi &\to& \hchi
 + (\hg\,\hH + \hM)\,\alpha
 - \frac{\hM^2}{\Lambda^2}\,\Bigl[ \mra^6_{\upphi\,\mrA}\,\hM + (\mra^6_{\upphi\,\mrA} + 
   \mra^6_{\upphi\,\Box})\,\hg\,\hH \Bigr]\,\alpha
 + \frac{1}{2}\,\frac{\hM^4}{\Lambda^4}\,\Bigl\{
    \hM\,\Bigl[ 3\,(\mra^6_{\upphi\,\mrA})^2 - \mra^6_{\upphi\,\mrD}\,\mra^6_{\upphi\,\mrA} - 
    \mra^8_{\upphi\,\mrA} \Bigr]
\nl
{}&+& \Bigl[ 3\,(\mra^6_{\upphi\,\mrA})^2 + 6\,\mra^6_{\upphi\,\Box}\,\mra^6_{\upphi\,\mrA} + 
    3\,(\mra^6_{\upphi\,\Box})^2 - \mra^6_{\upphi\,\mrD}\,\mra^6_{\upphi\,\mrA} - 
    2\,\mra^6_{\upphi\,\mrD}\,\mra^6_{\upphi\,\Box} - \mra^8_{\upphi\,\mrA} - 2\,\mra^8_{\upphi\,\Box}
          \Bigr]\,\hg\,\hH \Bigr\}\,\alpha \spc
\nl\nl
\hA_{\mu} &\to& \hA_{\mu} + \Bigl\{
 1 - \frac{\hM^2}{\Lambda^2}\,\mra^6_{\upphi\,\mrA}
 + \frac{1}{2}\,\frac{\hM^4}{\Lambda^4}\,\Bigl[ 3\,(\mra^6_{\upphi\,\mrA})^2 -
  \mra^6_{\upphi\,\mrA}\,\mra^6_{\upphi\,\mrD} - \mra^8_{\upphi\,\mrA} \Bigr]
  \Bigr\}\,\pdmu \alpha
\label{cngi8}
\eqa
The gauge fixing term will have the same form \hyperref[gfcn]{as given in Eq.(\ref*{gfcn})}.
\subsection{AHEFT: including fermions \label{ferm}}
The fermion Lagrangian is
\bq
\Lag^{(4)}_{f} = - \baruppsi\,(\sla{\mrD} + m_{\uppsi})\,\uppsi \spp
\eq
It is not our goal to discuss the full list of higher dimensional operators containing fermions; for instance
there are $\mrdim = 5$ and $\mrdim = 7$ operators,
\bq
\Ope^{(5)}_{\upphi\,\uppsi} = \upphi^*\,\upphi\,\baruppsi\,\uppsi \spc
\qquad
\Ope^{(7)}_{\uppsi\,\mrA} = \baruppsi\,\uppsi\,\mrF_{\mu\nu}\,\mrF_{\mu\nu} \spc
\eq
as well as $\mrdim = 6$ operators, \eg
\bq
\Ope^{(6)}_{\uppsi} = (\baruppsi\,\uppsi)^2 \spc \quad
\Ope^{(6)}_{\upphi\,\uppsi} = i\,(\upphi^*\,\Ddmu \upphi)\,\baruppsi\,\gamma^{\mu}\,\uppsi \spc
\quad \mbox{\etc}
\eq
Here we have defined $\upphi^*\,\Ddmu \upphi = \upphi^*\,\mrD_{\mu} \upphi - (\mrD_{\mu} \upphi)^*\,\upphi$.
As an example, we discuss one redundant operator,
\bq
\Ope^{(6)}_{\uppsi\,\mrR} = i\,\baruppsi\,\gamma^{\mu}\,\uppsi\,\pdmu (\upphi^*\,\upphi) =
- i\,\upphi^*\,\upphi\,\Bigl[ \baruppsi\,(\stackrel{\rightarrow}{\sla{\mrD}} + 
                                        \stackrel{\leftarrow}{\sla{\mrD}})\,\uppsi \Bigr] \spc 
\label{fredO}
\eq
where a total derivative has been neglected. We use
\bq
\stackrel{\rightarrow}{\sla{\mrD}}\,\uppsi = \sla{\partial}\,\uppsi - i\,g\,\sla{\mrA}\,\uppsi \spc
\qquad
\baruppsi\,\stackrel{\leftarrow}{\sla{\mrD}} = (\pdmu \baruppsi)\,\gamma^{\mu}\,\uppsi + 
i\,g\,\baruppsi\,\sla{\mrA}\,\uppsi \spc
\eq
to write
\bq
\Lag^{(4)}_f + g^2\,\frac{\mra^6_{\uppsi\,\mrR}}{\Lambda^2}\,\Ope^{(6)}_{\uppsi\,\mrR} =
- \frac{1}{2}\,\baruppsi\,( \stackrel{\rightarrow}{\sla{\mrD}} -
                            \stackrel{\leftarrow}{\sla{\mrD}})\,\uppsi -
 m_{\uppsi}\,\baruppsi\,\uppsi +
i\,g^2\,\frac{\mra^6_{\uppsi\,\mrR}}{\Lambda^2}\,\baruppsi\,\gamma^{\mu}\,\uppsi\,\pdmu (\upphi^*\,\upphi) \spc
\eq
where we have neglected total derivatives. Next we perform the transformation
\bq
\baruppsi \to - i\,g^2\,\frac{\mra^6_{\uppsi\,\mrR}}{\Lambda^2}\,\baruppsi\,\upphi^*\,\upphi \spc
\qquad
\uppsi \to + i\,g^2\,\frac{\mra^6_{\uppsi\,\mrR}}{\Lambda^2}\,\upphi^*\,\upphi\,\uppsi \spc
\eq
which has the effect of eliminating the \hyperref[fredO]{redundant operator of Eq.(\ref*{fredO})} and to 
introduce $\ord{1/\Lambda^{4}}$ compensations, \eg an operator of $\mrdim = 7$,
\bq
- \bigl( g^2\,\frac{\mra^6_{\uppsi\,\mrR}}{\Lambda^2} \bigr)^2\,m_{\uppsi}\,\baruppsi\,\uppsi\,(\upphi^*\,\upphi)^2 \spp
\eq
\section{The standard model: SMEFT  \label{SMEFT}}
Discussion of SMEFT follows from \autoref{AHEFT} but is made more complicate by the presence of many fields.
The complex scalar field $\upphi$ \hyperref[Hdoublet]{is replaced by the $SU(2)$ doublet of Eq.(\ref*{Hdoublet})} and
$\mrA_{\mu}$ by a $2\,\times\,2$ Hermitean matrix
\bq
\mrA_{\mu} = \frac{1}{2}\,\mrB^a_{\mu}\,\tau_a + \frac{1}{2}\,g_1\,\mrB^0_{\mu}\,\tau_0 \spc
\eq
with $g_1 = - \stW/\ctW$ and where $\tau_a$ are Pauli matrices while $\stW(\ctW)$ is the sine(cosine) of the weak-mixing
angle; $\tau_0$ is the $2\,\times\,2$ unit matrix.
The covariant derivative is replaced by
\bq
(\mrD_{\mu}\,\upPhi)_{ij} = (\delta_{ij}\,\pdmu - i\,g\,\mrA_{\mu\,;\,ij})\,\upPhi_j \spc
\eq
but $\upPhi^{\dag}\,(\mrD_{\mu}\,\mrD_{\mu}\,\upPhi) = - (\mrD_{\mu}\,\upPhi)^{\dag}\,\mrD_{\mu} \upPhi$ still holds.
Furthermore,
\bq
\mrF^a_{\mu\nu} = \pdmu\,\mrB^a_{\nu} - \pdnu\,\mrB^a_{\nu} + g\,\ep^{a b c}\,\mrB^b_{\mu}\,\mrB^c_{\nu} \spc
\qquad
\mrF^0_{\mu\nu} = \pdmu\,\mrB^0_{\nu} - \pdnu\,\mrB^0_{\nu} \spp
\eq
SMEFT at $\mrdim = 6$ is constructed using the Warsaw basis~\cite{Grzadkowski:2018ohf} and we refer to
their Tab.~$2$ for dimension-six operators other than the four-fermion ones and to Tab.~$3$ for the four-fermion
operators. A complete extension to $\mrdim = 8$ does not exist but a subset has been presented in \Bref{Hays:2018zze};
to be precise a $\mrdim = 8$ basis is constructable by using Hilbert series but the output of the Hilbert series 
is the number of invariants at each mass dimension, and the field content of the invariants. Therefore, one must 
translate the output into a format which tells us how the various indices carried by each of the fields should be 
contracted. This has been done explicitly in \Bref{Hays:2018zze} for only a subset of $\mrdim = 8$ operators.

In sect.~$5$ (bosonic) and sect.~$6$ (single-fermionic-current) of \Bref{Grzadkowski:2018ohf} the operator 
classification is given with examples of how redundant operators can be eliminated. In this section we want to 
discuss examples of the $\mrdim = 8$ compensations of SMEFT $\mrdim = 6$ redundant operators.
\paragraph{Example: class $\upPhi^4\,\mrD^2$} \hspace{0pt} \\
Consider the operator
\bq
\Ope^{(6)}_{\ssR} = (\upPhi^{\dag}\,\upPhi)\,(\mrD_{\mu}\,\upPhi)^{\dag}\,(\mrD_{\mu}\,\upPhi) \spp
\eq
The term containing
\bq
g^2\,\frac{\mra^6_{\ssR}}{\Lambda^2}\,\Ope^{(6)}_{\ssR} \spc
\eq
is eliminated by the transformation
\bq
\upPhi \to \upPhi - g^2\,\frac{\mra_{\ssR}}{\Lambda^2}\,(\upPhi^{\dag}\,\upPhi)\,\upPhi
\spc \qquad
\upPhi^{\dag} \to \upPhi^{\dag} - g^2\,\frac{\mra_{\ssR}}{\Lambda^2}\,(\upPhi^{\dag}\,\upPhi)\,\upPhi^{\dag} \spp
\label{REDT}
\eq
This transformation generates higher order compensations, \eg
\bq
(\upPhi^{\dag}\,\upPhi)^2 \to (\upPhi^{\dag}\,\upPhi)^2 -
4\,\frac{g^2\,\mra^6_{\ssR}}{\Lambda^2}\,(\upPhi^{\dag}\,\upPhi)^3 +
6\,(\frac{g^2\,\mra^6_{\ssR}}{\Lambda^2})^2\,(\upPhi^{\dag}\,\upPhi)^4 \spc
\eq
as well as
\bq
\Ope^{(6)}_{\upphi\,\sPB} = (\upPhi^{\dag}\,\upPhi)\,\mrF^0_{\mu\nu}\,\mrF^0_{\mu\nu} \to
\Ope^{(6)}_{\upphi\,\sPB} - \frac{g^2\,\mra^6_{\ssR}}{\Lambda^2}\,(\upPhi^{\dag}\,\upPhi)^2\,
\mrF^0_{\mu\nu}\,\mrF^0_{\mu\nu} \spp
\label{O8HB}
\eq
The last operator in \eqn{O8HB} is the $\Ope_{8\,,\,\sPH \sPB}$ operator of Tab.~$4$ in \Bref{Hays:2018zze}. 
Similar results hold for $\Ope^{(6)}_{\upphi\,\sPW}$ (Tab.~$2$ of \Bref{Grzadkowski:2018ohf}), generating 
$\Ope_{8\,,\,\sPH \sPW}$ \etc
\paragraph{Example: class $\uppsi^2\,\upPhi^2\,\mrD^2$} \hspace{0pt} \\
Here we introduce fermions, \eg quarks,
\bq
\mrQ_{\ssL} = \Bigl(
\begin{array}{c}
\mru \\ \mrd
\end{array}
\Bigr)_{\ssL}
\spc \qquad
\mru_{\ssR}
\spc \qquad
\mrd_{\ssR} \spc
\eq
where $\mrf_{\ssL\,,\,\ssR} = 1/2\,(1 \pm \gamma^5)\,\mrf$. Consider the operator
\bq
\Ope^{(6)}_{\ssR} = {\overline{\mrQ}}_{\ssL}\,(\mrD_{\mu}\,\mrD_{\mu}\,\upPhi)\,\mru_{\ssR} +
{\overline{\mru}}_{\ssR}\,(\mrD_{\mu}\,\mrD_{\mu} \upPhi)^{\dag}\,\mrQ_{\ssL} \spp
\eq
The redundant term
\bq
g\,\frac{\mra^6_{\ssR}}{\Lambda^2}\,\Ope^{(6)}_{\ssR} \spc
\eq
is eliminated by performing the transformation
\bq
\upPhi \to \upPhi - g\,\frac{\mra^6_{\ssR}}{\Lambda^2}\,{\overline{\mru}}_{\ssR}\,\mrQ_{\ssL} \spc
\quad
\upPhi^{\dag} \to \upPhi^{\dag} - g\,\frac{\mra^6_{\ssR}}{\Lambda^2}\,{\overline{\mrQ}}_{\ssL}\,\mru_{\ssR} \spp
\eq
The same transformation, applied to $(\upPhi^{\dag}\,\upPhi)^2$ generates additional terms, \eg 
$(\upPhi^{\dag}\,\upPhi)\,{\overline{\mrQ}}_{\ssL}\,\upPhi\,\mru_{\ssR}$ which is the operator $\Ope^{(6)}_{\mru\,\upphi}$ of
the Warsaw basis. There are also $\mrdim = 8$ compensations, \eg
\bq
\Ope^{(6)}_{\mrd\,\upphi} \to ({\overline{\mrQ}}_{\ssL}\,\upPhi\,\mru_{\ssR})\,
                              ({\overline{\mrQ}}_{\ssL}\,\upPhi^c\,\mrd_{\ssR}) \spc
\eq
where $\upPhi^c_i = \ep_{ij}\,\upPhi^*_j$ is given by charge conjugation.
\begin{remark}
The construction of the $\mrdim = 6$ basis through the elimination of redundant operators requires field redefinitions 
with higher order compensations (say, at $\mrdim = 8$ level) which can be absorbed into the coefficients of operators 
that are already present at any given order. However, an independent construction of the $\mrdim = 8$ basis could
include some IBP/EoM reduction which eliminates exactly these operators. As discussed in 
\Brefs{Hays:2018zze,Gripaios:2018zrz} there is a procedure for removing EoM-reducible terms in Hilbert series output 
(put forth in \Bref{Lehman:2015via}) and, for instance, there is reduction at $\ord{\mrD^2}$ where the presence of 
derivatives implies that we have redundancies due to integration by parts, which can shift the covariant derivative
from one field to another, and the equations of motion. Some preliminary work was also carried out in 
\Bref{Degrande:2013kka}.

Our observations indicate that $\mrdim = 6$ and $\mrdim = 8$ should be treated together and consistently when explicitly 
constructing non-redundant sets of higher dimensional operators in the SMEFT and beyond.   
\end{remark}
\subsection{SM, SMEFT and \texorpdfstring{\SMP}{SM'}  \label{interpre}}
We consider SM${}^{\prime}$, an extension of the SM whose Lagrangian contains both SM and BSM parameters; 
by explicitly integrating out the heavy fields from the \SMP generating 
functional~\cite{Gaillard:1985uh,Cheyette:1987qz,Dittmaier:1995cr,Dittmaier:1995ee,Henning:2014wua} (or equivalently, by 
computing the relevant diagrams) we obtain
\bq
\Lag_{\SM^{\prime}} \mapsto 
\frac{\mra_{\upphi}}{\Lambda^2_{\mrh}}\,\Ope^{(6)}_{\upphi} +
\frac{\mrb_{\upphi\,\Box}}{\Lambda^2_{\mrh}}\,\Ope^{(6)}_{\upphi\,\Box} +
\frac{\mrc_{\upphi\,\mrD}}{\Lambda^2_{\mrh}}\,\Ope^{(6)}_{\upphi\,\mrD\,;\,1} +
\;\;\dots\;\; +
\frac{\mrd_{\upphi}}{\Lambda^4_{\mrh}}\,\Ope^{(8)}_{\upphi} + \;\;\dots \quad
\mrE \muchless \Lambda_{\mrh}
\eq
where 
\bq
\Ope^{(6)}_{\upphi\,\mrD\,;\,1} = (\upPhi^{\dag}\,\mrD_{\mu} \upPhi)^{\dag}\,(\upPhi^{\dag}\,\mrD_{\mu} \upPhi) \spc
\quad
\Ope^{(6)}_{\upphi\,\Box} = (\upPhi^{\dag}\,\upPhi)\,\Box\,(\upPhi^{\dag}\,\upPhi) \spc
\eq  
and $\mra\,\dots\,\mrd$ and the heavy scale $\Lambda_{\mrh}$ are functions of $\{\mrp\}_{\mySM}$ and $\{\mrp\}_{\BSM}$.
A simple example of EFT is provided by QED at very low energies, $E_{\PGg} \muchless m_{\Pe}$.
In this limit, one can describe the light-by-light scattering using an effective Lagrangian in terms of the 
electromagnetic field only. Gauge, Lorentz, Charge Conjugation and Parity invariance constrain the possible
structures present in the effective Lagrangian:
\bq 
\Lag_{\QEDEFT} = - \frac{1}{4}\,\mrF_{\mu\nu}\,\mrF_{\mu\nu} +
   \frac{\mra}{m^4_{\Pe}}\,(\mrF_{\mu\nu}\,\mrF_{\mu\nu})^2 +
   \frac{\mrb}{m^4_{\Pe}}\,\mrF_{\mu\nu}\,\mrF_{\nu\rho}\,\mrF_{\rho\sigma}\,\mrF_{\rho\mu} +
   \ord{\mrF^6/m^8_{\Pe}} \spp
\eq
An explicit calculation gives
\bq
\mra =  \frac{1}{36}\,\alpha^2 \spc
\qquad
\mrb = \frac{7}{90}\,\alpha^2 \spp
\eq
Another example is provided by the matching of SMEFT into the so-called Weak Effective Theory 
(WET or LEFT)~\cite{Jenkins:2017jig} where the SM heavy degrees of freedom have been integrated out.

However, imagine a situation where our EFT basis contains
\bq
\mbox{EFT basis} \quad \ni \quad \Ope^{(6\,,\,8)}_{\upphi}, \;
\Ope^{(6)}_{\upphi\,\mrD\,;\,1}, \; \Ope^{(6)}_{\upphi\,\mrD\,;\,2} \spc
\eq
where
\bq
\Ope^{(6)}_{\upphi\,\mrD\,;\,2} = (\upPhi^{\dag}\,\upPhi)\,(\mrD_{\mu} \upPhi)^{\dag}\,(\mrD_{\mu} \upPhi) \spp
\eq
There is a considerable amount of literature on global SMEFT fits, see 
\Brefs{Buchalla:2015qju,Reina:2015yuh,Bjorn:2016zlr,Berthier:2016tkq,Buckley:2016cfg,Mariotti:2016owy,Brivio:2017btx,Murphy:2017omb,Alte:2017pme,Ellis:2018gqa,Hays:2018zze,Aebischer:2018iyb,AguilarSaavedra:2018nen}.
For a Bayesian parameter estimation for EFT, see \Bref{Wesolowski:2015fqa}.
In order to compare the fitted (constrained) values of the $\mra^6$ and $\mra^8$ Wilson coefficients
with the ``computed'' values of $\mra\,\dots\,\mrd$ we have to perform a ``translation'' containing the
following steps:
\begin{enumerate}

\item neglecting total derivatives, we write
\bq
\Ope^{(6)}_{\upphi\,\mrD\,;\,2} = \frac{1}{2}\,\Ope^{(6)}_{\upphi\,\Box} -
(\upPhi^{\dag}\,\upPhi)\,\Bigl[
\upPhi^{\dag}\,\mrD_{\mu}\,\mrD_{\mu} \Upphi + (\mrD_{\mu}\,\mrD_{\mu} \upPhi)^{\dag})\,\upPhi \Bigr] \spp
\eq
\item The second term is eliminated performing the field transformation
\bq
\upPhi \to \upPhi - g^2\,\frac{\mra^6_{\upphi\,\mrD\,;,\,2}}{\Lambda^2}\,(\upPhi^{\dag}\,\upPhi)\,\upPhi \spc
\eq
which will induce an higher order compensation.

\end{enumerate}
\paragraph{{\bf{Fitting versus interpreting}}} \hspace{0pt} \\
\textsl{Critical to note: the higher order compensation is such that the Wilson coefficient of $\Ope^{(8)}_{\upphi}$ 
(and others) will be a combination of $\mra^8_{\upphi}$ (linear) and $\mra^{(6)}_{\upphi\,\mrD\,;\,2}$ (quadratic).
This combination is the one to be ``compared'' with $\mrd_{\upphi}$ (for the sake of simplicity we are neglecting the
mixing among Wilson coefficients).}

To repeat the argument we follow \Bref{Gripaios:2018zrz}: there is a vector space ($\mrV$) containing all possible
$X\,$-Lagrangian invariants. Let $\mrR \subset \mrV$ a subspace of redundant operators; the space of physical operators, 
the quotient space $\mrQ$, may be used to form a basis of physical operators. Given two bases (\eg the one 
containing $\Ope^{(6)}_{\upphi\,\Box}$ and the one containing $\Ope^{(6)}_{\upphi\,\mrD\,;\,2}$), 
taking into account that there is a large freedom in the choice, we need the transformation rules for a change of basis.  
This freedom is useful when comparing $\mrQ$ with the vector space generated by the low energy behavior of $X^{\prime}$.

Clearly, we have provided an ad hoc example. The general case is better described in terms of the classification
provided in \Bref{Einhorn:2013kja}: in any given model extending the SM to higher energy scales, some operators
may arise from tree diagrams in an underlying theory, while others may only emerge from loop corrections. 
The equivalence theorem relates some operators arising from loops to operators arising from trees~\footnote{For the
apparent puzzle of equivalence of $\Ope_{\PTG}$, $\Ope_{\LG}$ see sect.~$4$ of \Bref{Einhorn:2013kja}.};
imagine a situation where the SMEFT is defined by choosing as basis vectors (the quotient space) PTG operators while 
the underlying theory generates LG operators; this is exactly the scenario under discussion: the equivalence
of two operators is a property of SMEFT while it is possible for the BSM theory to generate one but not the other.
From this point of view we must be careful not to omit any basis operators without good reason, since their contributions
to Green’s functions can be very different. An incomplete basis set may lead to spurious relations among observables.
An example~\cite{Einhorn:2013kja} is as follows:
\bq
\Lag = - \frac{1}{2}\,\pdmu \upphi\,\pdmu \upphi - \frac{1}{2}\,m^2 \upphi^2  
       - \frac{1}{4}\,g \upphi^4 - {\overline{\uppsi}}\,( \sla{\partial} - \lambda\,\upphi )\,\uppsi +
       \frac{\mra^1_{\PTG}}{\Lambda^2}\,\upphi^3\,{\overline{\uppsi}}\,\uppsi +
       \frac{\mra^2_{\PTG}}{\Lambda^2}\,({\overline{\uppsi}}\,\uppsi)^2 \spc
\eq
where the $\mrdim = 6$ operators are PTG. Imagine that the underlying theory generates 
$\Ope_{\LG} = ({\overline{\uppsi}}\,\uppsi)\,\Box \upphi$ which is LG. However, 
$\Ope_{\LG}$ and $- g\,\Ope^1_{\PTG} + \lambda\,\Ope^2_{\PTG}$ are equivalent. Furthermore, \Bref{Einhorn:2013kja} shows 
that for this example there are $10$ $\mrdim = 6$ operators and $5$ independent equivalence classes, so there must 
be $5$ independent basis operators and we need the set of transformations and the set of equivalence relations. 
The field transformation needed to eliminate $\Ope_{\LG}$ is of the form
\bq
\upphi \to \upphi + \eta\,\frac{\mra_{\LG}}{\Lambda^2}\,{\overline{\uppsi}}\,\uppsi \spc
\eq
generating $\mrdim = 8$ compensations, \eg $\upphi^2\,({\overline{\uppsi}}\,\uppsi)^2$.

To summarize: the underlying strategy is as follows. When sufficient Wilson coefficients have been fitted, we need 
to connect to UV complete models: therefore, we integrate out the heavy states of the UV completion,
run resulting Wilson coefficients of BSM theory and SMEFT theory to same scale and compare consistency of 
(non) vanishing Wilson coefficients and general self-consistency. Hopefully, this will allow us to reject or support UV
complete theories; however, we must carry out the procedure in a consistent way testing the various approximations
which are often made due to technical difficulties in performing calculations in the SMEFT.
\subsection{SMEFT, HEFT and mixings  \label{SMEFTmix}}
In this subsection we compare again the SMEFT scenario with the one where the Higgs is a dominantly $J^{\ssP} = 0^+$ 
scalar, low-energy remnant of an underlying theory with more scalars and mixings.
First we consider the $\PH\,\PWp\,\PWm$ and $\PH\,\PZ\,\PZ$ couplings as derived in SMEFT at $\ord{1/\Lambda^2}$ 
(see also \Bref{Kanemura:2018yai}). After canonical normalization we obtain
\bq
\PH\,\PV^{\mu}(p_1)\,\PV^{\nu}(p_2) = \mrF^{\sPVV}_{\ssD}\,\delta^{\mu\nu} + \mrF^{\sPVV}_{\ssT}\,\mrT^{\mu\nu} \spc
\qquad
\mrT^{\mu\nu} = \spro{p_1}{p_2}\,\delta^{\mu\nu} - p^{\nu}_1\,p^{\mu}_2 \spp
\label{HVVstru}
\eq
SMEFT prediction is
\bqa
\mw\,\mrF^{\sPWW}_{\ssD} &=&
 - g\,\mws\,\Bigl[ 1 + \frac{g_6}{\sqrt{2}}\,
(\mra_{\upphi\,\sPW} + \mra_{\upphi\,\Box} - \frac{1}{4}\,\mra_{\upphi\,\sPD} ) \Bigr] \spc
\nl
\mw\,\mrF^{\sPZZ}_{\ssD} &=&
 - g\,\mzs\,\uprho\,\Bigl[ 1 + \frac{g_6}{\sqrt{2}}\,
(\mra_{\upphi\,\sPW} + \mra_{\upphi\,\Box} + \frac{1}{4}\,\mra_{\upphi\,\sPD} ) \Bigr] \spc
\nl
\mw\,\mrF^{\sPWW}_{\ssT} =
- 2\,g\,\frac{g_6}{\sqrt{2}}\,\mra_{\upphi\,\sPW} \spc
&\qquad&
\mw\,\mrF^{\sPZZ}_{\ssT} =
- 2\,g\,\frac{g_6}{\sqrt{2}}\,\mra_{\sPZZ} \spc
\label{HVVc}
\eqa
where $\mra_{\sPZZ}$ is defined in \eqn{aZZdef}, $\uprho= \mws/(\ctWs\,\mzs)$ and $\sqrt{2}\,g_6 = 1/(\myGF\,\Lambda^2)$.
As a consequence, SMEFT predicts a change in the normalization of the $SM\,$-like term and the appearance of the
transverse term. 

The action of $\Ope_{\upphi\,\mrD}$ deserves a comment: custodial symmetry $SU(2)_c$ is the diagonal 
subgroup, after electroweak symmetry breaking, of an accidental $SU(2)_{\ssL}\,\times\,SU(2)_{\ssR}$ global symmetry of 
the SM Lagrangian~\cite{Veltman:1977kh,Einhorn:1981cy}. The custodial symmetry is linked to the fact that the Higgs 
potential is invariant under $SO(4)$ 
which mixes the $4$ real components of the Higgs doublet. Since $SO(4) \sim SU(2)_{\ssL}\,\times\,SU(2)_{\ssR}$ the Higgs VEV 
breaks it down to the diagonal subgroup. As it is well known custodial symmetry is only an approximate symmetry (\eg due
to Yukawa interactions). Assuming that the Higgs sector consists of one electroweak doublet, the SMEFT term 
$\Ope_{\upphi\,\mrD}$ violates custodial symmetry as shown explicitly in \eqn{HVVc} where, however, we can set $\uprho = 1$.
A possible way out would be to introduce the Higgs doublet as a real representation $(2,2)$ of 
$SU(2)_{\ssL}\,\times\,SU(2)_{\ssR}$. 
Inclusion of $\mrdim = 8$ operators in \eqn{HVVc} involves several terms. Contributions to the transverse part
(using the terminology of \Bref{Hays:2018zze}) are coming from $\mra^8_{\sPH\,\sPW}, \mra^8_{\sPH\,\mrD\,\sPH\,\sPW}$ and  
$\mra^8_{\sPH\,\mrD\,\sPH\,\sPW\,2}$.

As discussed in \Bref{Low:2012rj}, tree level couplings to $\PW$ and $\PZ$ bosons of a scalar charged under
electroweak symmetry can be classified using the quantum number of the scalar under custodial symmetry. Therefore,
a pair of $\PW/\PZ$ bosons can only couple to a CP even neutral scalar that is either a custodial singlet
or a custodial $5$-plet. We follow the conventions of \Bref{Brivio:2016fzo} (for a detailed comparison of SMEFT and HEFT
operators see their Tab.~$2$),
\bq
\Lag_{\HEFT} = \Lag_0 + \Delta \Lag \spc
\eq
where $\Lag_0$ contains the leading order operators and the second one accounts for new interactions and for deviations
from LO. The LO parametrization of the effective couplings of the singlet to $\PV\PV$ is
proportional to the SM couplings through a global factor, as shown in Eq.(33) of \Bref{Alonso:2015fsp} or 
Eq.(1) of \Bref{Low:2012rj}. A contribution to $\mrF^{\sPVV}_{\ssT}$ of \eqn{HVVstru} enters through $\Delta \Lag$, 
for instance through the operators ${\mathcal{P}}_{\sPW}(\Ph)$ and $\mathcal{P}_1$ defined in sect.~$2.2.1$ of 
\Bref{Brivio:2016fzo}, with a suppression factor determined using the so-called NDA master formula, see
appendix~$4$ of \Bref{Gavela:2016bzc}.   

As an example of the underlying theory we consider the singlet extension of the SM. In the mass eigenbasis we have the
light Higgs boson ($\Ph$) and the heavy one ($\PH$) with a mixing angle $\alpha$ such that
\bq
\sin \alpha = \ord{\Lambda^{-1}} \spc
\qquad
\cos \alpha = 1 + \ord{\Lambda^{-2}} \spc
\eq
which implies that also the mixing angle must be expanded, not only heavy propagators and loops.
Integration of the heavy degree of freedom in loops requires a careful inclusion of heavy{-}light 
contributions~\cite{Bilenky:1993bt,delAguila:2016zcb,Boggia:2016asg,Henning:2016lyp,Ellis:2016enq,Fuentes-Martin:2016uol,Braathen:2018htl}. The result for the low energy limit of the $\Ph \PZ \PZ$ vertex can be summarized as 
follows~\cite{Boggia:2016asg}: 
\begin{itemize}

\item[\snitem] there are SM-like terms, $\mrV^0_{\sPh\sPZZ}$ of $\ord{g}$, $\mrV^{\TG}_{\sPh\sPZZ}$ of $\ord{g/\Lambda^2}$ (tree
generated) and $\mrV^{\LG}_{\sPh\sPZZ}$ of $\ord{g^3/\pi^2}$ (loop generated) containing both $\ord{1}$ and 
$\ord{1/\Lambda^2}$ terms

\item[\snitem] contributions to the $\mrF^{\sPZZ}_{\mrT}$ term can arise only from mixed (light{-}heavy) loops and are of
$\ord{1/\Lambda^4}$.

\end{itemize} 
Relevant quantities to be constrained, \eg in Higgs into four leptons, are
\bq
2\,\sqrt{2}\,\bigl( \frac{\mrF^{\sPWW}_{\ssD}}{\mw} - \frac{\mrF^{\sPZZ}_{\ssD}}{\mz} \bigr) \spc
\eq
a ``measure'' of $\mra_{\upphi\,\mrD}/\Lambda^2$ and $\mrF^{\sPVV}_{\mrT}$, a ``measure'' of a non-SM tensor structure at
$\ord{1/\Lambda^2}$ (\ie a ``measure'' of $\mra_{\upphi\,\sPW}$ and $\mra_{\sPZZ}$ in SMEFT or of the corresponding
operators in HEFT).
It is interesting to compare $\PH\PV\PV$ and $\PH\PH\PVV$ vertices in SMEFT; we introduce
\bq
\upkappa^{\sPWW}_{\sPH} = - g\,\mw \spc
\quad
\upkappa^{\sPZZ}_{\sPH} = - g\,\frac{\mzs}{\mw}\,\uprho \spc
\quad
\delta \kappa^{\sPWW}_{\sPH} = \mra_{\upphi\,\sPW} + \mra_{\upphi\,\Box} - \frac{1}{4}\,\mra_{\upphi\,\mrD} \spc
\quad
\delta \kappa^{\sPZZ}_{\sPH} = \mra_{\upphi\,\sPW} + \mra_{\upphi\,\Box} + \frac{1}{4}\,\mra_{\upphi\,\mrD} \spc
\eq
to derive
\bqa
\PH\,\PW^-_{\mu}(p_1)\,\PW^+_{\nu}(p_2) &=& \upkappa^{\sPWW}_{\PH}\,(1 + \frac{g_6}{\sqrt{2}}\,
                                            \delta \upkappa^{\sPWW}_{\PH})\,\delta_{\mu\nu} -
                                            \sqrt{2}\,\frac{g}{\mw}\,g_6\,\mra_{\upphi\,\sPW}\,\mrT_{\mu\nu} \spc
\nl
\PH\,\PZ_{\mu}(p_1)\,\PZ_{\nu}(p_2) &=& \upkappa^{\sPZZ}_{\PH}\,(1 + \frac{g_6}{\sqrt{2}}\,
                                            \delta \upkappa^{\sPZZ}_{\PH})\,\delta_{\mu\nu} -
                                            \sqrt{2}\,\frac{g}{\mw}\,g_6\,\mra_{\sPZZ}\,\mrT_{\mu\nu} \spc
\nl\nl 
\PH\,\PH\,\PW^-_{\mu}(p_1)\,\PW^+_{\nu}(p_2) &=& \frac{1}{2}\,\frac{g}{\mw}\,
  \upkappa^{\sPWW}_{\PH}\,(1 + \sqrt{2}\,g_6\,
  \delta \upkappa^{\sPWW}_{\sPH})\,\delta_{\mu\nu} -
  \frac{g^2}{\mws}\,\frac{g_6}{\sqrt{2}}\,\mra_{\upphi\,\sPW}\,\mrT_{\mu\nu} \spc
\nl
\PH\,\PH\,\PZ_{\mu}(p_1)\,\PZ_{\nu}(p_2) &=& \frac{1}{2}\,\frac{g}{\mw}\,
  \upkappa^{\sPZZ}_{\PH}\,\Bigl[ 1 + \sqrt{2}\,g_6\,(\delta \upkappa^{\sPZZ}_{\PH} - \mra_{\upphi\,\mrD})
  \Bigr]\,\delta_{\mu\nu} -
  \frac{g^2}{\mws}\,\frac{g_6}{\sqrt{2}}\,\mra_{\sPZZ}\,\mrT_{\mu\nu} \spp
\label{oHtH}
\eqa
Comparing \eqn{oHtH} with
\bq
( 1 + \frac{\mrc^{\sPVV}_1}{\Lambda^2}\,\PH + \frac{\mrc^{\sPVV}_2}{\Lambda^2}\,\PH^2 + \;\dots )\,
 \PV_{\mu}\,\PV_{\mu} + \; \dots
\eq
gives some information on the doublet structure of the scalar field, \ie
\bq
\frac{\mrc^{\sPWW}_2}{\mrc^{\sPWW}_1} = \frac{1}{2}\,\frac{g}{\mw}\,
     (1 - \frac{g_6}{\sqrt{2}}\,\delta \upkappa^{\sPWW}_{\PH}) \spc
\quad
\frac{\mrc^{\sPZZ}_2}{\mrc^{\sPZZ}_1} = \frac{1}{2}\,\frac{g}{\mw}\,
     \Bigl[ 1 + g_6\,(\mra_{\upphi\,\mrD} - \frac{1}{\sqrt{2}}\,\delta \upkappa^{\sPZZ}_{\sPH}) \Bigr] \spc
\eq
with $\delta \upkappa^{\sPWW}_{\sPH} = \delta \upkappa^{\sPZZ}_{\sPH}$ if $\mra_{\upphi\,\mrD} = 0$.
The explicit expressions for $\mrc^{\sPVV}_{1,2}$ given by the HSESM Lagrangian, including TG/LG generated and tadpoles,
are too long to be reported here and can be derived from Eqs.($110{-}111$) of \Bref{Boggia:2016asg}.

Similar results can be derived for $\PH \Pg \Pg$ and $\PH \PH \Pg \Pg$,
\bqa
\PH\,\Pg^a_{\mu}(p_1)\,\Pg^b_{\nu}(p_2) &=&
- \frac{g_{\ssS}}{\mw}\,\frac{g_6}{\sqrt{2}}\,\delta^{ab}\,\mra_{\upphi\,\sPg}\,\mrT_{\mu\nu} \spc
\nl
\PH\,\PH\,\Pg^a_{\mu}(p_1)\,\Pg^b_{\nu}(p_2) &=&
- \frac{g\,g_{\ssS}}{\mws}\,\frac{g_6}{\sqrt{2}}\,\delta^{ab}\,\mra_{\upphi\,\sPg}\,\mrT_{\mu\nu} \spc
\eqa
suggesting that di-Higgs production (as compared to di-boson decay~\cite{Zheng:2015dua}) could show the fingerprints 
of mixing effects~\cite{Gorbahn:2015gxa}.

To summarize: SM $\,\subseteq\,$ SMEFT $\,\subseteq\,$ HEFT and HEFT can be written as a SMEFT if there is an
$O(4)$ fixed point~\cite{Alonso:2016oah}. 
\subsection{SMEFT: field transformations, fits and constraints \label{ftfc}}
A final comment on fits is the following: the complexity of SMEFT has led many authors to propose that 
we can place a bound on the coefficient of a particular operator by assuming that all the other operators in that 
basis have vanishing coefficients. However, this is an ad hoc assumption which cannot be justified as discussed 
in \Bref{Willenbrock:2014bja} and explicitly shown in \Bref{Jiang:2016czg}: their results contradict the practice of 
neglecting operators that induce canonical normalization effects, a practice based on the misleading motivation that 
we are studying a specific set of data while these operators are better constrained by another set of data.

To expand on this point we start by considering canonical normalization in SMEFT; the
corresponding $\mrdim = 6$ transformations have been given in sect.~$2.4$ of \Bref{Ghezzi:2015vva} and
partially extended to $\mrdim = 8$ in appendix~D of \Bref{Hays:2018zze}, \eg
\bq
\PH \to \Bigl[ 1 - \frac{1}{4}\,\frac{M^2}{\Lambda^2}\,(\mra_{\upphi\,\mrD} - 4\,\mra_{\upphi\,\Box})
\Bigr]\,\PH \spp
\eq
From the point of view of gauge invariance the situation is similar to what we have derived in the AHEFT,
\hyperref[cngi8]{see Eq.(\ref*{cngi8}}; furthermore, transformations that realize canonical normalization of the 
Lagrangian should be performed in an arbitrary gauge and they should not be restricted to the unitary one.

Consider now the process $\PAQu + \PQu \to \PH + \PZ$. The SMEFT LO amplitude is constructed by assembling the following
vertices:
\bqa
\PAQu(p_1)\,\PQu(p_2)\,\PZ(-p_{\sPZ}) &=&
i\,\frac{g}{4\,\ctW}\,\gamma^{\mu}\,(1 - \frac{8}{3}\,\stWs + \gamma^5)
\nl
{}&+& i\,\frac{g}{48\,\ctW}\,\frac{g_6}{\sqrt{2}}\,\Bigl[
(13 - 8\,\stWs)\,(\mra_{\upphi\,\sPD} - 4\,\mra_{\upphi\,\sPW} + 4\,\ctWs\,\mra_{\sPZZ}) +
4\,(3 + 8\,\stWs)\,\ctWs\,\mra_{\sPZZ} 
\nl
{}&+&
24\,(\mra_{\upphi\,\PQu} + \mra^{(1)}_{\upphi\,\PQq} + \mra^{(3)}_{\upphi\,\PQq}) +
3\,(4\,\mra_{\upphi\,\sPW} - 8\,\mra_{\upphi\,\PQu} + 8\,\mra^{(1)}_{\upphi\,\PQq} + 8\,\mra^{(3)}_{\upphi\,\PQq} -
\mra_{\upphi\,\sPD})\,\gamma^5 \Bigr] \spc
\nl\nl
\PZ^{\nu}(p_{\sPZ})\,\PH(-p_{\sPH})\,\PZ^{\mu}(p_{\sPH} - p_{\sPZ}) &=&
- \frac{g\,\mw}{\ctWs}\,\delta^{\mu\nu} -
\frac{1}{4}\,\frac{g \mw}{\ctWs}\,\frac{g_6}{\sqrt{2}}\,(
\mra_{\upphi\,\sPD} + 4\,\mra_{\upphi\,\Box} + 4\,\mra_{\upphi\,\sPW})\,\delta^{\mu\nu} 
\nl
{}&-& 2\,\frac{g}{\mw}\,\frac{g_6}{\sqrt{2}}\,(\spro{p_{\sPH}}{p_{\sPZ}}\,\delta^{\mu\nu} -
p^{\mu}_{\sPZ}\,p^{\nu}_{\sPH})\,\mra_{\sPZZ} \spc 
\nl\nl
\PAQu(p_1)\,\PQu(p_2)\,\PH(-p_{\sPH})\,\PZ^{\mu}(-p_{\sPZ}) &=&
i\,\frac{g}{2\,\mw\,\ctW}\,\frac{g_6}{\sqrt{2}}\,\gamma^{\mu}\,\Bigl[
\mra_{\upphi\,\PQu} + \mra^{(1)}_{\upphi\,\PQq} + \mra^{(3)}_{\upphi\,\PQq} -
(\mra_{\upphi\,\PQu} - \mra^{(1)}_{\upphi\,\PQq} - \mra^{(3)}_{\upphi\,\PQq})\,\gamma^5 \Bigr] \spc
\eqa
where $p_{\sPZ} = p_1 + p_2$ and where we have introduced
\bq
\mra_{\sPZZ} = \stWs\,\mra_{\upphi\,\sPB} + \ctWs\,\mra_{\upphi\,\sPW} -
\stW\,\ctW\,\mra_{\upphi\,\sPWB} \spc 
\quad
g_6 = \frac{1}{\sqrt{2}\,\myGF\,\Lambda^2} \spp
\label{aZZdef}
\eq
To give an example, the operator $\Ope_{\upphi\,\Box}$ enters only through canonical normalization and changes the
overall normalization of the SM-like $\delta_{\mu\nu}$ term but does not enter into the coefficient of
$p^{\mu}_{\sPZ}\,p^{\nu}_{\sPH}$. Therefore, when squaring the amplitude up to and including terms of $\ord{g^2_6}$ it will 
affect the shape of distributions and not only their overall normalization. 
To be more precise, let us consider
\bq
\PAQu(x_1\,p_1) + \PQu(x_2\,p_2) \to \PH(-p_{\sPH}) + \PZ(-p_{\sPZ}) \spp
\eq
We introduce
${\hat{s}} = - 2\,x_1 x_2\,\spro{p_1}{p_2} \spc
\;\;
{\hat{t}} = 2\,x_1\,\spro{p_1}{p_{\sPZ}} + \mzs \spc
$
and compute the part amplitude squared depending on $\mra_{\upphi\,\Box}$, obtaining
\bqa
\sum_{\spin}\,\mid \mrA \mid^2 &=&
\frac{3}{4}\,\frac{g^4\,\mrv^2_{\PQu}}{\ctWq}\,\mzs\,{\hat{s}}\,\mid \Delta_{\sPZ}\mid^2\,
(1 + 2\,\frac{g_6}{\sqrt{2}}\,\mra_{\upphi\,\Box}) + \frac{g^4}{\ctWq}\,g^2_6\,\mra_{\upphi\,\Box}\,
      (\mrA_2\,\mid \Delta_{\sPZ}\mid^2 + \mrA_1\,\Re\,\Delta_{\sPZ}) \spc
\nl\nl
\mrA_1 &=& \frac{1}{4}\,\mrv_{\PQu}\,(\mra_{\upphi\,\PQu} + \mra^{(1)}_{\upphi\,\PQq} + \mra^{(3)}_{\upphi\,\PQq})\,
\Bigl[
2\,\hats - \mhs +
(\hats - \mhs - \mzs)\,\frac{\hatt}{\mzs} +
\frac{\hatt^2}{\mzs} \Bigr] \spc
\nl\nl
\mrA_2 &=& 
\frac{3}{16}\,\mrv^2_{\PQu}\,(4\,\mra_{\upphi\,\sPW} + \mra_{\upphi\,\sPD})\,\mzs\,\hats -
\frac{1}{16}\,(10 + 3\,\mrv_{\PQu})\,\mrv_{\PQu}\,(4\,\mra_{\upphi\,\sPW} - \mra_{\upphi\,\sPD})\,\mzs\,\hats
\nl
{}&+& \frac{3}{2}\,\mrv_{\PQu}\,(\mra_{\upphi\,\PQu} + \mra^{(1)}_{\upphi\,\PQq} + \mra^{(3)}_{\upphi\,\PQq})\,\,
      \mzs\,\hats
\nl
{}&+& 4\,\ctWs\,\mrv_{\PQu}\,\mra_{\sPZZ}\,\mzs\,\hats +
\frac{1}{8}\,\mrv^2_{\PQu}\,\Bigl\{
7\,\mhs\,\mzs + (\hats - \mhs)\,(6\,\hats + 6\,\mzs + \mhs) 
\nl
{}&+& \Bigl[ (\hats - \mhs)^2 - \mzq \Bigr]\,\frac{\hatt}{\mzs} +
(\hats - \mhs + \mzs)\,\frac{\hatt^2}{\mzs}
\Bigr\}\,\mra_{\sPZZ} + \frac{3}{8}\,\mrv^2_{\PQu}\,\mra_{\upphi\,\Box}\,\mzs\,\hats \spc
\eqa
where
\bq
\mrv_{\PQu} = 1 - \frac{8}{3}\,\stWs \spc
\qquad
\Delta_{\sPZ} = \frac{1}{{\hat{s}} - \mzs} \spp
\eq
The conclusion is that at $\ord{1/\Lambda^2}$ the Wilson coefficient $\mra_{\upphi\,\Box}$ modifies the normalization 
of the $\hats\,$ (the $\PH{-}\PZ$ invariant mass squared) distribution; at $\ord{1/\Lambda^4}$ the shape of the 
$\hatt\,$-distribution is modified. Statements like ``only two operators of the Warsaw basis'' contribute to the
top decay are, at the very least, questionable.   
\subsubsection{Linear  vs.  quadratic representation} 
Most of the SMEFT calculations include the extra term, \ie
\bq
\mid \mrA^{(4)} + \frac{1}{\Lambda^2}\,\mrA^{(6)} \mid^2 \to
\mid \mrA^{(4)} \mid^2 + 2\,\frac{1}{\Lambda^2}\Re\,\bigl[ \mrA^{(4)} \bigr]^*\,\mrA^{(6)} + 
\frac{1}{\Lambda^4}\,\mid \mrA^{(6)} \mid^2 \spc
\label{quad}
\eq
making positive definite (by construction) all the observables. 
Consider the transverse momentum spectra of the
$\PZ$ boson from $\PZ\PH$ production: it has been shown that the linear SMEFT for $\Lambda = 1\,\UTeV$
starts to become negative above $p_{\ssT} \approx 200 \UGeV$ which can cured by the inclusion of
squared terms~\cite{Brehmer:2015rna,Boggia:2017hyq}. However, both approaches deviate significantly at higher 
$p_{\ssT}$; as a matter of fact, this behavior signals the breakdown of the EFT approximation.
For additional attempts to assess the impact of the squared EFT terms see \Bref{Degrande:2016dqg}.  

Obviously, \eqn{quad} is missing the (yet) unavailable $\mrdim = 8$ operators. As a matter of fact, there is more 
than neglecting the $\mrdim = 4\,/\,\mrdim = 8$ interfence: the point is that we construct $\mrS\,$-matrix
elements at $\ord{1/\Lambda^4}$ using a canonically transformed Lagrangian truncated at $\ord{1/\Lambda^2}$.
What we have is 
\bq
\Lag= - \frac{1}{2}\,(1 + \frac{M^2}{\Lambda^2}\,\delta \mrZ^6_{\sPH} + 
\fbox{$\frac{M^4}{\Lambda^4}\,\delta \mrZ^8_{\sPH}$} )\,\pdmu \PH\,\pdmu \PH +
\;\dots\; +
\frac{1}{\Lambda^2}\,\Bigl[ \mra\,M^3\,\PH\,\PZ_{\mu}\,\PZ_{\mu} + \;\dots \Bigr] +
\fbox{$\frac{1}{\Lambda^4}\,\sum_i\,\mra^8_i\,\Ope^{(8)}_i$} \spc
\eq
where the frame box indicates that the terms are not available. We should write
\bq
\PH = (1 + \frac{M^2}{\Lambda^2}\,\eta^6_{\sPH} + \frac{M^4}{\Lambda^4}\,\eta^8_{\sPH} )\,\hH \spc
\eq
select 
\bq
\eta^6_{\sPH} = - \frac{1}{2}\,\delta \mrZ^6_{\sPH} \spc
\qquad
\eta^8_{\sPH} = \frac{3}{8}\,\bigl[ \delta \mrZ^6_{\sPH} \bigr]^2 \spc
\eq
obtaining
\bq
{\hat{\Lag}} = - \frac{1}{2}\,\pdmu \hH\,\pdmu \hH +
\mra\,\frac{M^3}{\Lambda^2}\,(1  - \ovalbox{$\frac{1}{2}\,\frac{M^2}{\Lambda^2}\,\delta \mrZ^6_{\sPH}$} )\,
\hH\,\PZ_{\mu}\,\PZ_{\mu} + \;\dots 
\eq
where the oval box gives terms that are neglected in the quadratic approach of \eqn{quad}.

SMEFT effects in Vector Boson Scattering (VBS) have been analyzed in \Bref{Gomez-Ambrosio:2018pnl} with the
conclusion that a global study of the set of $\mrdim = 6$ operators is necessary.

It must also be said that there are theoretical issues and experimental issues. From the experimental point of view, 
keeping the squared terms has clear advantages: cross sections in all phase space points are always positive.
In the case of only linear EFT terms, any negative cross section value in any phase space point gives an unphysical 
configuration, \eg a negative probability density, for which the fit cannot calculate how it compares to data.
Technically speaking, the only chance for the fit is randomly trying new parameter configurations until we are back 
in the lands of positive cross section and this can be difficult in high dimensional parameter 
spaces~\footnote{I am grateful to M.~Duehrssen for this comment.}. 

From the theoretical point of view, if the final measured parameter combination is valid for both the just linear 
terms and including the squared terms, the distance (if not too large) will tell us something about theory uncertainties. 
Otherwise, the measured point (negative cross section) is invalid in a pure linear EFT.
In other words, we can treat the linear fit as an actual EFT expansion and the quadratic fit as an estimate of the 
truncation error, thus defining a validity scale measurement by measurement.

It remains true that none of established approaches scales well to high-dimensional problems with many parameters and 
observables, such as the SMEFT measurements, \ie individual processes at the LHC are sensitive to several operators 
and require simultaneous inference over a multi-dimensional parameter space. While a naive parameter
scan works well for one or two dimensions, it becomes 
computationally demanding for more than a few parameters.
Alternative, recent progress in putting significantly stronger bounds on 
effective $\mrdim = 6$ operators than the traditional approach is described in \Brefs{Brehmer:2016nyr,Brehmer:2018eca}.
\paragraph{Quadratic vs. linear: summary} \hspace{0pt} \\
To summarize, the proper definition of ``quadratic'' EFT is as follows: given a ``truncated'' Lagrangian
\bq
\Lag = \Lag^{(4)} + \frac{1}{\Lambda^2}\,\Lag^{(6)} + \frac{1}{\Lambda^4}\,\Lag^{(8)} \spc
\eq
we distinguish between redundant and non-redundant operators:
\bq
\Lag^{(6,8)} = \Lag^{(6,8)}_{\ssN\ssR} + \sum_{i \in \ssR}\,\Ope^{(6,8)}_i\,\frac{\delta \Lag^{(4)}}{\delta \upphi} \spc
\eq
and redefine fields according to
\bq
\upphi \to \upphi - \sum_{n=2,4}\,\frac{1}{\Lambda^n}\,\sum_{i \in \ssR}\,\Ope^{(n + 4)}_i \spp
\eq
The corresponding shift in $\Lag$ will eliminate redundant operators but leave a term
\bq
\Delta\Lag = - \frac{1}{\Lambda^4}\,\Bigl[
\frac{\delta \Lag^{(4)}}{\delta \upphi}\,\sum_{i \in \ssR}\,\Ope^{(8)}_i +
\frac{\delta \Lag^{(6)}}{\delta \upphi}\,\sum_{i \in \ssR}\,\Ope^{(6)}_i +
\frac{1}{2}\,\frac{\delta^2 \Lag^{(4)}}{\delta \upphi^2}\,\sum_{i,j \in \ssR}\,\Ope^{(6)}_i\,\Ope^{(6)}_j \Bigr] \spp
\eq
Once again, $\Delta\Lag$ will never generate terms that are not present in $\Lag^{(8)}$ (symmetry); however, when it comes
to the interpretation of ``fitted'' Wilson coefficients in terms of the low-energy behavior of some high-{-}energy 
(possibly complete) theory including or not all possible sources of higher-dimensional terms will make a difference.

Once redundant operator are eliminated we will have
\bq
\Lag = - \frac{1}{2}\,\mrZ^{ij}_{\upphi}\,\pdmu \upphi_i\,\pdmu \upphi_j -
\frac{1}{2}\,\mrZ^{ij}_{m}\,\upphi_i\,\upphi_j + \Lag_{\rest} \spc
\eq
where the non-canonical normalization is
\bqa
\mrZ^{ij}_{\upphi} &=& \delta^{ij} + \frac{1}{\Lambda^2}\,\delta \mrZ^{(6)\,;\,ij}_{\upphi}
+ \frac{1}{\Lambda^4}\,\delta \mrZ^{(8)\,;\,ij}_{\upphi} \spc
\nl
\mrZ^{ij}_m &=& m^2_i\,\delta^{ij} + \frac{1}{\Lambda^2}\,\delta \mrZ^{(6)\,;\,ij}_m
+ \frac{1}{\Lambda^4}\,\delta \mrZ^{(8)\,;\,ij}_m \spp
\eqa
We therefore rescale fields and masses (and possibly couplings) in order to reestablish canonical
normalization. This additional transformation will affect $\Lag_{\rest}$. Actually, this is not the end of the story
since we have to link the parameters of the Lagrangian to a given set of experimental data, the so-called 
IPS~\cite{Bardin:1999gt}.
These relations will, once again, change $\Lag_{\rest}$. 
In any extension of the SM the Higgs- as well as the gauge-boson masses can be chosen on-shell and the SM parameters
defined via $\myGF, \mw, \mz$, where $\myGF$ is the Fermi coupling constant. BSM parameters are treated as 
$\MSB$ parameters; the various renormalization schemes differ in the treatment of tadpole 
contributions~\cite{Altenkamp:2017kxk,Boggia:2018yzv}.  

Once we have obtained the Lagrangian, up to $\ord{1/\Lambda^4}$, we can obtain Feynman rules and amplitudes. When
we say that a given LO amplitude contains terms up to $\ord{1/\Lambda^4}$ we mean single and double insertions of
higher dimensional operators in the LO diagrams obtained from $\Lag^{(4)}$ (plus set of diagrams having new
structures, not originating from $\Lag^{(4)}$). Given
\bq
\mrA= \mrA^{(4)} + \frac{1}{\Lambda^2}\,\mrA^{(6)} + \frac{1}{\Lambda^4}\,\mrA^{(8)} \spc    
\eq
linear means including the interference between $\mrA^{(4)}$ and $\mrA^{(6)}$, quadratic means the complete inclusion
of all terms giving $1/\Lambda^4$ (not only the square of $\mrA^{(6)}$).
\section{A two scale problem  \label{tsp}}
In this section, we will consider a $\mrdim = 4$ Lagrangian containing two heavy, almost degenerate, degrees of freedom
and derive an effective Lagrangian taking into account their mass difference. The simple example is given by
the following Lagrangian~\footnote{For more realistic models see \Bref{Haber:2018iwr}}:
\bqa
\Lag &=& - \frac{1}{2}\,\pdmu \upphi\,\pdmu \upphi - \frac{1}{2}\,\Lag_{\inte}(\upphi)
         - \frac{1}{2}\,\sum_{i=1,2}\,( \pdmu \upchi_i\,\pdmu \upchi_i
         + M^2_i\,\upchi^2_i )
\nl
{}&-& \upphi^2\,(\lambda_1\upchi^2_1 + \lambda_2\,\upchi^2_2)
      - \lambda_{12}\,\upchi^2_1\,\upchi^2_2
      - \lambda_3\,\upchi^4_1 - \lambda_4\,\upchi^4_2
      - \lambda_5\,\upphi^2\,\upchi_1\,\upchi_2 \spp
\eqa
We assume a scenario where $m \muchless M_i$ but $\mid M^2_1 - M^2_2 \mid \muchless M^2_1 + M^2_2$, with $M_1 > M_2$.
We expand $\upchi_i = \upchi^c_i + \upchi$ and apply the background field method (BFM) 
formalism~\cite{tHooft:1973bhk,Abbott:1981ke,Barvinsky:1988ds,Barvinsky:1990up}, integrating over $\upchi_1, \upchi_2$. 
We are interested in Green's functions with external $\upphi\,$-lines and, for the sake of simplicity, we will neglect
the mixed heavy-light contributions to the effective action. The BFM Lagrangian is
\bq
\Lag_2 = \frac{1}{2}\,\upchi^{\dag}\,X\,\upchi \spc
\quad 
X =  (\Box - M^2_+)\,\tau_0 - M^2_-\,\tau_3 + Y \spc
\eq
where $M^2_{\pm} = 1/2(M^2_1 \pm M^2_2)$, $\tau_a$ are the Pauli matrices, $\tau_0$ is the identity matrix and
\bq
Y = - \Bigl(
\begin{array}{cc}
2\,\lambda_1 & \lambda_5 \\
\lambda_5 & 2\,\lambda_2 \\
\end{array} \Bigr)\,\upphi^2 \spp
\eq
A key problem in relation to the degenerate case, $M_1 = M_2$ is represented by the fact that
the mass matrix does not commute with $Y$. We can derive a triple expansion, in $1/M^2_{1,2}$ and $M^2_-$ following
the general result obtained in \Bref{Osipov:2001th} and based on a generalized heat kernel expansion,
\bq
\ln \mid \mathrm{det}\,\mrD \mid = - \frac{1}{2}\,\int_0^{\infty}\,\frac{d t}{t}\,\rho(t)\,
\mathrm{Tr}\,\exp( - t\,\mrD^{\dag}\,\mrD ) \spp
\eq
The kernel $\rho(t)$ formally defines the regularization procedure. We define
\bq
2\,\mrI_n = \mrJ_n(M^2_1) + \mrJ_n(M^2_2) \spc
\qquad
\mrJ_n(\mu^2) = \int_0^{\infty}\,d t\,t^{n - 2}\,\rho(t)\,\exp( - \mu^2\,t ) \spc
\eq
and obtain
\bq 
\mrJ_1 = [\,\mathrm{div}\,] - 1 \spc
\quad
\mrJ_2 = \frac{1}{\mu^2} \spc
\quad
\mrJ_3 = \frac{1}{\mu^4} \spc
\quad
\mrJ_4 = \frac{2}{\mu^6} \spc
\eq
\etc 
The divergent part ($[\,\mathrm{div}\,]$) is regularization dependent. The effective Lagrangian can be written as
\bq
\Lag{\eff} = \frac{1}{32\,\pi^2}\,\sum_{n=0}^{\infty}\,(-1)^n\,2^{n - 4}\,\mrI_{n-1}\,\mathrm{Tr}\,\Ope_i \spc
\eq
where $\Ope_0 = 1$ and
\bq
\Ope_1 = - Y \spc
\quad
\Ope_2 = \frac{1}{2}\,Y^2 - M^2_-\,Y_3 \spc
\quad
\Ope_3 = - \frac{1}{3\,!}\,Y^3 + \frac{1}{12}\,\pdmu Y\,\pdmu Y + \frac{1}{2}\,M^2_-\,Y_3\,Y \spc
\eq
\bq
\Ope_4 = \frac{1}{4\,!}\,Y^4 - \frac{1}{12}\,Y\,\pdmu Y\,\pdmu Y + 
         \frac{1}{120}\,(\Box\,Y)^2 -
         \frac{1}{6}\,M^2_-\,Y_3\,Y^2 +
         \frac{1}{12}\,M^2_-\,\tau_3\,\pdmu Y\,\pdmu Y -
         \frac{1}{12}\,M^4_-\,(Y^2 - Y^2_3) +
         \frac{1}{3}\,M^6_-\,Y_3 \spc 
\eq
and where $Y_3 = \tau_3\,Y$. Therefore
\bq
\upphi^6 \spc \quad (\upphi\,\pdmu \upphi)^2 \spc \quad \upphi^8 \spc \quad
\upphi^2\,(\upphi\,\pdmu \upphi)^2 \spc \quad(\Box\,\upphi^2)^2 \spc \quad \dots
\eq
operators are generated and the (small) mass difference between the two heavy fields is properly taken into
account. The problem with more, almost degenerate, scales can be treated according to the formulation of
\Bref{Osipov:2002gr}. 
\section{Mixed heavy-light effects  \label{mixhl}}
In this section we consider a $\mrdim = 4$ Lagrangian containing one light field $\upphi$ and one heavy field
$\upPhi$,
\bq
\Lag = \frac{1}{2}\,\upphi\,(\Box - m^2)\,\upphi + \frac{1}{2}\,\upPhi\,(\Box - M^2)\,\upPhi +
       \Lag_{\inte}(\upphi\,,\,\upPhi) \spp
\eq
We would like to integrate the heavy field in a manner that includes both heavy and heavy-light (loop) contribution.
The procedure is standard; the light field is decomposed in $\upphi = \upphi_c + \upphi$ and there is no classical
part for $\upPhi$. According to the BFM formalism we write
\bq
\Lag = \Lag(\upphi_c) + \frac{1}{2}\,\uppsi^{\dag}\,X\,\uppsi + \ord{\uppsi^3} \spc
\qquad
\uppsi^{\dag} = (\upphi\,,\,\upPhi) \spc
\eq
where $X = \tau_0\,\Box - \mrM + Y$, $\tau_0$ being the identity, $\mrM$ is the (squared) mass matrix and $Y$
depends on $\upphi_c$. The relevant object is
\bq
\mathrm{Tr}\,\ln \Bigl[ \tau_0\,\Box_x - \mrM + Y(x) \Bigr]\,\delta^4(x - y) =
\int d^4 x\,\int \frac{d^4 q}{(2\,\pi)^4}\,\,\mathrm{tr}\,\ln \Bigl[ - \tau_0\,q^2 - \mrM + 
        \Box + 2\,i\,\spro{p}{\partial} + Y \Bigr] \spp
\label{fromS}
\eq
When $m$ and $M$ are the same we rewrite the square bracket in \eqn{fromS} as
\bq
- (q^2 + M^2)\,(\tau_0 + \mrK) \spc
\eq
and expand $\ln (\tau_0 + \mrK)$ in powers of the matrix $\mrK$ obtaining the large $M$ expansion. 
\paragraph{Matrix logarithm} \hspace{0pt} \\
One should observe that
\bq
\mathrm{tr}\,\ln(\mrA + \mrB) = \mathrm{tr}\,\ln \mrA +  \mathrm{tr}\,\ln \mrB \spc \qquad
\mathrm{if} \;\; \mrA, \mrB \quad \mathrm{are\; both\; positive-definite} \spc
\eq
\bq
\ln(\mrA\,\mrB) = \ln \mrA + \ln \mrB \spc \qquad
\mathrm{if} \;\; \mrA, \mrB \quad \mathrm{commute} \spp
\eq
To be more precise: let $\mrA, \mrB \in \Cf^{n \times n}$ commute and have no eigenvalues on $\Rf^-$; if for every
eigenvalue $\lambda_i$ of $\mrA$ and the corresponding eigenvalue $\mu_i$ of $\mrB$, 
$\mid \mathrm{arg} \lambda_i + \mathrm{arg} \mu_i \mid < \pi$, then $\ln \mrA \mrB = \ln \mrA + \ln \mrB$, the
principal logarithm of $\mrA \mrB$  

It is easily seen that the degenerate matrix $\mrM$ does not commute with $Y$ and they are not both 
positive-definite~\footnote{A symmetric $n\,\times\,n$ real matrix $\mrM$ is said to be positive definite if the scalar 
$(\mrz\,,\,\mrM \mrz)$ is strictly positive for every non-zero column vector $\mrz$ of $n$ real numbers.}. 
A solution to this problem can be found by observing that there is a theorem involving the matrix 
logarithm~\cite{Habernote} stating that
\bq
\ln (\mrA + \mrB) - \ln \mrA = \int_0^{\infty}\,d\mu^2\,\Bigl[ (\mrA + \mu^2\,\mrI)^{-1} -
  (\mrA + \mrB + \mu^2\,\mrI)^{-1} \Bigr] \spc
\label{Mlog}
\eq
where $\mrI$ is the unit matrix. A quite similar approach has been developed in \Bref{Ellis:2016enq}.
\paragraph{Taylor expansion for matrix logarithms} \hspace{0pt} \\
The correct Taylor expansion in \eqn{Mlog} is~\cite{Adlernote,Lashkari:2018tjh} 
\bq
\ln (\mrA + \mrB) - \ln \mrA = \int_0^{\infty}\,d\mu^2\,\Bigl[
\mrA^{-1}_+\,\mrB\,\mrA^{-1}_+ -
\mrA^{-1}_+\,\mrB\,\mrA^{-1}_+\,\mrB\,\mrA^{-1}_+ \;+\, \dots \Bigr] \spc 
\eq
where $\mrA_+ = \mrA + \mu^2\,\mrI$.
We will not care about the requirement of boundedness since it involves mathematical machinery~\cite{Lashkari:2018tjh} 
going beyond the scope of the present work. 
\paragraph{Self{-}adjoint operators} \hspace{0pt} \\
The other restriction that should be mentioned here is that the standard BFM result, $\mathrm{Det}\,\mid \mrD \mid$, 
follows only if the operator $\mrD$ is self-adjoint~\cite{Vassilevich:2003xt}.
\section{Extensions of the Yukawa model, including heavy{-}light contributions  \label{eYm}}
In this section we consider a Lagrangian containing one massless fermion, one light scalar and one 
heavy particle:
\bei
\item[a)] heavy scalar:
\bqa
\Lag_{\mrY \mrS} &=& - \frac{1}{2}\,\pdmu \upphi\,\pdmu \upphi - \frac{1}{2}\,m^2 \upphi^2  
       - \frac{1}{4}\,g \upphi^4 - {\overline{\uppsi}}\,( \sla{\partial} - \lambda_1\,\upphi )\,\uppsi 
\nl
{}&-&    \frac{1}{2}\,\pdmu \mrS\,\pdmu \mrS - \frac{1}{2}\,M^2\,\mrS^2 
       - \frac{1}{4}\,\lambda_4\,\mrS^4 + \lambda_2\,{\overline{\uppsi}}\,\uppsi\,\mrS
       - \lambda_3\,\upphi^2\,\mrS^2 \spc
\label{LYS}
\eqa
\item[b)] heavy vector:
\bqa
\Lag_{\mrY \mrV} &=& - \frac{1}{2}\,\pdmu \upphi\,\pdmu \upphi - \frac{1}{2}\,m^2 \upphi^2  
       - \frac{1}{4}\,g \upphi^4 - {\overline{\uppsi}}\,( \sla{\partial} - \lambda_1\,\upphi )\,\uppsi 
\nl
{}&-&  \frac{1}{2}\,( \pdmu \mrW_{\nu}\,\pdmu \mrW_{\nu} -
                      \pdmu \mrW_{\nu}\,\pdnu\,\mrW_{\mu} +
                      M^2\,\mrW_{\mu}\,\mrW_{\mu} ) +
       \frac{1}{4}\,\lambda_2\,\upphi^2\,\mrW_{\mu}\,\mrW_{\mu} +
       \lambda_3\,{\overline{\uppsi}}\,\gamma^{\mu}\,\uppsi\,\mrW_{\mu} \spp
\label{LYV}
\eqa
\eei
The Lagrangians are invariant under $\uppsi \to \gamma^5\,\uppsi$, $\upphi \to - \upphi$. The goal is to derive the
$\mrdim = 6$ Lagrangian obtained by integrating the $\mrW(\mrS)$ field when the heavy-light mixing is not neglected.  
In terms of loops we will consider only those diagrams where there is at least one internal heavy line.
Case b) will be used to illustrate problems that should not be underestimated, \eg we have written an explicit mass 
term for the vector field. When extending the SM with new vectors the masses can arise from vacuum expectation 
values of extra scalar fields (the ``mixing'' problem will arise), but this is usually neglected in the literature.
\paragraph{TG operators} \hspace{0pt} \\
There are TG operators, \eg \hyperref[LYS]{from Eq.(\ref*{LYS})} we have
\bqa
\mrdim = 6 \quad &\mid& \quad \frac{\lambda^2_2}{M^2}\,({\overline{\uppsi}}\,\uppsi)^2 \spc
\nl
\mrdim = 8 \quad &\mid& \quad - \frac{\lambda^2_2}{M^4}\,{\overline{\uppsi}}\,\uppsi\,\Box\,{\overline{\uppsi}}\,\uppsi
\spc \quad - \frac{\lambda^2_2 \lambda_3}{M^4}\,({\overline{\uppsi}}\,\uppsi)^2\,\upphi^2 \spp
\eqa
\paragraph{LG operators, heavy scalar} \hspace{0pt} \\
To derive LG operators we use the BFM formalism: we split the light fields as $\upphi = \upphi_c + \upphi$ and 
$\uppsi= \uppsi_c + \uppsi$ while there is no classical part for the heavy fields, \ie we are not interested in 
Green's functions with external heavy degrees of freedom. The BFM Lagrangian becomes the sum of two terms
\bq
\Lag_{\mrf} = - {\overline{\uppsi}}\,X_{\mrf}\,\uppsi + {\overline{\upeta}}\,\uppsi +
{\overline{\uppsi}}\,\upeta \spc
\qquad
\Lag_{\mrb} = \frac{1}{2}\,\upPhi^{\mrt}_i\,X_{ij}\,\upPhi_j \spc
\label{fbLags}
\eq
where $i=1, 2$ $\upPhi^{\mrt} = (\mrS\,,\,\upphi)$, and the matrix $X$ is block diagonal
\bq
X_{11} = \Box - M^2 - 2\,\lambda_3\,\upphi^2_c \spc
\qquad
X_{22} = \Box - m^2 - 3\,g\,\upphi^2_c \spc
\eq
while $X_{\mrf} = \sla{\partial} - \lambda_1\,\upphi_c$. Furthermore the source $\upeta$ is defined by
\bq
\upeta = ( \lambda_1\,\upphi + \lambda_2\,\mrS)\,\uppsi_c \spp
\label{etasS}
\eq
The standard result for the integration over the fermion field would be
\bq
\exp\bigl\{ - \mathrm{Tr}\,\ln X_{\mrf} - i\,\int d^4x d^4y\,\oeta(x)\,X^{-1}_{\mrf}(x,y)\,\upeta(y) \bigr\} \spp
\eq
Instead of dealing with the inverse of $X_{\mrf}$ we prefer to use
\bqa
\mrZ_{\mrf} &=& \int\,\bigl[ \mathcal{D}\,\uppsi \bigr]\,\bigl[ \mathcal{D}\,\baruppsi \bigr]\,
\exp \bigl\{ i\,\int d^4 x \Bigl[ \baruppsi\,X_{\mrf}\,\uppsi + \oeta\,\uppsi + \baruppsi\,\upeta \Bigr] \bigr\} =
\exp \bigl\{ - i\,\lambda_1\,\int d^4z\,\upphi_c(z)\,\frac{\delta^2}{\delta\upeta(z)\,\delta\oeta(z)} \bigr\}
\nl
{}&\times&
\int\,\bigl[ \mathcal{D}\,\uppsi \bigr]\,\bigl[ \mathcal{D}\,\baruppsi \bigr]\,
\exp \bigl\{ i\,\int d^4 x \Bigl[ \baruppsi\,\sla{\partial}\,\uppsi + 
               \oeta\,\uppsi + \baruppsi\,\upeta \Bigr] \bigr\} \spp
\eqa
After introducing $\mrS_{\mrF}$,
\bq
\sla{\partial}_x\,\mrS_{\mrF}(x - y) = \delta^4(x - y) \spc
\eq
we conclude that
\bq
\mrZ_{\mrf} = \exp \bigl\{ - \mathrm{Tr}\,\ln\,\sla{\partial} \bigr\}\,
\Bigl[ 1 - i\,\lambda_1\,\int d^4z\,\upphi_c(z)\,\frac{\delta^2}{\delta\upeta(z)\,\delta\oeta(z)} + \dots \Bigr]\,
\exp \bigl\{ i\,\int d^4x d^4y\,\oeta(x)\,\mrS_{\mrF}(x - y)\,\upeta(y) \bigr\} \spp
\eq
The effective Lagrangian can be derived from the following expression
\bqa
\mrZ &=& \exp \bigl\{ - \mathrm{Tr}\,\ln (\sla{\partial}) \bigr\}\,
\int \bigl[ \mathcal{D}\,\mrS \bigr]\,\bigl[ \mathcal{D}\,\upphi \bigr]\,
\exp \bigl\{ i\,\int d^4x d^4y\,\Lag_{\upPhi}(x,y) \bigr\}
\nl
{}&\times& 
\Bigl[ 1 + \;\; \mathrm{loops} \;\; + i\,\int d^4x_1 d^4x_2\,\oeta(x_1)\,\Gamma(x_1,x_2)\,\upeta(x_2) + \dots \Bigr] \spc
\label{nlexp}
\eqa
where we have introduced
\bq
\Lag_{\upPhi} =  \frac{1}{2}\,\upPhi^{\dag}(x)\,X(x)\,\upPhi(x)\,\delta^4(x - y) +
\oeta(x)\,\mrS_{\mrF}(x - y)\,\upeta(y) \spc
\qquad
\Gamma(x,y) = \sum_{i,j}\,\lambda^i_1\,\lambda^j_2\,\Gamma_{ij}(x,y) \spc
\label{Gammaexp}
\eq
where $\Gamma_{ij}$ are open strings of $\gamma\,$-matrices and of propagators $\mrS_{\mrF}$ while ``loops'' indicates
closed strings, generating loop diagrams with internal fermion lines; the latter will be discarded while
the open strings are exponentiated to the desired order. Results will be given at a fixed order in the couplings,
\ie $\lambda^i_1\,\lambda^j_2\,\lambda^k_3$ with $i + j + k \le 4$. The effective Lagrangian becomes
\bq
\exp \bigl\{ i\,\int d^4x\,\Lag_{\eff} \bigr\} \propto
\int \bigl[ \mathcal{D}\,\mrS \bigr]\,\bigl[ \mathcal{D}\,\upphi \bigr]\,
\exp \bigl\{ i\,\int d^4x\,\Lag_{d}(x) + i\,\int d^4x d^4y\,\Lag_{nd}(x,y) \bigr\} \spc
\eq
containing a diagonal and a non-diagonal part,
\bq
\Lag_{d} = \frac{1}{2}\,\upPhi^{\dag}(x)\,\mrD_{l}(x)\,\upPhi(x) \spc
\qquad
\Lag_{nd} = \upPhi^{\dag}(x)\,\mrD_{nl}(x,y)\,\upPhi(y) \spp
\eq
When we insert $\upeta$ \hyperref[etas]{from Eq.(\ref*{etas})} \hyperref[nlexp]{into Eq.(\ref*{nlexp})}
we obtain
\bqa
\int d^4x d^4y\,\oeta(x)\,\Gamma(x,y)\,\upeta(y) &=&
\int d^4x d^4y\,\Bigl[ 
\mrS(x)\,\mcS_{11}(x\,,\,y)\,\mrS(y) + 
\upphi(x)\,\mcS_{22}(x\,,\,y)\,\upphi(y) 
\nl
{}&+&  \mrS(x)\,\mcS_{12}(x\,,\,y)\,\upphi(y) + 
\upphi(x)\,\mcS_{21}(x\,,\,y)\,\mrS(y) \Bigr] + \ord{\mbox{field}^3} \spc
\eqa
where the elements of symmetric matrix $\mcS$ are
\bqa
{\mathcal{S}}_{1\,1}(x\,,\,y) &=& - \lambda^2_2\,\baruppsi_c(x)\,\mrS(x,y)\,\uppsi_c(y) +
\lambda_1\,\lambda^2_2\,\int d^4 \mrz_1\,\baruppsi_c(x)\,\mrS(x,\mrz_1,y)\,\uppsi_c(y)\,\upphi_c(\mrz_1)
\nl
{}&-& 
\lambda^2_1\,\lambda^2_2\,\int d^4 \mrz_1\,d^4 \mrz_2\,\baruppsi_c(x)\,\mrS(x,\mrz_1,\mrz_2,y)\,\uppsi_c(y)\,
\upphi_c(\mrz_1)\,\upphi_c(\mrz_2) \spc
\nl\nl
{\mathcal{S}}_{2\,2}(x\,,\,y) &=& - \lambda^2_1\,\baruppsi_c(x)\,\mrS(x,y)\,\uppsi_c(y) +
\lambda^3_1\,\int d^4 \mrz_1\,\baruppsi_c(x)\,\mrS(x,\mrz_1,y)\,\uppsi_c(y)\,\upphi_c(\mrz_1)
\nl
{}&-& 
\lambda^4_1\,\int d^4 \mrz_1\,d^4 \mrz_2\,\baruppsi_c(x)\,\mrS(x,\mrz_1,\mrz_2,y)\,\uppsi_c(y)\,
\upphi_c(\mrz_1)\,\upphi_c(\mrz_2) \spc
\nl\nl
{\mathcal{S}}_{1\,2}(x\,,\,y) &=& - \lambda_1\,\lambda_2\,\baruppsi_c(x)\,\mrS(x,y)\,\uppsi_c(y) +
\lambda^2_1\,\lambda_2\,\int d^4 \mrz_1\,\baruppsi_c(x)\,\mrS(x,\mrz_1,y)\,\uppsi_c(y)\,\upphi_c(\mrz_1)
\nl
{}&-& 
\lambda^3_1\,\lambda_2\,\int d^4 \mrz_1\,d^4 \mrz_2\,\baruppsi_c(x)\,\mrS(x,\mrz_1,\mrz_2,y)\,\uppsi_c(y)\,
\upphi_c(\mrz_1)\,\upphi_c(\mrz_2) \spc
\eqa
where higher orders in the couplings have been neglected. The strings $\mrS$ are given by
\bq
\mrS(x\,,\,\mrz_1\,,\,\dots\,,\,\mrz_n\,,\,y) = 
\mrS_{\mrF}(x - \mrz_1)\,\dots\,\mrS_{\mrF}(\mrz_n - y) \spp
\eq
We now assemble the various pieces with no attempt at mathematical rigor but avoiding highly questionable steps that
sometimes appear in the literature. Putting together the various ingredients we obtain
\bq
\exp \bigl\{ i\,\int d^4x\,\Lag_{\eff} \propto
\int \bigl[ \mathcal{D}\,\mrS \bigr]\,\bigl[ \mathcal{D}\,\upphi \bigr]\,
\exp \bigl\{ - \frac{i}{2}\,\int d^4x d^4y\,\upPhi^{\dag}(x)\,\Delta(x\,,\,y\,,\,\partial_x)\,\upPhi(y) \bigr\} \spc
\eq
where $\Delta$ is the sum of two pieces,
\bq
\Delta(x\,,\,y\,,\,\partial_x) = \Delta^{\loc}(x\,,\,\partial_x)\,\delta^4(x - y) + \Delta^{\nloc}(x\,,\,y) \spc
\eq
\bqa
\Delta^{\loc}_{1\,1} = - \Box_x + M^2 + 2\,\lambda_3\,\upphi_c(x)  \spc
&\qquad&
\Delta^{\loc}_{2\,2} = - \Box_x + m^2 + 3\,g\,\upphi_c(x)  \spc
\nl
\Delta^{\nloc}_{ij}(x\,,\,y) &=& - 2\,\mcS_{ij}(x\,,\,y) \spp
\eqa
Using
\bq
\mrS_{\mrF}(x - y) = \frac{1}{(2\,\pi)^4\,i}\,\int d^4q\,\exp\{i\,\spro{q}{(x-y)}\}\,\frac{- i\,\sla{q}}{q^2} \spc
\eq
the result of the functional integration is
\bqa
\exp \bigl\{ i\,\int d^4x\,\Lag_{\eff} \bigr\} &\propto&
\exp \Bigl\{ - \frac{1}{2}\,\mbox{Tr}\,
\int \frac{d^4q}{(2\,\pi)^4}\,\ln\; \exp\{i\,\spro{q}{(x-y)}\}\,\Bigl[
\Delta^{\loc}(x\,,\partial + i\,q) + \frac{1}{i}\,\Delta^{\nloc}(q\,,\,x\,,\,y) \Bigr] \Bigr\} 
\nl
{}&=&
\exp \Bigl\{ - \frac{1}{2}\,
\int d^4x\,\frac{d^4q}{(2\,\pi)^4}\,\mathrm{tr}\,\ln \Bigl[
\Delta^{\loc}(x\,,\partial + i\,q) + \frac{1}{i}\,\Delta^{\nloc}(q\,,\,x\,,\,x) \Bigr] \Bigr\} \spp 
\label{functint}
\eqa
For instance, we have
\bqa
\mcS_{2\,2}(x\,,\,y) &=& - \lambda^2_1\,\baruppsi_c(x)\,\mrS(x\,,\,y)\,\uppsi_c(y) = 
- \frac{\lambda^2_1}{(2\,\pi)^4\,i}\,\int d^4q\,
\exp\{i\,\spro{q}{(x-y)}\}\,\baruppsi_c(x)\,\frac{- i\,\sla{q}}{q^2}\,\uppsi_c(y) 
\nl
{}&=& \frac{1}{(2\,\pi)^4\,i}\,\int d^4q\,\exp\{i\,\spro{q}{(x-y)}\}\,\mcS_{2\,2}(q\,,\,x\,,\,y)
\eqa
The matrix $\Delta$ is transformed into
\bqa
\Delta^{\loc}_{1\,1}(q\,,\,x) &=& 
q^2 + M^2 - \Box - 2\,i\,\spro{q}{\partial} + 2\,\lambda_2\,\upphi_c(x)  \spc
\nl\nl
\Delta^{\loc}_{2\,2}(q\,,\,x) &=& q^2 + m^2 - \Box - 2\,i\,\spro{q}{\partial} + 3\,g\,\upphi_c(x) \spc
\nl\nl
\Delta^{\nloc}_{ij}(q\,,\,x) &=& 2\,\mcS_{ij}(q\,,\,x) \spp
\eqa
The non-local part is given by
\bqa
\mcS_{1\,1}(q\,,\,x) &=& 
\lambda^2_2\,\baruppsi_c(x)\,\frac{\sla{q}}{q^2}\,\uppsi_c(x) + \lambda_1\,\lambda^2_2\,
\int d^4p\,\exp\{i\,\spro{p}{x}\}\,\frac{1}{q^2\,(q - p)^2}\,\baruppsi_c(x)\,( q^2 +
\sla{q}\,\sla{p} )\,\uppsi_c(x)\,\upphi_c(p)
\nl
{}&+& \frac{1}{2}\,\lambda^2_1\,\lambda^2_2\,\int d^4p_1 d^4p_2\,\exp\{i\,\spro{(p_1 + p_2)}{x}\}\,
\frac{1}{q^2\,(q - p_1 - p_2)^2}\,\baruppsi_c(x)
\nl
{}&\times&
\Bigl\{
\frac{1}{(q - p_1)^2}\,\Bigl[ \sla{q}\,\sla{p_1}\,\sla{p_2} + (q - p_1)^2\,\sla{q} - q^2\,\sla{p_2} \Bigr] 
+ \frac{1}{(q - p_2)^2}\,\Bigl[ \sla{q}\,\sla{p_2}\,\sla{p_1} + (q - p_2)^2\,\sla{q} - q^2\,\sla{p_1} \Bigr] 
\Bigr\}
\nl
{}&\times& \uppsi_c(x)\,\upphi(p_1)\,\upphi(p_2) + \mbox{h.o.} \spc
\label{nlpart}
\eqa
with similar terms for the other entries of the matrix.
In \eqn{nlpart} $\upphi_c(p)$ is the Fourier transform of $\upphi_c(x)$.
While it is well known that heavy contributions can be solved in terms of tadpole integrals the non-local nature
of the mixed heavy{-}light contributions requires the introduction of quasi-tadpoles.
In performing the functional integration \hyperref[functint]{in Eq.(\ref*{functint})}
we have used the fact that the operator $\mcS^{\nloc}$ is self-adjoint, \ie
\bq
\langle\,\upphi\,,\,\mcS\,\mrS\,\rangle =
\int d^4x d^4y\,\upphi(x)\,\mcS_{1\,2}(x\,,\,y)\,\mrS(y) =
\langle\,\mcS\,\upphi\,,\,\mrS\,\rangle =
\int d^4x d^4y \upphi(x)\,\mcS^{\dag}_{1\,2}(y\,,\,x)\,\mrS(y) \spp
\eq
The matrix $\Delta$ is rewritten as 
\bq
\Delta_{ij} = \mrP_{ij} + \mrK_{ij} \spc
\eq
where $\mrP$ is diagonal, \ie
\bq
\mrP_{1\,1} = q^2 + M^2 \spc
\qquad
\mrP_{2\,2} = q^2 + m^2 \spc
\eq
and we use
\bq
\ln (\mrP + \mrK) - \ln \mrP = \int_0^{\infty}\,d\mu^2\,\Bigl[ (\mrP + \mu^2\,\mrI)^{-1} -
  (\mrP + \mrK + \mu^2\,\mrI)^{-1} \Bigr] \spc
\eq
to obtain
\bq
\mrL = \int d^4q\,\mathrm{tr}\,\ln\,\Delta \mid_{\partial = \partial + i\,q\;,\;x = y} =
\int d^4q \mathrm{tr}\,\Bigl\{ \ln \mrP + \int_0^{\infty} d\mu^2 \Bigl[ 
(\mrP + \mu^2\,\mrI)^{-1} - (\mrP + \mrK + \mu^2\,\mrI)^{-1} \Bigr] \Bigr\} \spc 
\eq
With $\mrP_+ = \mrP + \mu^2\,\mrI$ we expand as follows:
\bq
(\mrP_+ + \mrK)^{-1} = \mrP^{-1}_+\,(\mrI + \mrK\,\mrP^{-1}_+)^{-1} =
\mrP^{-1}_+ + \mrP^{-1}_+\,\sum_{n=1}^{\infty}\,\bigl( - \mrK\,\mrP^{-1}_+ \bigr)^n \spc
\eq
\bq
\mrL = \;\;(\mbox{field independent})\;\; - \int_0^{\infty} d\mu^2\,\int d^4q\,\mbox{tr}\,
\mrP^{-1}_+\,\sum_{n=1}^{\infty}\,\bigl( - \mrK\,\mrP^{-1}_+ \bigr)^n \spp
\label{expan}
\eq
\paragraph{LG operators, heavy vector} \hspace{0pt} \\
In the \hyperref[fbLags]{Lagrangian of Eq.(\ref*{fbLags})} the indices are
$i=0,\,\dots\,,5$ and the fields are $\upPhi^{\mrt} = (\mrW_{\mu}\,,\,\upphi)$, where the  block diagonal matrix $X$ 
becomes
\bq
X_{\mu\nu} = (\Box - M^2 + \frac{1}{2}\,\lambda_2\,\upphi^2_c)\,\delta_{\mu\nu} - \pdmu \pdnu \spc
\qquad
X_{55} = \Box - m^2 - 3\,g\,\upphi^2_c \spc
\eq
with $X_{\mrf} = \sla{\partial} - \lambda_1\,\upphi$. The source $\upeta$ is now defined by
\bq
\upeta = (\lambda_1\,\upphi + \lambda_3\,\gamma^{\mu}\,\mrW_{\mu})\,\uppsi_c \spp
\label{etas}
\eq
The effective Lagrangian can now be derived from the following expression
\bqa
\mrZ &=& \exp \bigl\{ - \mathrm{Tr}\,\ln (\sla{\partial}) \bigr\}\,
\int \bigl[ \mathcal{D}\,\mrW_{\mu} \bigr]\,\bigl[ \mathcal{D}\,\upphi \bigr]\,
\exp \bigl\{ i\,\int d^4x d^4y\,\Lag_{\upPhi}(x,y) \bigr\}
\nl
{}&\times& 
\Bigl[ 1 + \;\; \mathrm{loops} \;\; + i\,\int d^4x_1 d^4x_2\,\oeta(x_1)\,\Gamma(x_1,x_2)\,\upeta(x_2) + \dots \Bigr] \spp
\label{nlexpV}
\eqa
The effective Lagrangian becomes
\bq
\exp \bigl\{ i\,\int d^4x\,\Lag_{\eff} \bigr\} \propto
\int \bigl[ \mathcal{D}\,\mrW_{\mu} \bigr]\,\bigl[ \mathcal{D}\,\upphi \bigr]\,
\exp \bigl\{ i\,\int d^4x\,\Lag_{l}(x) + i\,\int d^4x d^4y\,\Lag_{nl}(x,y) \bigr\} \spc
\eq
containing a local and a non-local part.
When we insert $\upeta$ \hyperref[etas]{from Eq.(\ref*{etas})} \hyperref[nlexp]{into Eq.(\ref*{nlexp})}
we obtain
\bqa
\int d^4x d^4y\,\oeta(x)\,\Gamma(x,y)\,\upeta(y) &=&
\int d^4x d^4y\,\Bigl[ \upphi(x)\,\upphi(y)\,{\mathcal{S}}_{5\,5}(x\,,\,y) +
\upphi(x)\,\mrW_{\mu}(y)\,{\mathcal{S}}_{5\,\mu}(x\,,\,y) 
\nl
{}&+& \mrW_{\mu}(x)\,\upphi(y)\,{\mathcal{S}}_{\mu\,5}(x\,,\,y) +
\mrW_{\mu}(x)\,\mrW_{\nu}(y)\,{\mathcal{S}}_{\mu\,\nu}(x\,,\,y) + \ord{\mbox{field}^3} \spc
\eqa
where the elements of $\mcS$ are
\bqa
\mcS_{\mu\,\nu}(x\,,\,y) &=&
- \lambda^2_3\,\baruppsi_c(x)\,\mrS_{\mu\nu}(x\,,\,y)\,\uppsi_c(y)
\nl
{}&+& \frac{1}{2}\,\lambda_1\,\lambda^2_3\,\int d^4 \mrz_1\,\Bigl[
\baruppsi_c(x)\,\mrS_{\mu\nu}(x\,,\,\mrz_1\,,\,y)\,\uppsi_c(y) +
\baruppsi_c(y)\,\mrS_{\nu\mu}(y\,,\,\mrz_1\,,\,x)\,\uppsi_c(x) \Bigr]\,\upphi_c(\mrz_1)
\nl
{}&-& \frac{1}{2}\,\lambda^2_1\,\lambda^2_3\,\int d^4\mrz_1 d^4\mrz_2\,\Bigl[
\baruppsi_c(x)\,\mrS_{\mu\nu}(x\,,\,\mrz_1\,,\,\mrz_2\,,\,y)\,\uppsi_c(y) +
\baruppsi_c(y)\,\mrS_{\nu\mu}(y\,,\,\mrz_1\,,\,\mrz_2\,,\,x)\,\uppsi_c(y) \Bigr]\,\upphi_c(\mrz_1)\,\upphi_c(\mrz_2) + 
\mbox{h.o.}
\nl\nl
\mcS_{5\,5}(x\,,\,y) &=&
- \lambda^2_1\,\baruppsi_c(x)\,\mrS(x\,,\,y)\,\uppsi_c(y)
\nl
{}&+& \frac{1}{2}\,\lambda^3_1\,\int d^4 \mrz_1\,\Bigl[
\baruppsi_c(x)\,\mrS(x\,,\,\mrz_1\,,\,y)\,\uppsi_c(y) +
\baruppsi_c(y)\,\mrS(y\,,\,\mrz_1\,,\,x)\,\uppsi_c(x) \Bigr]\,\upphi_c(\mrz_1)
\nl
{}&-& \frac{1}{2}\,\lambda^4_1\,\int d^4\mrz_1 d^4\mrz_2\,\Bigl[
\baruppsi_c(x)\,\mrS(x\,,\,\mrz_1\,,\,\mrz_2\,,\,y)\,\uppsi_c(y) +
\baruppsi_c(y)\,\mrS(y\,,\,\mrz_1\,,\,\mrz_2\,,\,x)\,\uppsi_c(y) \Bigr]\,\upphi_c(\mrz_1)\,\upphi_c(\mrz_2) + 
\mbox{h.o.}
\nl\nl
\mcS_{\mu\,5}(x\,,\,y) &=&
- \lambda_1\,\lambda_3\,\baruppsi_c(x)\,\mrS_{\mu\,1}(x\,,\,y)\,\uppsi_c(y)
\nl
{}&+& \frac{1}{2}\,\lambda^2_1\,\lambda_3\,\int d^4 \mrz_1\,\Bigl[
\baruppsi_c(x)\,\mrS_{\mu\,1}(x\,,\,\mrz_1\,,\,y)\,\uppsi_c(y) +
\baruppsi_c(y)\,\mrS_{1\,\mu}(y\,,\,\mrz_1\,,\,x)\,\uppsi_c(x) \Bigr]\,\upphi_c(\mrz_1)
\nl
{}&-& \frac{1}{2}\,\lambda^3_1\,\lambda_3\,\int d^4\mrz_1 d^4\mrz_2\,\Bigl[
\baruppsi_c(x)\,\mrS_{\mu\,1}(x\,,\,\mrz_1\,,\,\mrz_2\,,\,y)\,\uppsi_c(y) +
\baruppsi_c(y)\,\mrS_{1\,\mu}(y\,,\,\mrz_1\,,\,\mrz_2\,,\,x)\,\uppsi_c(y) \Bigr]\,\upphi_c(\mrz_1)\,\upphi_c(\mrz_2) 
\nl
{}&+&  \mbox{h.o.}
\eqa
and the strings $\mrS$ are now given by
\bqa
\mrS(x\,,\,\mrz_1\,,\,\dots\,,\,\mrz_n\,,\,y) &=& 
\mrS_{\mrF}(x - \mrz_1)\,\dots\,\mrS_{\mrF}(\mrz_n - y) \spc
\nl
\mrS_{\mu\,1}(x\,,\,\mrz_1\,,\,\dots\,,\,\mrz_n\,,\,y) &=&
\gamma_{\mu}\,\mrS_{\mrF}(x - \mrz_1)\,\dots\,\mrS_{\mrF}(\mrz_n - y) \spc
\nl
\mrS_{1\,\mu}(x\,,\,\mrz_1\,,\,\dots\,,\,\mrz_n\,,\,y) &=&
\mrS_{\mrF}(x - \mrz_1)\,\dots\,\mrS_{\mrF})(\mrz_n - y)\,\gamma_{\mu} \spc
\nl
\mrS_{\mu\nu}(x\,,\,\mrz_1\,,\,\mrz_2\,,\,y) &=& 
\gamma_{\mu}\,\mrS_{\mrF}(x - \mrz_1)\,\mrS_{\mrF}(\mrz_1 - \mrz_2)\,
\mrS_{\mrF}(\mrz_2 - y)\,\gamma_{\nu} \spp
\eqa
Putting together the various ingredients we obtain
\bq
\exp \bigl\{ i\,\int d^4x\,\Lag_{\eff} \propto
\int \bigl[ \mathcal{D}\,\mrW_{\mu} \bigr]\,\bigl[ \mathcal{D}\,\upphi \bigr]\,
\exp \bigl\{ - \frac{i}{2}\,\int d^4x d^4y\,\upPhi^{\dag}(x)\,\Delta(x\,,\,y\,,\,\partial_x)\,\upPhi(y) \bigr\} \spc
\eq
where $\Delta$ is the sum of two pieces,
\bq
\Delta(x\,,\,y\,,\,\partial_x) = \Delta^{\loc}(x\,,\,\partial_x)\,\delta^4(x - y) + \Delta^{\nloc}(x\,,\,y) \spc
\eq
\bqa
\Delta^{\loc}_{\mu\,\nu} &=& 
\Bigl[ - \Box_x + M^2 - \frac{1}{2}\,\lambda_2\,\upphi_c(x) \Bigr]\,\delta_{\mu\,\nu} +
\partial_{\mu}\,\partial_{\nu}
\nl
\Delta^{\loc}_{5\,5} &=& \Bigl[ - \Box_x + m^2 + 3\,g\,\upphi_c(x) \Bigr] \spc
\nl
\Delta^{\nloc}_{ij}(x\,,\,y) &=& -2\,{\mathcal{S}}_{ij}(x\,,\,y) \spp
\eqa
The matrix $\Delta$ is transformed into
\bqa
\Delta^{\loc}_{\mu\,\nu}(q\,,\,x) &=& 
\Bigl[ q^2 + M^2 - \Box - 2\,i\,\spro{q}{\partial} - \frac{1}{2}\,\lambda_2\,\upphi_c(x) \Bigr]\,\delta_{\mu\,\nu} -
q_{\mu} q_{\nu} + \partial_{\mu}\,\partial_{\nu} + i\,(q_{\mu} \pdnu + q_{\nu} \pdmu) \spc
\nl\nl
\Delta^{\loc}_{5\,5}(q\,,\,x) &=& q^2 + m^2 - \Box - 2\,i\,\spro{q}{\partial} + 3\,g\,\upphi_c(x) \spc
\nl\nl
\Delta^{\nloc}_{ij}(q\,,\,x) &=& 2\,\mcS_{ij}(q\,,\,x) \spp
\eqa
For the non-local part we present only one term, 
\bq
\mcS_{5\,5}(q\,,\,x) = \lambda^2_1\,\baruppsi_c(x)\,\frac{\sla{q}}{q^2}\,\uppsi_c(y) 
+ \lambda^3_1\,\int d^4p\,\exp\{i\,\spro{p}{x}\}\,\frac{1}{(q - p_1)^2}\,\baruppsi_c(x)\,\uppsi_c(x)\,\upphi_c(p)
+ \ord{\lambda^4_1} \spp
\eq
Also in this case we have used the fact that the operator $\mcS^{\nloc}$ is
self-adoint, \ie
\bq
\langle\,\upphi\,,\,\mcS_{\mu}\,\mrW^{\mu}\,\rangle =
\int d^4x d^4y\,\upphi(x)\,\mcS_{\mu}(x\,,\,y)\,\mrW^{\mu}(y) =
\langle\,\mcS_{\mu}\,\upphi\,,\,\mrW^{\mu}\,\rangle =
\int d^4x d^4y \upphi(x)\,\mcS^{\dag}_{\mu}(y\,,\,x)\,\mrW^{\mu}(y) \spp
\eq
\subsection{LG operators,  prelims}
In order to understand local and non-local contributions to the higher-dimensional Lagrangians we consider a simple
example:
\bq
\mrI= \int d^{\mrd}q\,d^4p\,\frac{1}{q^2\,(q^2 + M^2)\,(q + p)^2}\,\exp\{i\,\spro{p}{x}\}\,
\upphi_c(p)\,\baruppsi_c(x)\,\uppsi(x) \spp
\eq
Note that we compute the $q\,$-integral in dimension regularization. Since
\bq
\mrI= \frac{1}{M^2}\,\int d^{\mrd}q\,d^4p\,(\frac{1}{q^2} - \frac{1}{q^2 + M^2})\,\frac{1}{(q + p)^2}\,
\exp\{i\,\spro{p}{x}\}\,\upphi_c(p)\,\baruppsi_c(x)\,\uppsi(x) \spc
\eq
and writing
\bq
\int d^{\mrd}q\,\frac{1}{(q^2 + M^2)\,(q  + p)^2} =
\int d^{\mrd}q\,\frac{1}{q^2\,(q^2 + M^2)}\,\bigl( 1 - \frac{p^2 - 2\,\spro{q}{p}}{q^2 + M^2} + \dots \bigr) \spc
\eq
we obtain a non-local term of $\ord{1/M^2}$,
\bq
\mrI_{\nloc} = \frac{i\,\pi^2}{M^2}\,\baruppsi_c(x)\,\uppsi(x)\,\upphi_c(x)\,\mrB_0(p^2\,;\,0\,,\,0) \spc
\eq
and a local one, also of $\ord{1/M^2}$,
\bq
\mrI_{\loc} = \frac{i\,\pi^2}{M^2}\,\baruppsi_c(x)\,\uppsi(x)\,\upphi_c(x)\,\mra_0(M) \spc
\eq
where
\bq
\mra_0(M) = - \frac{1}{\epb} + \ln\frac{M^2}{\muRs} - 1 \spc
\quad
\mrB_0(p^2\,;\,0\,,\,0) = \frac{1}{\epb}  - \ln\,\frac{p^2 - i\,0}{\muRs} + 2 \spc
\eq
with $1/\epb= 2/(4 - \mrd) - \gamma - \ln \pi$ and where $\muR$ is the renormalization scale; as shown $\mrB_0$ has 
a branch cut along the negative $p^2\,$-axis. The origin of the non-local term is in the heavy{-}light loops.  

More significant examples are the following: 
\bqa
\mrI_{\mrC} &=& \int d^{\mrd}q\,\frac{1}{(q^2 + M^2)\,((q + p_1)^2 + M^2)\,((q + p_1 + p_2)^2 + M^2)} \spc
\nl
\mrI_{\mrD} &=& \int d^{\mrd}q\,\frac{1}{(q^2 + M^2)\,((q + p_1)^2 + m^2)\,((q + p_1 + p_2)^2 + m^2)\,
                                         ((q + p_1 + p_2 + p_3)^2 + m^2)} \spp
\eqa
The iteration of
\bq
\frac{1}{(q + p)^2 + M^2} = \frac{1}{q^2 + M^2}\,\Bigl[ 1 - \frac{p^2 + 2\,\spro{p}{q}}{(q + p)^2 + M^2} \Bigr] \spc
\label{vvexp}
\eq
and suitable changes of the loop momentum give:
\bq
\mrI_{\mrC} = \frac{i\,\pi^2}{M^2}\,\Bigl[ \frac{1}{2} - \frac{1}{6}\,\frac{p^2_1 + (p_1 + p_2)^2}{M^2} +
\ord{\frac{1}{M^4}} \Bigr] \spp
\eq
Results that are polynomial in external momenta (contact interactions) are, by definition, local. For the $4\,$-point 
function we derive
\bqa
\mrI_{\mrD} &=& {\underbrace{- \frac{i\,\pi^2}{M^4}}_{\loc}} \quad + i\,\pi^2\;[\,\mbox{non-loc}\,] \spc
\nl
\mbox{non-loc} &=& \frac{1}{M^2}\,\mrC_0(p_2\,,\,p_3\,;\,m\,,\,m\,,\,m) +
            \frac{1}{M^4}\,\Bigl[ (m^2 - p^2_1)\,\mrC_0(p_2\,,\,p_3\,;\,m\,,\,m\,,\,m) 
\nl
{}&+& 2\,\spro{p_1}{p_2}\,\mrC_{11}(p_2\,,\,p_3\,;\,m\,,\,m\,,\,m) +
      2\,\spro{p_1}{p_3}\,\mrC_{12}(p_2\,,\,p_3\,;\,m\,,\,m\,,\,m) -
      \mrB^{\fin}_0(p_3\,;\,m) \Bigr] + \ord{\frac{1}{M^6}} \spc
\label{nlc}
\eqa
where the UV divergent parts have been shifted to the local part. Non-local terms are given by loops with less internal
(light) lines and show the characteristic pattern of singularities (\eg normal or anomalous~\cite{Passarino:2018wix}
thresholds) of $2(3\,,\,\dots)\,$-point functions. 

\begin{figure}[t]
   \centering
   \vspace{-3.cm}
   \includegraphics[width=0.9\textwidth, trim = 30 250 50 80, clip=true]{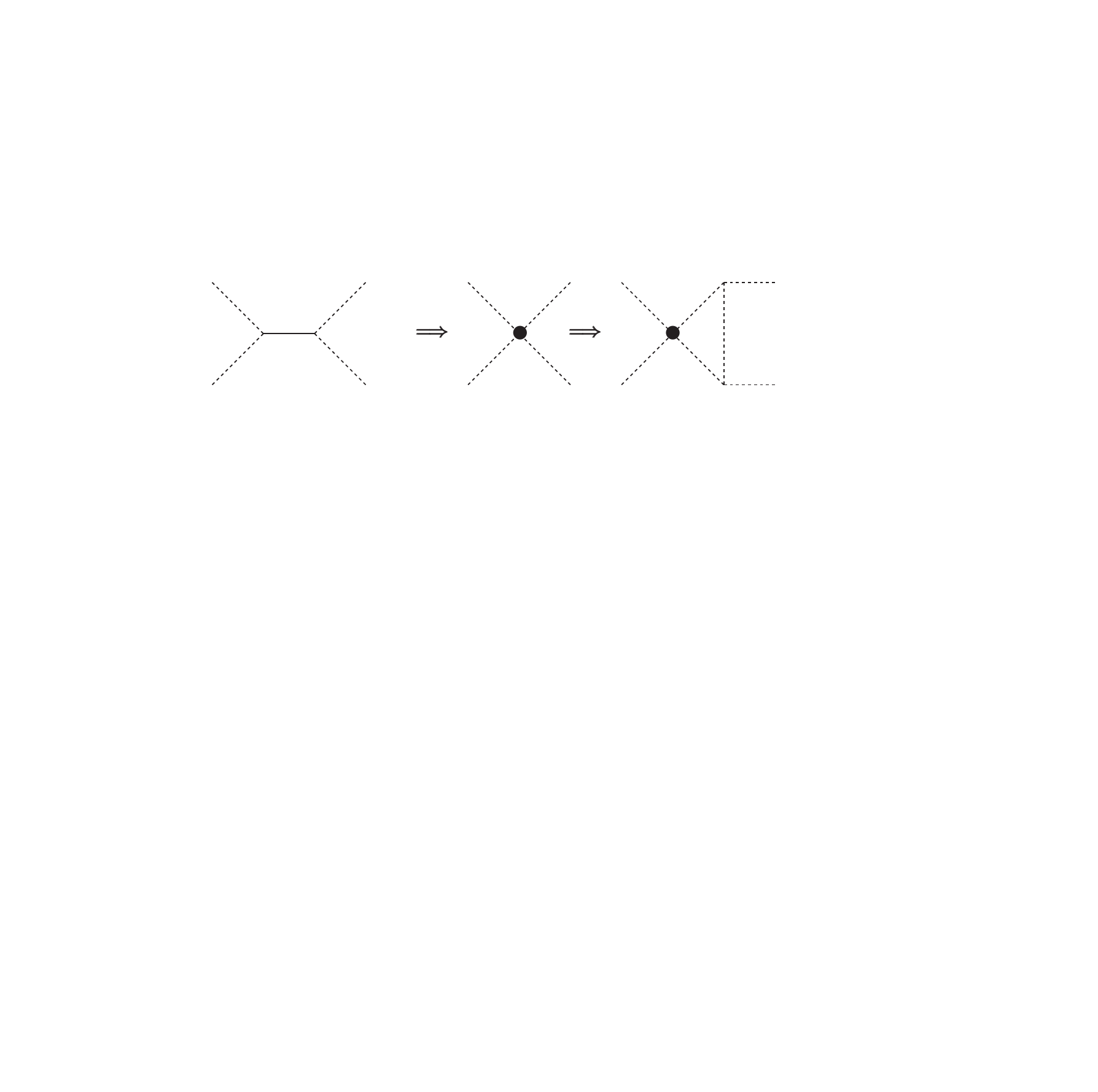}
\vspace{-9.cm}
\caption[]{Tree generated $\mrdim = 6$ operator and its insertion in a loop diagram.}
\label{lfig}
\end{figure}

The procedure based on \eqn{vvexp} is the one used in \Bref{vanderBij:1983bw} to compute two-loop large Higgs mass 
correction to the $\uprho\,$-parameter. To first order it amounts to replace
\bq
\frac{1}{(q + p)^2 + M^2} \to \frac{1}{q^2 + M^2}\,\Bigl[ 1 - \frac{p^2 + 2\,\spro{p}{q}}{q^2 + M^2} \Bigr] \spc
\label{vexp}
\eq
which reproduces the correct result for one-loop diagrams but fails at the two-loop level where both $q$ and $p$ are
loop momenta (the $p$ integral becomes more divergent); of course \eqn{vexp} cannot be used for $p^2 = - M^2$.
More details are given in \Bref{Veltman:1976rt} but one additional example is as follows. Consider the function 
\bq
\mrF(P^2\,,\,M) = \int d^{\mrd}q\,\frac{1}{q^2\,((q + p_1)^2 + M^2)\,(q + p_1 + p_2)^2} \spc
\eq
where $p^2_i = 0$ and $P = p_1 + p_2$. The explicit results is
\bq
\mrF(P^2\,,\,M) = \frac{1}{P^2}\,\Bigl[ \li{2}{1} - \li{2}{1 - \frac{P^2 - i\,0}{M^2}} \Bigr] \spc
\eq
where $\li{2}{\mrz}$ is the di-logarithm of $\mrz$. Using well-known properties of the di-logarithm we obtain the
following expansion
\bqa
\li{2}{1 - \frac{P^2 - i\,0}{M^2}} &=& \li{2}{1} 
\nl
{}&-& \frac{P^2}{M^2} - \frac{1}{4}\,\bigl(\frac{P^2}{M^2}\bigr)^2 + \dots
\nl
{}&+& \frac{P^2}{M^2}\,\Bigl[ 1 + \frac{1}{2}\,\frac{P^2}{M^2} + \dots \Bigr]\,\ln \frac{P^2}{M^2} \spp
\label{nlFexp}
\eqa
The function $\mrF$ has a branch cut along the negative $P^2\,$-axis (normal threshold) which is the origin of the 
non-local terms, third line in \eqn{nlFexp}. The corresponding logarithm could never be represented by a local effective 
lagrangian and is a distinctive feature of long distance (low energy) quantum loops.
Using \eqn{vexp} we obtain
\bqa
\mrF(P^2\,,\,M) &=& \int d^{\mrd}q\,\frac{1}{q^2\,(q^2 + M^2)\,(q + P)^2}\,\Bigl[1 - 
2\,\frac{\spro{P}{q}}{q^2 + M^2} \Bigr] + \dots
\nl
{}&=& \frac{1}{M^2}\,\int d^{\mrd}q\,\Bigl( \frac{1}{q^2} - \frac{1}{q^2 + M^2} \Bigr)\,\frac{1}{(q + P)^2} +
\dots
\nl
&=& \frac{1}{M^2}\,\Bigl\{
\Bigl[ \frac{1}{\epb} - \ln\frac{P^2}{\muRs} + 2 \Bigr] -
\Bigl[ \frac{1}{\epb} - \ln\frac{M^2}{\muRs} + 2 - ( 1 + \frac{M^2}{P^2} )\,\ln( 1 + \frac{P^2 - i\,0}{M^2} ) \Bigr]
\Bigr\} + \dots
\nl
{}&=& \frac{1}{M^2}\,\Bigl( 1 + \ln\frac{M^2}{P^2} \Bigr) + \dots
\eqa
as expected. If the internal masses are $(M\,,\,0\,,\,M)$ instead of $(0\,,\,M\,,\,0)$ the normal threshold is
at $P^2 = - 4\,M^2$, \ie we can Taylor expand around $P^2 = 0$.
\paragraph{Heavy{-}light LO and light NLO} \hspace{0pt} \\
Consider a one-loop diagram in the high-energy theory with one heavy internal line and several light internal
lines. Deriving the corresponding heavy{-}light contribution is equivalent to shrink the heavy line to a point which, on the
other hand, is equivalent to the insertion of one $\mrdim = 6$, tree-generated, operator inside a one-loop diagram of 
the theory where the heavy fields have been removed: the latter is nothing but a part of NLO EFT. Therefore, when going to 
NLO EFT one has to be careful in avoiding double counting if heavy{-}light contributions have already been included.
The presence of non-local terms within the BFM formalism has been nicely illustrated in \Bref{Donoghue:2017pgk}, see their
sect.~$5.1$ and sect.~$7.3$.
To give an example, consider a scalar theory with a light field $\upphi$ and an heavy one $\PH$ (of mass $M$).  
\hyperref[lfig]{In fig.(\ref*{lfig})} we show the tree diagram generating a $\mrdim = 6$ operator, $\upphi^4$ (denoted by
a black box). This operator is inserted into a loop diagram of the light theory reproducing the term of
$\ord{1/M^2}$ in the non-local part of the box \hyperref[lfig]{in Eq.(\ref*{nlc})}. 
\subsubsection{Loops, matching, and all that}
The upshot of this is that the EFT Lagrangian is determined by the local part of the loops, \ie
the local effective action is the one which can be expanded in terms of local operators (only has terms with 
a bounded number of derivatives). 
Diagrams of the underlying theory with light external legs and heavy internal ones admit a local low{-}energy limit. 
As anticipated, diagrams of the underlying theory with light external legs and mixed internal 
legs~\cite{delAguila:2016zcb,Boggia:2016asg,Henning:2016lyp,Jiang:2018pbd} may show normal-threshold singularities in the 
low{-}energy region and give inherently non-local parts which can be matched to one-loop EFT Green's functions
(the two theories are identical in the IR, so non-analytic terms depending on the light fields must be the same).

\begin{figure}[t]
   \centering
   \vspace{-3.cm}
   \includegraphics[width=0.9\textwidth, trim = 30 250 50 80, clip=true]{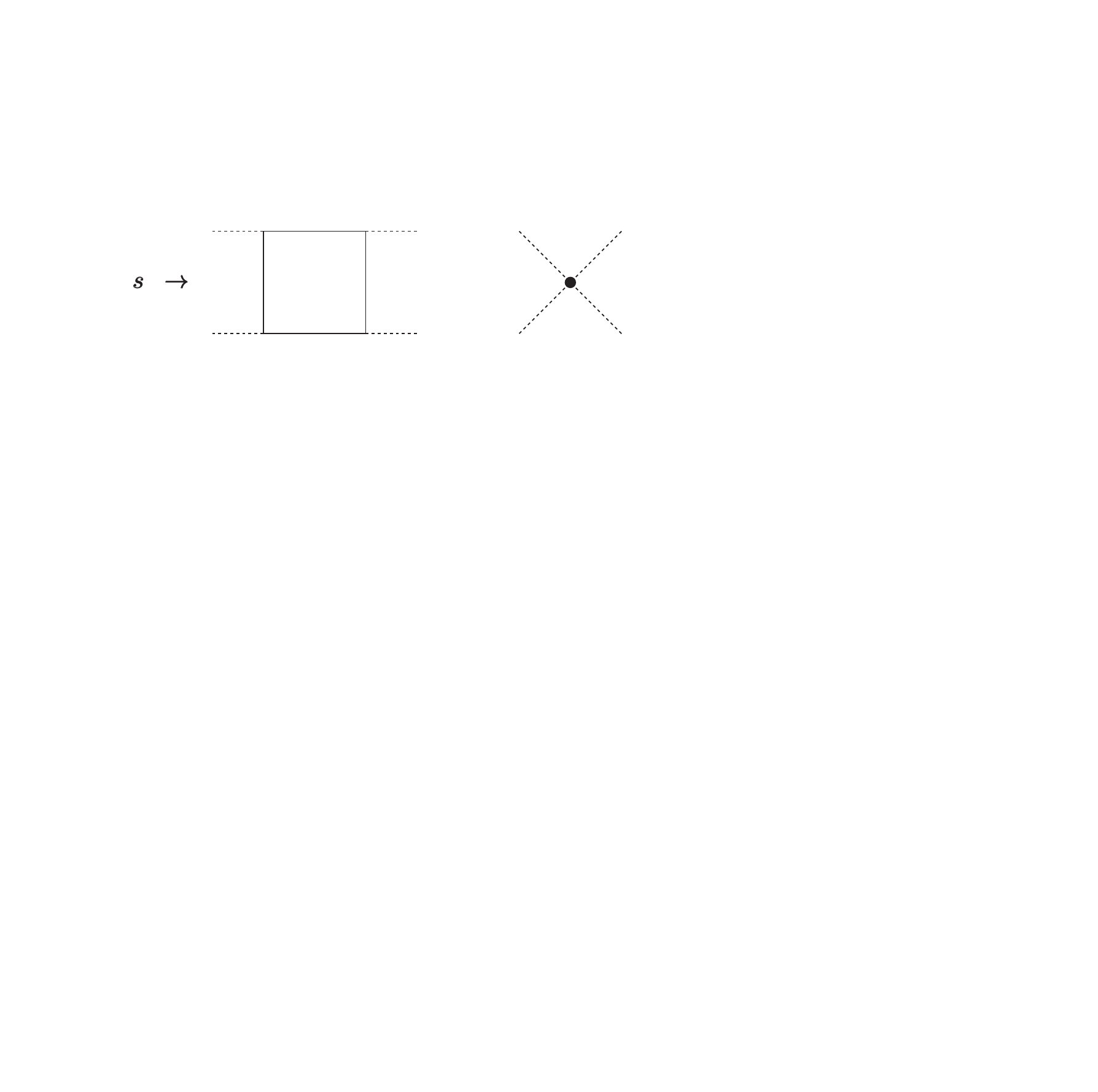}
\vspace{-10.cm}
\caption[]{Correspondence between $X^{\prime}$ and XEFT: the local case.}
\label{locfig}
\end{figure}

\begin{figure}[b]
   \centering
   \vspace{-3.cm}
   \includegraphics[width=0.9\textwidth, trim = 30 250 50 80, clip=true]{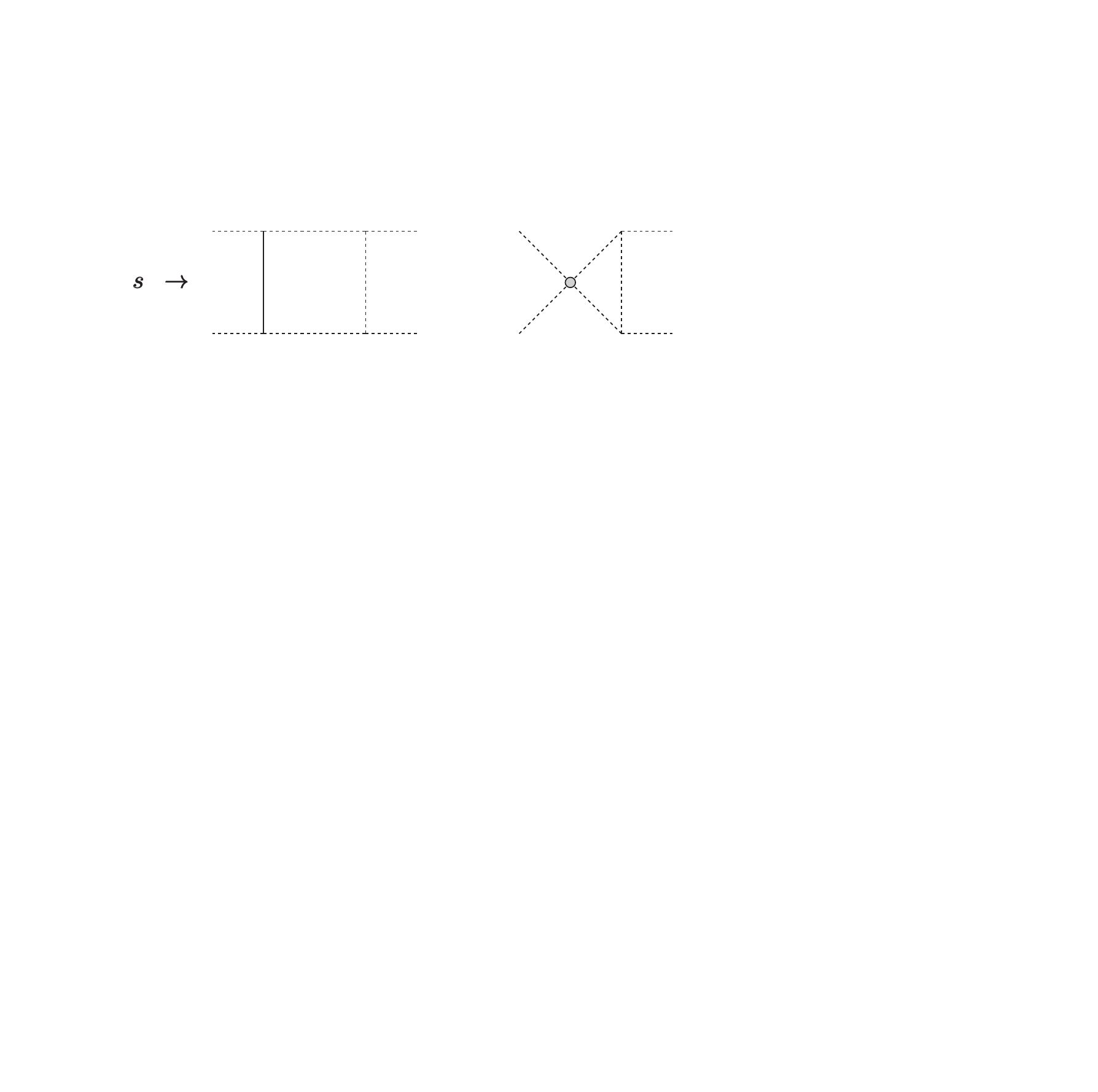}
\vspace{-10.cm}
\caption[]{Correspondence between $X^{\prime}$ and XEFT: the non local case.}
\label{nlocfig}
\end{figure}

The correspondence is shown \hyperref[locfig]{in fig.(\ref*{locfig})} and \hyperref[nlocfig]{in fig.(\ref*{nlocfig})}.
In both figures light particles (of mass $m$) are represented by dashed lines, heavy particles (of mass $M$) by
straight lines. We are interested in the region $\sqrt{s}\,,\,m \muchless M$ and $s > 4\,m^2$. 
The left diagram (underlying theory) \hyperref[locfig]{in fig.(\ref*{locfig})} has a normal threshold
at $s = 4\,M^2$ and corresponds to a loop-generated operator (denoted by a black circle) in the corresponding 
tree-level EFT.
The left diagram (underlying theory) \hyperref[nlocfig]{in fig.(\ref*{nlocfig})} has a normal threshold
at $s = 4\,m^2$ and corresponds (one-loop matching) to the right diagram which is one-loop EFT with the insertion of 
a tree-generated operator (denoted by a grey circle). Therefore, to match amplitudes we also need to compute one-loop 
scattering amplitude in the effective theory, very much in the same way as indicated in \Bref{Ellis:2016enq}.

The full theory is renormalizable and will yield finite predictions in terms of its renormalized (mixed on-shell 
and $\MSB$) parameters~\cite{Denner:2016etu,Altenkamp:2017kxk}.
The EFT will have quite different UV properties because it is missing the heavy degrees of freedom
(but the divergencies will disappear through the renormalization of Wilson coefficients), however the low
energy effects will be similar in both calculations. In particular, the non-local behavior is exactly what is found 
by taking the low{-}energy limit of the full theory, expanded to this order~\cite{Donoghue:1998dd,Donoghue:2017pgk}.
The key advantages are a more precise matching and the appearance of some important kinematic dependence, \eg also
the coefficients of ``kinematic'' logarithms match~\cite{Donoghue:2017pgk}; an explicit example, 
$\PGpp \PGpz \to \PGpp \PGpz$ in the linear sigma{-}model, is shown in sect.~$7.3$ of \Bref{Donoghue:2017pgk}.   
As far as normal threshold singularities are concerned we mention that loop calculations should be performed in
the so-called complex-mass scheme~\cite{Denner:2006ic,Actis:2006rb,Denner:2014zga}.
As a matter of fact, \Bref{Donoghue:2017pgk} underlines that the most important predictions of the EFT are related to 
non–analytic in momenta loop contributions which modify tails of distributions.
For a covariant treatment of logarithms in the EFT expansion \Bref{Barvinsky:1994cg} points out that we can use a spectral 
representation, \ie
\bq
\ln\bigl( -\,\frac{\Box}{\muRs} \bigr) = \int_0^{\infty} dm^2\,\Bigl( \frac{1}{m^2 + \muRs} -
\frac{1}{m^2 - \Box} \Bigr) \spp
\eq
The traditional approach splits the non-local contributions into one-loop EFT diagrams with TG operator 
insertions (\eg the ``non-loc'' term \hyperref[nlc]{in Eq.(\ref*{nlc})}), which are not inserted into the EFT 
Lagrangian and LG operators used at tree level (\eg first term \hyperref[nlc]{in Eq.(\ref*{nlc})}) which constitute 
the heavy–light contributions to matching.
This splitting should be carefully described since triangles, boxes etc have both local and non-local parts,
as shown \hyperref[nlFexp]{in Eq.(\ref*{nlFexp})}.
The authors of \Bref{Donoghue:2017pgk} draw our attention to the fact that non–local effects correspond to long 
distance propagation and hence to the reliable predictions at low energy. The local terms by contrast summarize 
the unknown effects from high energy. Having both local and non–local terms allows us to implement
the full EFT program.

One way or the other, once the matching is done, all processes can be calculated using 
the ``fitted'' (renormalized) Wilson coefficients without the need to match again for each process.

The authors of \Bref{Fuentes-Martin:2016uol} use a different language, based on 
``expansion by regions''~\cite{Beneke:1997zp,Jantzen:2011nz}, separating the hard region contribution 
from the soft one:
\bei

\item[hard] $\mid\,q^2\,\mid\,\sim\,M^2\,\gg\,\mid\,p^2_i\,\mid$

\item[soft] $M^2\,\gg\,\mid\,q^2\,\mid\,\sim\,\mid\,p^2_i\,\mid$

\eei
and reach the conclusion that the EFT Lagrangian at one-loop is then containing only the hard part of the loops.


\paragraph{Low{-}energy limit, Mellin{-}Barnes expansion }\hspace{0pt} \\
If we deal with the low{-}energy limit from a diagrammatic point of view~\cite{Passarino:2012cb} the best technique 
is given by Mellin{-}Barnes expansion~\cite{HTF}. Consider a simple example,
\bq
\mrI = \int d^{\mrd} q\,\frac{1}{(q^2 + m^2)\,((q + p_1)^2 + M^2)\,((q + p_1 + p_2)^2 + m^2)} \spc
\eq
where $p^2_i = - m^2$ and $s = - (p_1 + p_2)^2$. We introduce
\bq
M^2 = \lambda\,m^2 \spc \qquad
s = 4\,m^2\,\mrr \spc
\eq
and derive the following representation in Feynman parameter space,
\bq
\mrI = \frac{1}{m^2}\,\int_0^1 dx\,\int_0^x dy\,\bigl[ x^2 - 4\,\mrr\,y\,(x - y) + \lambda\,(1 - x) \Bigr]^{-1} \spc
\eq
where $\mrr$ must be understood as $\mrr + i\,0$ and $\lambda$ as $\lambda - i\,0$. It follows
\bq
\mrI = \frac{1}{2\,\pi\,i}\,\int_{-\,i\,\infty}^{+\,i\,\infty} d\mrt\,\mrB(\mrt\,,\,1 - \mrt)\,
\int_0^1 dx\,\int_0^x dy\,\bigl[ \lambda\,(1 - x) \bigr]^{\mrt - 1}\,\bigl[ x^2 - 4\,\mrr\,y\,(x - y) \bigr]^{- \mrt} \spc
\eq
which is valid in the vertical strip $0 < \Re \mrt < 1$ and where $\mrB$ is the Euler beta{-}function. 
The $x{-}y$ integral can be expressed as
\bq
\mrB(2 - 2\,\mrt\,,\,\mrt)\,\int_0^1 dy\,\bigl[ 1 - 4\,\mrr\,y\,(1 - y) \bigr]^{- \mrt} \spc
\eq
giving 
\bq
\mrI = \frac{1}{2\,\pi\,i}\,\,\int_{-\,i\,\infty}^{+\,i\,\infty} d\mrt\,\lambda^{\mrt - 1}\,\mrB(\mrt\,,\,1 - \mrt)\,
\mrB(2 - 2\,\mrt\,,\,\mrt)\,
\,\int_0^1 dy\,\bigl[ 1 - 4\,\mrr\,y\,(1 - y) \bigr]^{- \mrt} \spp
\eq
Since we are interested in the limit $\lambda \to \infty$, the $\mrt\,$-integral will be closed over the left-hand 
complex half-plane at infinity, with double poles at $\mrt = 0\,,\,- 1\,,\,\dots$. Using the well know Laurent and 
Taylor expansions of the Euler gamma-function we obtain the result summing over the poles; the leading term in 
the expansion, $\ord{1/\lambda}$, gives
\bq
\mrI = \frac{1}{M^2}\,\Bigl\{ 1 + \ln \lambda - 
\int_0^1 dy\,\ln \Bigl[ 1 - 4\,\mrr\,y\,(1 - y) \Bigr] \Bigr\} + \ord{\frac{1}{M^4}} \spc
\eq
showing the large momentum logarithm.
For $s < 4\,m^2$ (unphysical region) we can expand the second logarithm in powers of $\mrr$ (Taylor expansion); 
for $ s > 4\,m^2$ we are above the normal threshold and the $y\,$-integral is the UV finite part of a two-point 
functions, showing the non-local, kinematic, logarithm
\bq
- \beta\,\ln\frac{\beta + 1}{\beta - 1} \spc \qquad
\beta^2 = 1 - \frac{1}{\mrr} \spp
\label{blog}
\eq
To summarize, for $\sqrt{s}\,,\,m \muchless M$, we can Taylor expand only in the region $s < 4\,m^2$,
Note the presence of an imaginary part in \eqn{blog}, $\pi\,\beta$, when $s > 4\,m^2$. 
Non-local contributions to the effective Lagrangian usually contain the function
\bq
\mrL(\mrz) = \int\,\frac{d^4q}{(2\,\pi)^4}\,\exp\{i\,\spro{q}{\mrz}\}\,\ln\frac{q^2}{\mu^2} \spc
\eq
inside a term $\upphi(x)\,\mrL(x - y)\,\upphi(y)$.

The next term in the expansion is given by the residue of the pole at $\mrt = - 1$ and gives
\bq
\frac{1}{M^4}\,\frac{2}{\beta^2 - 1}\,\Bigl[ 1 + \beta^2 - 2\,(1 - 3\,\beta^2)\,\ln\frac{M^2}{m^2} -
4\,\beta^3\,\ln\frac{\beta + 1}{\beta - 1} \Bigr] \spc
\eq
showing again large momentum and kinematic logarithms. 
\subsection{LG operators,  results}
\paragraph{Heavy scalar, explicit results} \hspace{0pt} \\
Before presenting the list of $\mrdim = 6$ operators and their coefficients, functions of the parameters in the
high-energy theory, some considerations are needed. Following sect.~7 of \Bref{tHooft:1973bhk} we use the fact that,
within the BFM algorithm, one may make use of the (BFM) EoMs to simplify the counter-Lagrangian. In our case they
read as follows:
\bq
\sla{\partial}\,\uppsi_c = \lambda_1\,\upphi_c\,\uppsi_c \spc
\qquad 
\sla{\partial}\,\baruppsi_c = - \lambda_1\,\upphi_c\,\baruppsi_c \spc
\label{bfmeq}
\eq
where we have taken $\mrS_c = 0$. \hyperref[expan]{From Eq.(\ref*{expan})} we find UV poles proportional to
\bq
\baruppsi_c\,\gamma^{\mu}\,\uppsi_c\,\pdmu \upphi^2_c =
- \upphi^2_c\,\Bigl[ (\pdmu\,\baruppsi_c)\,\gamma^{\mu}\,\uppsi_c +
                     \baruppsi_c\,\gamma^{\mu}\,\pdmu \uppsi_c \Bigr] + \mbox{t.d.}
\eq
which is zero by virtue of \eqn{bfmeq}.
Staring \hyperref[expan]{from Eq.(\ref*{expan})} we derive UV divergent terms of $\mrdim = 4$, \eg
\bq
2\,\lambda_1\,\lambda^2_2\,\bigl( \frac{1}{\epb} - \ln\frac{M^2}{\muRs} \bigr)\,
\baruppsi_c\,\uppsi_c\,\upphi_c \spc
\eq
which are canceled by a counter-Lagrangian, \ie we start with
\bqa
\Lag_{\mrY \mrS} &=& - \mrZ_{\upphi}\,\frac{1}{2}\,\pdmu \upphi\,\pdmu \upphi 
 - \frac{1}{2}\,m^2\,\mrZ_m\,\mrZ_{\upphi}\,\upphi^2  
       - \frac{1}{4}\,g\,\mrZ_g\,\mrZ^2_{\upphi}\,\upphi^4 
- \mrZ_{\uppsi}\,{\overline{\uppsi}}\,( \sla{\partial} - \mrZ_1\,\mrZ^{1/2}_{\upphi}\,\lambda_1\,\upphi )\,\uppsi 
\nl
{}&-&    \frac{1}{2}\,\mrZ_{\mrS}\,\pdmu \mrS\,\pdmu \mrS 
       - \frac{1}{2}\,M^2\,\mrZ_M\,\mrZ_{\mrS}\,\mrS^2 
       - \frac{1}{4}\,\lambda_4\,\mrZ_4\,\mrZ^2_{\mrS}\,\mrS^4 
       + \lambda_2\,\mrZ_2\,\mrZ_{\uppsi}\,\mrZ^{1/2}_{\mrS}\,{\overline{\uppsi}}\,\uppsi\,\mrS
       - \lambda_3\,\mrZ_3\,\mrZ_{\upphi}\,\mrZ_{\mrS}\,\upphi^2\,\mrS^2 \spc
\label{LYSren}
\eqa
instead \hyperref[LYS]{ofEq.(\ref*{LYS})}. The remaining $\mrdim = 6$ terms are as follows:
\bei
\item[{\underline{Local}}] Local operators are shown in tab.~\ref{Sdim6list}.
\eei
We can conceive a scenario where the fit has been performed using the basis~\cite{Einhorn:2013kja}
\bq
\mbox{Basis} \quad = \quad \{ \upphi^6\,,\,\baruppsi\,\uppsi\,\upphi^3\,,\,(\baruppsi\,\uppsi)^2\,,\,
(\baruppsi\,\gamma^5\,\uppsi)^2\,,\,(\baruppsi\,\gamma^{\mu}\uppsi)^2 \} \spc
\label{basisO}
\eq
of potentially tree generated operators. The explicit calculation returns the following set of operators:
\bq
\{ \upphi^6\,,\,\baruppsi\,\uppsi\,\upphi^3\,,\,(\baruppsi\,\uppsi)^2\,,\,(\baruppsi\,\gamma^{\mu}\uppsi)^2\,,\,
\upphi^3\,\Box \upphi\,,\,\baruppsi\,\uppsi\,\Box \upphi \} \spc
\label{genO}
\eq
with equivalence relations
\bq
\upphi^3\,\Box \upphi \equiv ( \upphi^6\,,\,\baruppsi\,\uppsi\,\upphi^3 ) \spc 
\qquad
\baruppsi\,\uppsi\,\Box \upphi \equiv ( \baruppsi\,\uppsi\,\upphi^3\,,\,(\baruppsi\,\uppsi)^2 ) \spp
\label{equiO}
\eq
We stress that the equivalence of operators in \eqn{equiO} is a property of the EFT and can be used to change basis
in the EFT while the high{-}energy theory generates the operators in \eqn{genO} with explicit Wilson coefficients given
in tab.~\ref{Sdim6list}. Mapping the parameters of the high{-}energy theory into the fitted Wilson coefficients
corresponding to the operators listed in \eqn{basisO} requires field transformations at the EFT level 
with $\mrdim = 8$ compensations.
\begin{table}
\begin{center}
\begin{tabular}{ll}
\hline
Coefficient & Operator \\
\hline
& \\
$\frac{\lambda^2_2}{M^2}\,( \lambda^2_1\,\ln\frac{m^2}{M^2} - \frac{1}{2}\,\lambda^2_2 )$ &
$\baruppsi_c\,\gamma^{\mu}\,\uppsi_c\,\baruppsi_c\,\gamma^{\mu}\,\uppsi_c$ \\
& \\
$- \frac{8}{9}\,\frac{\lambda^2_3}{M^2}$ & $\upphi_c^3\,\Box\,\upphi_c$ \\
& \\
$\frac{4}{3}\,\frac{\lambda^3_3}{M^2}$ & $\upphi^6_c$ \\
& \\
$\frac{\lambda_1\,\lambda^2_2}{M^2}\,( \frac{13}{16} + \ln\frac{M^2}{\muRs} )$ &
$\baruppsi_c\,\uppsi_c\,\Box\,\upphi_c$ \\
& \\
$- 4\,\frac{\lambda_1\,\lambda^2_2\,\lambda_3}{M^2}$ & $\baruppsi_c\,\uppsi_c\,\upphi^3_c$ \\
& \\
\hline
\end{tabular}
\end{center}
\label{Sdim6list}
\end{table}
\bei
\item[{\underline{Non-local}}] The corresponding contributions can be written as follows:
\eei
\bqa
\mbox{non-local} &=& (2\,\pi)^4\,\mrN_1\,\int d^4p d^4q_1 d^4q_2\,\delta^{(4)}(p + q_1 + q_2)\, 
\baruppsi_c(q_1)\,\uppsi_c(q_2)\,\upphi_c(p) 
\nl\nl
{}&+&
(2\,\pi)^4\,\sum_{i=1,2}\,\mrN_{2i}\,\int d^4p_1 d^4p_2 d^4q_1 d^4q_2\,\delta^{(4)}(p_1 + p_2 + q_1 + q_2)
\nl
{}&\times& \baruppsi_c(q_1)\,\sla{p_i}\,\uppsi_c(q_2)\,\upphi_c(p_1)\,\upphi_c(p_2) \spc
\eqa
where
\bq
\mrN_1 = \lambda_1\,\lambda^2_2\,\frac{p^2}{M^2}\,\ln(p^2) \spc
\qquad
\mrN_{2i} = \lambda^2_1\,\lambda^2_2\,\frac{\mrF_i(p_1\,,\,p_2)}{M^2} \spp
\eq
The functions $\mrF_i$ are combinations of three-point tensor integrals, 
\bqa
\mrF_1(p_1\,,\,p_2) &=& \frac{1}{2}\,\ln\frac{p^2_1}{\muRs} - \ln\frac{p^2_2}{\muRs} 
\nl
{}&-& \mrC_{11}(p_1\,,\,p_2\,;\,0\,,\,0\,,\,0)\,p^2_1 -
      \mrC_{11}(p_2\,,\,p_1\,;\,0\,,\,0\,,\,0)\,p^2_2 
\nl
{}&-& \mrC_{12}(p_1\,,\,p_2\,;\,0\,,\,0\,,\,0)\,p^2_2 -
      \mrC_{11}(p_2\,,\,p_1\,;\,0\,,\,0\,,\,0)\,(p^2_2 + 2\,\spro{p_1}{p_2})
\nl
{}&-& 2\,\mrC_{21}(p_1\,,\,p_2\,;\,0\,,\,0\,,\,0)\,p^2_1 -
      2\,\mrC_{22}(p_2\,,\,p_1\,;\,0\,,\,0\,,\,0)\,\spro{p_1}{p_2}
\nl
{}&-& 2\,\mrC_{23}(p_1\,,\,p_2\,;\,0\,,\,0\,,\,0)\spro{p_1}{p_2} -
      2\,\mrC_{23}(p_2\,,\,p_1\,;\,0\,,\,0\,,\,0)\,p^2_2
\nl
{}&-& 2\,\mrC^{\fin}_{24}(p_1\,,\,p_2\,;\,0\,,\,0\,,\,0)\spro{p_1}{p_2} \spc
\label{nlcfun}
\eqa
where $\mrC^{\fin}_{24}$ denotes the UV finite part of $\mrC_{24}$ (for a definition of the $\mrC\,$-functions see
ch.~$5.1.4$ of \Bref{Bardin:1999ak}). Finally, $\mrF_2(p_1\,,\,p_2) = \mrF_1(p_2\,,\,p_1)$.
It is obvious from \eqn{nlcfun} that there is no advantage in deriving the non-local terms with the expansion 
\hyperref[expan]{of Eq.(\ref*{expan})}; it is actually more convenient to use a diagram-by-diagram calculation 
in the high-energy theory, followed by an expansion in $M$.

More details on matching can be found in \Brefs{Skiba:2010xn,Henning:2016lyp,Zhang:2016pja}: for the matching procedure, 
we can use any choice of external momenta, \ie the momenta can be either on-shell or off-shell and for any small external 
momenta the full and effective theories must be identical.
\paragraph{Heavy vectors, explicit results} \hspace{0pt} \\
We have deliberately chosen this example to show the kind of problems that arise when the high-energy theory is
not UV complete. This fact should not be confused with a negative statement on renormalizability of the (interacting) 
Proca theory where one introduces the Stueckelberg ghost~\cite{Stueckelberg:1938zz}, into the action of a vector field with 
a ``naive mass term''.
In this way the Lagrangian leads to a manifestly gauge invariant action; this gauge invariance remains an exact 
symmetry, if matter fields are introduced and the vector field is minimally coupled to conserved currents of 
these fields. This, however, is not the case that we are considering. 
It is worth mentioning that we are dealing with an accidental property of massive electrodynamics. Massive gauge fields
are nonrenormalizable in general, \ie for a gauge group other than $U(1)$.

In our example the local terms give rise to the following Lagrangian:
\bqa
\mbox{local boson} &=&
\frac{1}{2}\,\lambda_2\,\bigl( \frac{3}{\epb} - 3\,\ln\frac{M^2}{\muRs} + 1 \bigr)\,M^2\,\upphi^2_c
\nl
{}&+& \frac{1}{8}\,\lambda^2_2\,\bigl( \frac{3}{\epb} - 3\,\ln\frac{M^2}{\muRs} - 2 \bigr)\,M^2\,\upphi^4_c
\nl
{}&-& \frac{3}{8}\,\frac{\lambda^3_2}{M^2}\,\bigl( \frac{1}{\epb} - \ln\frac{M^2}{\muRs} \bigr)\,\upphi^6_c
\nl
{}&+& \frac{1}{4}\,\frac{\lambda^2_2}{M^2}\,\bigl( \frac{9}{\epb} - 9\,\ln\frac{M^2}{\muRs} - \frac{3}{2} \bigr)\,
\upphi^2_c\,\pdmu \upphi_c\,\pdmu \upphi_c \spc
\nl\nl
\mbox{local fermion} &=& 
- \frac{1}{4}\,\frac{\lambda_2\,\lambda^2_3}{M^2}\,\baruppsi_c\,\sla{\partial}\,\uppsi_c\,\upphi^2_c
\nl
{}&-& 4\,\frac{\lambda^2_2\,\lambda^2_3}{M^2}\,\bigl( \frac{1}{\epb} + 3\,\ln\frac{M^2}{\muRs} -
        4\,\ln\frac{m^2}{\muRs} + \frac{3}{2} \bigr)\,
        (\baruppsi_c\,\uppsi_c)^2
\nl
{}&-& 3\,\frac{\lambda^4_3}{M^2}\,\bigl( \frac{1}{\epb} + \frac{190}{3}\,\ln\frac{M^2}{\muRs} + \frac{55}{36} \bigr)\,
          \baruppsi_c\,\gamma^{\mu}\,\uppsi_c\,\baruppsi_c\,\gamma^{\mu}\,\uppsi_c
\nl
{}&-& 3\,\frac{\lambda^4_3}{M^2}\,
          \baruppsi_c\,\gamma^{\mu}\,\gamma^5\,\uppsi_c\,\baruppsi_c\,\gamma^{\mu}\,\gamma^5\,\uppsi_c
\nl
{}&-& \frac{\lambda^2_1\,\lambda^2_3}{M^2}\,\bigl( \frac{1}{\epb} + 3\,\ln\frac{M^2}{\muRs} -
            4\,\ln\frac{m^2}{\muRs} + \frac{3}{2} \bigr)\,
          \baruppsi_c\,\sigma^{\mu\nu}\,\uppsi_c\,\baruppsi_c\,\sigma^{\mu\nu}\,\uppsi_c \spp
\eqa
In conclusion, Wilson coefficients show non renormalizable terms, as expected. This simple example shows the
riskiness of extending the SM with new vector bosons having ``naive mass terms''; the standard argument against 
is that longitudinal components do not have a falling off propagator, which means that loops involving longitudinal 
vectors blow up as a power at high momentum. In gauge theories, gauge invariance guarantees that longitudinal bosons 
are not produced, but this requires that gauge invariance is broken spontaneously not by the Lagrangian. 
\section{Interpreting derivative-coupled field theories \label{Matt}}
An old problem seen through the lens of EFT: consider the following Lagrangian,
\bq
\Lag = - \frac{1}{2}\,\pdmu \upphi\,\pdmu \upphi - \frac{1}{2}\,m^2 \upphi^2  
       - \frac{1}{4}\,\lambda \upphi^4 - \frac{1}{2}\,\frac{\mra}{\Lambda^2}\,\upphi\, \Box^2\, \upphi \spc
\label{olag}
\eq
where $\mra$ is a Wilson coefficient and $\Lambda >> m$ is a large scale. We can write
$\Lag = \Lag_0 + \Lag_{\mri}$ with two options
\bqa
\Lag^{(a)}_0 = \frac{1}{2}\,\upphi\,\lpar \Box - m^2 - \frac{\mra}{\Lambda^2}\,\Box^2 \rpar\,\upphi \spc 
&\qquad&
\Lag^{(a)}_{\mri} = - \frac{1}{4}\,\lambda\,\upphi^4 \spc
\nl
\Lag^{(b)}_0 = \frac{1}{2}\,\upphi\,\lpar \Box - m^2 \rpar\,\upphi \spc 
&\qquad&
\Lag^{(b)}_{\mri} = - \frac{1}{2}\,\frac{\mra}{\Lambda^2}\,\upphi\,\Box^2\,\upphi- \frac{1}{4}\,\lambda\,\upphi^4 \spp
\label{twoo}
\eqa 
The equation of motion (EoM) gives
\bq
\lpar \Box - m^2 - \frac{\mra}{\Lambda^2}\,\Box^2 \rpar\,\upphi - \lambda\,\upphi^3 = 0 \spp
\eq
The first point to be investigated is the question of separation of the quadratic, \ie propagator defining, part
in $\Lag$. Thus let there be given a Lagrangian
\bq
\Lag = \upphi\,\mrU\,\upphi + \upphi\,\mrV\,\upphi \spp
\eq
One can either say that one has a propagator $- (\mrU + \mrV)^{-1}$ for the $\upphi\,$-field or alternatively
a propagator $- \mrU^{-1}$ and a vertex $\mrV$. Obviously, the two cases give the same result; however this
requires summing over all possible insertions of the vertex $\mrV$, which is different from a truncated expansion 
in $\mrV$.
\subsection{Option a)  \label{opta}}
Consider $\Lag^{(a)}$ in \eqn{twoo}: the first point to be investigated is the validity of 
Matthews's theorem~\cite{Matt}, \ie the Feynman rules are just those obtained by using $\Lag_{\mri}$ to determine
the vertices and the covariant $\mrT^*$ product to determine the propagators (in other words, one can read Feynman
rules from the Lagrangian).
The validity of the theorem for the Lagrangian of \eqn{olag} has been proven long ago in \Bref{Bernard:1974st} (see also
\Bref{GrosseKnetter:1993td}) where the authors used an indirect implementation of canonical methods, finding an 
equivalent Lagrangian which contains only first derivatives but yields the same results (the standard method, 
originally due to Ostrogradsky~\cite{Ostro}).
Always in \Bref{Bernard:1974st} one can find the spectrum of the theory: there are two masses, solutions
of the equation
\bq
\frac{\mra}{\Lambda^2}\,\mu^4 - \mu^2 + m^2 = 0 \spp
\label{roots}
\eq
To exclude tachyons we must have
\bq
\mra > 0 \spc 
\qquad
\mra\,\frac{m^2}{\Lambda^2} < \frac{1}{4} \spp
\eq
However, there is a negative metric for the particle with the larger mass, \ie there is a ``ghost'' in the spectrum.

In the Ostrogradsky formalism the higher order derivatives are considered to be independent objects; however, this means 
that new degrees of freedom are introduced. These additional degrees of freedom involve the unphysical effects. 
To generalize the example consider a $\mrdim = 8$ operator
\bq
\Lag_8 = - \frac{1}{2}\,\frac{\mra}{\Lambda^4}\,\upphi\,\Box^3\,\upphi \spp
\eq
Two new fields are needed to write an equivalent first-order Lagrangian
\bq
\Lag_8 = \frac{1}{\sqrt{\mra}}\,\Lambda^2\,\uppsi_1\,\uppsi_2 - 
         \uppsi_1\,\Box\,\uppsi_1 + \frac{1}{\sqrt{2}}\,\uppsi_2\,\Box\,\upphi \spc
\eq
which means that there could be more ghosts in the spectrum; indeed the mass spectrum is given by the values
of $\mu^2$ that satisfy the cubic equation 
\bq
\frac{a}{\Lambda^4}\,\mu^6 - \mu^2 + m^2 \spc
\eq
as it can be seen by inserting the Fourier decomposition for $\upphi$ into
the corresponding (linear) equation of motion. As a matter of fact all roots are real only if 
$\mra$ is positive and less tha $4/27\,\Lambda^4/m^4$. However the product of the roots is $- m^2$; therefore,
there is at leat one tachyon in the spectrum. 
\subsection{Option b)  \label{optb}}
With option b) one works at first order in $1/\Lambda^2$. The ``dangerous'' term (second order derivatives) is 
substituted by using EoMs where terms of $\mcO(\Lambda^{-2})$ are neglected,
\bq
\Box\,\upphi = m^2\,\upphi^2 + \lambda\,\upphi^3 \spc
\eq
this can be achieved by using the transformation
\bq
\upphi \to \upphi + \frac{1}{2}\,\frac{\mra}{\Lambda^2}\,\Box \upphi =
\upphi + \frac{1}{\Lambda^2}\,\Ope^{(6)}_{\Box} \spp
\eq
Of course one could work at second order in $\Lambda^{-2}$, including
$\mrdim = 8$ operators, \etc. The result is
\bqa
\Lag &=& \Lag^{(4)} + \Lag^{(6)} + \Lag^{(8)} \spc 
\nl
\upphi &\to& \upphi + \frac{1}{\Lambda^2}\,\Ope^{(6)}_{\Box} \spc
\qquad
\Delta \Lag = \Bigl[ \frac{\delta \Lag^{(4)}}{\delta \upphi} + \frac{\delta \Lag^{(6)}}{\delta \upphi} \Bigr]\,
\frac{1}{\Lambda^2}\,\Ope^{(6)}_{\Box} + \frac{1}{2}\,\frac{\delta^2 \Lag^{(4)}}{\delta \upphi^2}\,
\frac{1}{\Lambda^4}\,\bigl[ \Ope^{(6)}_{\Box} \bigr]^2 \spc
\nl
\upphi &\to& \upphi + \frac{1}{\Lambda^4}\,\Ope^{(8)}_{\Box} \spc
\qquad
\Delta \Lag = \frac{\delta \Lag^{(4)}}{\delta \upphi}\,\frac{1}{\Lambda^4}\,\Ope^{(8)}_{\Box} \spc
\eqa
where $\Ope^{(8)}_{\Box}$ contains $\upphi\,\Box^3\,\upphi$.
There is a diagrammatic proof of these relations, see sect.~$10.4$ of \Bref{'tHooft:186259}.
\paragraph{To summarize} \hspace{0pt} \\
EoMs can be applied in order to covert terms in the interaction Lagrangian~\cite{Arzt:1993gz};
the statement is nontrivial because, in general, EoMs must not be inserted into the Lagrangian. However, one can find 
field transformations which have the same effect as the application of EoMs and yield Lagrangians that are
physically equivalent, see \Brefs{Kallosh:1972ap,Arzt:1993gz,Passarino:2016saj}.   
A field redefinition~\cite{Passarino:2016pzb,Passarino:2016saj} absorbs the first two terms in \eqn{EoM} 
in the $\mrdim = 4$ part of the Lagrangian while the rest is treated as an interaction term.
This is the strategy adopted in \Bref{Grzadkowski:2010es} to construct the Standard Model effective
field theory (SMEFT).

Therefore, option b) is an effective realization of the original Lagrangian, where one assumes that the effective
theory will be replaced by a ``well-behaved'' new theory at some larger scale, therefore justifying a truncated 
perturbative expansion in $1/\Lambda^2$, even in the quadratic part of the Lagrangian. 
To summarize, option b) is based on the supposed existence of a ``well-behaved'' UV completion
which justifies the omission of all unphysical effects.
It is worth noting, see \Bref{GrosseKnetter:1993td}, that the assumption
$\ep = (m^2/\Lambda^2)\,\mra \muchless 1$ alone is not sufficient since theories with higher derivatives have no
analytic limit for $\ep \to 0$. For more details on higher derivative Lagrangians see
\Brefs{Pais:1950za,Batalin:1987fx,Simon:1990ic}.

Nevertheless, \hyperref[opta]{option a) (\ref*{opta})} tells us something about the range of validity of 
the effective Lagrangian, \ie
\begin{enumerate}
\item $0 < \frac{m^2}{\Lambda^2}\,\mra < \frac{1}{4}$ (no tachyons),
\item $\mrE \muchless \mu_{+}$, where $\mu^2_{+}$ is the upper real (positive) root of \eqn{roots} and
$E$ is the scale at which we test the predictions of the effective Lagrangian, \ie $E$ must be well below
the region where the (resummed) theory develops ghosts.
\end{enumerate}
In the realistic SMEFT scenario, bases for effective field theories~\cite{Einhorn:2013kja}, the so-called
Warsaw basis~\cite{Grzadkowski:2010es} in particular, are constructed by using the fact 
that all possible operators in class $\upphi^2\,\mrD^4$ (where now $\mrD$ is a covariant derivative) reduce by 
EoMs either to operators containing fermions or to classes $X^3, X^2\,\upphi^2, \upphi^6$ and $\upphi^4\,\mrD^2$, 
where $X$ is any of the SM field strength tensors. Keeping instead the $\upphi^2\,\mrD^4$ class in the zeroth order
Hamiltonian would be equivalent to option a).

Theories obtained by canonical quantization of a second-order Lagrangian, option a), are
completely equivalent to the Pauli-Villars regularization of the corresponding first-order theories. However,
the regulator mass is not here taken to infinity~\cite{Bernard:1974st}. Other examples can be found in
\Brefs{Lee:1969fy,Lee:1970iw}, Abelian gauge theory with a $\mrF_{\mu\nu}\,\Box\,\mrF^{\mu\nu}$ term; in 
\Bref{Grinstein:2007iz}, with a whole extension of the Standard Model inserting quadratic differential operators;
\Bref{Solomon:2017nlh} discusses effective field theory for general scalar-tensor theories;

Finally, let us consider a top-down approach: in any complete theory a term $\upphi\,\Box^2\,\upphi$ 
is part of the $\upphi$ (light) two-point function with heavy-heavy or light-heavy internal lines. 
In this case we compute the light-light self-energy and then Dyson resum the light propagator.
At this point we expand in the heavy scale, perform mass renormalization for the light particle and get corrections 
to the light particle wave function. One could think of moving these terms from the EFT interaction Lagrangian to the
wave-function of the light degree of freedom; however, resumming a class of higher order EFT effects implies that
EFT has not been consistently used in the matching~\footnote{I am grateful to M.~Trott for this comment.}.
\section{Conclusions \label{Concu}}
In this work~\footnote{Some of the issues discussed in this paper partially overlap with the very recent work 
of \Bref{Criado:2018sdb}.} we have examined the relationship among a four-dimensional, weakly coupled, field theory 
$X$, its EFT extension $X\mbox{EFT}$ and the class of UV completions of $X$, say $X^{\prime}$.  
Technically speaking we are interested in constructing $X\mbox{EFT}$ up to and including $\mrdim = 8$ operators.

In developing the SM EFT there is a dual motivation for including $\mrdim = 8$ operators:
it may prove that an analysis purely at the $\mrdim = 6$ level is inadequate, or that $\mrdim = 8$ effects
should be accounted for as a systematic uncertainty. Indeed, given the plethora of 
new physics scenarios that may show up in future precision Higgs measurements only at the
percent level, it is an urgent issue to understand and reduce theory uncertainties so that they do not
become a limiting factor in our searches for new physics.

As stated in the Introduction, EFT is being widely used in an effort to interpret experimental measurements of 
SM processes. In this scenario, various consistency issues arise; we have critically examined the issues 
and argued for the necessity to learn more general lessons about new physics within the EFT approach.  

Every EFT extension of $X$ is a consistent QFT, although differing from $X$ in the UV, \ie 
we can calculate in the EFT without unnecessary reference to the UV physics that is is decoupled.
However, if the value of Wilson coefficients in broad UV scenarios could be inferred in general this would
be of significant scientific value. Therefore, the central question in top-down EFT studies is:
given a UV theory, what is the low-energy  EFT, and what are its observable consequences?
The answer to this question is of course UV theory specific.
We have discussed this scenario from the point of view of the higher order compensations arising from the removal 
of ``redundant'' operators.

The classical LSZ formalism~\cite{Lehmann:1957zz} is better handled by using the so-called ``canonical normalization''; 
we have discussed the interply between canonical normalization, gauge invariance and gauge fixing.
Actually, canonical normalization should be understood in a broad sense: $1/\Lambda$ corrections to $\mrdim = 4$
operators are reabsorbed into the parameters of the $\mrdim = 4$ Lagrangian. From this point of view it is more convenient
to work with the $\mrR_{\xi\xi}\,$-gauge~\cite{Actis:2006rb}, so that the gauge parameters can also be normalized.
Canonical normalization for the $\beta_h(\beta_{\mrt})$ tadpole parameter, which is not strictly needed for SMEFT at LO, will 
play an important role when constructing SMEFT at NLO~\cite{Ghezzi:2015vva,Passarino:2016pzb}.

When considering the SM the EFTs are further distinguished by the presence of a Higgs doublet (or not) in the 
construction. In the SMEFT the EFT is constructed with an explicit Higgs doublet, while in the 
HEFT (an Electroweak chiral Lagrangian with a dominantly $J^{\ssP} = 0^+$ scalar) no such doublet is included. 
Usability of SMEFT/HEFT in UV scenarios based on a proliferation of scalars and mixings has also been discussed. 

We have analyzed the impact on global fits of different effects: selection of a basis, transformation of basis,
truncation of basis, inclusion/exclusion of canonical normalization. All of them go in the same direction:
inconsistent results usually attributed to SMEFT are in fact the consequence of unnecessary further approximations
and multiple operators, after canonical normalization, should be retained in data analyses.
Effective field theories accomplish internal consistency; quite often, when EFTs are considered in the actual practice, 
only a limited set of terms is used destroying the mathematical consistency. 
Precision tests of the SM in all sectors and critical examination of the EFT framework have become more
important than ever.
  
A key ingredient of top-down EFT studies is matching a given UV theory onto its low-energy EFT. 
By this we mean that, after the Higgs boson discovery, we have a paradigm shift, \ie we use the EFT for fitting the data. 
The ``fitted'' Wilson coefficients become the pseudo{-}data and we may take any specific BSM model, compute the 
corresponding low{-}energy limit and choose the BSM parameters to be those appropriate for the pseudo{-}data.

There are elegant methods for obtaining operator coefficients, which avoid the need for computing complicated
correlation functions. They are based on direct evaluation of the functional path integral and we have analyzed in
details some of the subtle points which arise during the process, proper handling of matrix logarithms and 
non-local terms due to heavy{-}light contributions. We argue for the key advantages of having a more precise, one-loop, 
matching. 
An insurgence of non-local terms in the effective action is due to heavy{-}light effects, although not all 
of them have a non-local character (of the form of kinematic logarithms), which is controlled by loop diagrams having 
normal, as well as anomalous, thresholds in the low{-}energy region.     

Finally, we have reviewed the effective field theory interpretation of derivative-coupled field theories.


The BSM picture is still incomplete. Paraphrasing Freeman Dyson~\cite{BaF}, we can say that some physicists are 
birds, others are frogs. Birds fly high in the air (top-down approach) and survey broad vistas out to the far horizon. 
Frogs live in the mud below (bottom-up approach) and see only the flowers that grow nearby. They solve problems one 
at a time. BSM physics needs both birds and frogs. Physics is rich and beautiful because birds give it broad visions and 
frogs give it intricate details (this paper happens to be a frog). 


{\bf{Acknowledgments}}:
I gratefully acknowledge a constructive correspondence with A.~David.

\clearpage

\appendix
\section{Field transformations and their Jacobian  \label{Jac}}
Consider a theory with one scalar field and a transformation $\upphi \to \upphi - \mra/\Lambda^2\,\Box\,\upphi$.
As it is well known this transformation has several effects, including the presence of a Jacobian in the
functional integral. The latter shows up as a ghost Lagrangian (not the FP ghosts however) 
\bq
{\overline{c}}\,(1 - \frac{\mra}{\Lambda^2}\,\Box)\,c \spp
\eq
The reason why this part of the Lagrangian does note contribute is explained in sect.~$2$ of \Bref{Arzt:1994gp}.
Here we want to repeat the argument by giving the proper emphasis to a crucial aspect of any EFT,
local{-}analytic expansion in EFT as described in \Bref{Brivio:2017vri}. The ghost propagator is
\bq
\Delta_c = \frac{\Lambda^2}{\mra\,p^2 + \Lambda^2} \spp
\eq
Suppose that we want to use this propagator inside EFT loop diagrams, \eg for the $\upphi$ $n\,$-point function.
It is immediately seen that loop integration and $\Lambda \to \infty$ limit are non-commuting operations. 
In particular,
\bq
\Delta_c \sim 1 + \ord{\frac{p^2}{\Lambda^2}} \spc
\eq
which means that $c\,$-loops are zero in dimensional regularization when the limit is first. That one has to expand first
is inherent to EFT, as explained in sect.~$3.1$ of \Bref{Brivio:2017vri} and it is due to ``separation of scales'' and
to the fact that the result of integrating havy degrees of freedom is an effective action that describes non-local 
interactions between the low-energy degrees of freedom. To obtain a local effective action the effective action 
is expanded in terms of local operators~\cite{Polchinski:1992ed}.
\section{Higher dimensional operators and FP ghosts  \label{HDOFP}} 
If $X$ is the SM and SMEFT is constructed according to \eqn{EFTex}, what about FP ghost fields in $\Lag^{(d)}, d > 4$? 
This question is only relevant for NLO SMEFT, nevertheless let us examine the situation in more details.
Suppose that $X^{\prime}$ is the singlet extension of the SM; after fixing the gauge in $X^{\prime}$ and taking the mixing 
into account we will have couplings between the heavy Higgs, $\PH$, and FP-fields, \eg 
${\overline{\PX}}^{\pm}\,\PH\,\PX^{\pm}$. At the same time there are $\PH\,\Ph^3$ couplings so that we can generate 
(in $X^{\prime}$) a tree-level diagram with $\Ph\,\Ph\,\Ph\,{\overline{\PX}}^{\pm}\,\PX^{\pm}$ external lines and one 
$\PH$ internal line. After expanding the $\PH$ propagator we obtain an effective vertex with only light fields and 
FP fields; this vertex can be used to construct EFT loops contributing to $\Ph + \Ph \to \Ph + \Ph + \Ph + \Ph$.
This fact raises the question of whether $d\,$-dimensional operators containing FP ghosts should be included in the EFT
Lagrangian. 
\section{Tadpole and log-tadpole integrals  \label{Tint}}
Tadpole integrals are defined as follows:
\bq
\mrT^{l,j}_{\mu_1\,\dots\,\mu_l} = \int d^{\mrd} q\,\frac{1}{(q^2 + m^2)^j}\,\prod_{k=1,l}\,q_{\mu_k} \spc
\label{pcount}
\eq
which vanishes for odd values of $l$ and gives 
\bq
\mrT^{2\,i,j}_{\mu_1\,\dots\,\mu_{2i}} = \mrT_0^{2\,i,j}\,\delta_{\mu_1\,\dots\,\mu_{2i}} \spc
\qquad
\delta_{\mu_1\,\dots\,\mu_{2i}} = \sum_{j=2,2\,i}\,\delta_{\mu_1\,\mu_j}\,
\delta_{\mu_2\,\dots\,\mu_{j-1}\,\mu_{j+1}\,\dots\,\mu_{2i}} \spp
\eq
We need to evaluate
\bq
\mrI^{2\,i,j,k}_{\mu_1\,\dots\,\mu_{2i}} = \int d^{\mrd}q\,\int_0^{\infty} d\mu^2\,
\frac{1}{(q^2 + M^2_+)^j\,(q^2 + m^2_+)^k}\,\,\prod_{l=1,2i}\,q_{\mu_l} = 
\mrI_0^{2\,i,j,k}\,\delta_{\mu_1\,\dots\,\mu_{2i}} \spc
\eq
with $M^2_+ = M^2 + \mu^2$ and $m^2_+ = m^2 + \mu^2$. We derive
\bqa
\mrI_0^{j,k} &=& \int d^{\mrd}q\,\int_0^{\infty} d\mu^2\,
\frac{1}{(q^2 + M^2_+)^j\,(q^2 + m^2_+)^k} 
\nl
{}&=& 
\sum_{i=0}^{j-1}\,\frac{(-1)^i}{(m^2 - M^2)^{k+i}}\,
\Bigl(
\begin{array}{c}
k + i - 1 \\
k - 1 \\
\end{array}
\Bigr)\,\mrJ_{j - i}(M^2)
+
\sum_{i=0}^{k-1}\,\frac{(-1)^i}{(m^2 - M^2)^{j+i}}\,
\Bigl(
\begin{array}{c}
j + i - 1 \\
j - 1 \\
\end{array}
\Bigr)\,\mrJ_{k - i}(m^2) \spc
\label{pfd}
\eqa
where we have defined
\bq
\mrJ_n(M^2)= \int d^{\mrd}q\,\int_0^{\infty} d\mu^2\,\frac{1}{(q^2 + M^2_+)^n} \spp
\eq
From \eqn{pfd} we see that the $\mrJ_1$ functions always appear in the combination
\bqa
\mrJ_1(M^2) - \mrJ_1(m^2) &=& \int d^{\mrd}q\,\int_0^{\infty}\,(
\frac{1}{q^2 + M^2_+} - \frac{1}{q^2 + m^2_+} )
\nl
{}&=& \int d^{\mrd}q\,\Bigl[ \ln(q^2 + m^2) - \ln(q^2 + M^2) \Bigr] =
\frac{i\,\pi^2}{2}\,\Bigl[ M^4\,\mra_0(M) - m^4\,\mra_0(m) \Bigr] \spc
\nl
\mra_0(m) &=& \frac{1}{\epb} - \ln\frac{m^2}{\muRs} + \frac{3}{2} \spc
\eqa
with $1/\epb= 2/(4 - \mrd) - \gamma - \ln\pi$, $\gamma$ id the Euler{-}Mascheroni constant and $\muR$ 
is the 't Hooft scale. 
\bei
\item[{\bf{Log{-}tadpoles}}]
\eei
In deriving the result we have used~\cite{Kleinert:2001ax}
\bqa
\int d^{\mrd}q\,\ln(q^2 + m^2) &=& - \frac{i \pi^2}{2}\,\mra_0(m) = - \frac{i \pi^2}{2}\,
( \frac{1}{\epb} - \ln\frac{m^2}{\muRs} + \frac{3}{2} ) \spc
\nl
\int d^{\mrd}q\,q^2\,\ln(q^2 + m^2) &=& - \frac{i \pi^2}{3}\,\mra_1(m) = - \frac{i \pi^2}{3}\,
( \frac{1}{\epb} - \ln\frac{m^2}{\muRs} + \frac{4}{3} ) \spc
\eqa
\etc More log-tadpole integrals are need;
\bq
\mrL_t(1)= \int d^{\mrd}q\,\frac{1}{q^2}\,\ln(q^2 + M^2) \spc
\eq
is defined in the strip $2 < \Re \mrd < 4$. The explicit result is
\bq
\mrL_t(1) = i\,\pi^2\,M^2\,\bigl( \frac{1}{\epb} - \ln\frac{M^2}{\muRs} + 2 \bigr) \spp
\eq
We also need $\mrL_t(k)$ with $k > 1$. The integral $\mrL_t(2)$ is defined for $4 < \Re \mrd < 6$ \etc
By analytic continuation we obtain 
\bq
\mrL_t(3) = - \frac{i\,\pi^2}{M^2}\,\bigl( \frac{1}{\epb} - \ln\frac{M^2}{\muRs} + \frac{1}{2} \bigr) \spc
\qquad
\mrL_t(4) =  \frac{i\,\pi^2}{2\,M^4}\,\bigl( \frac{1}{\epb} - \ln\frac{M^2}{\muRs} + \frac{1}{2} \bigr) \spc
\eq
while $\mrL_t(2)$ shows a double pole.

The remaining integrals can be computed using~\cite{Veltman:1994wz}
\bqa
\int \frac{d^{\mrd}q}{(q^2 + m^2)^n} &=& i\,\pi^2\,\mrF_n(m^2) \spc
\nl
\int d^{\mrd}q\,\frac{q_{\mu} q_{\nu}}{(q^2 + m^2)^n} &=& i\,\pi^2\,\mrG_n(m^2)\,\delta_{\mu\nu} \spc
\nl
\int d^{\mrd}q\,\frac{q_{\mu} q_{\nu} q_{\alpha} q_{\beta}}{(q^2 + m^2)^n} &=& i\,\pi^2\,
\mrH_n(m^2)\,\delta_{\mu\nu\alpha\beta} \spc
\eqa
\begin{itemize}

\item[{\bf{$\mrF$ family}}]

\bqa
\mrF_1(m^2) &=& m^2\,(\frac{1}{\epb} - \ln\frac{m^2}{\muRs} - 1) \spc
\nl
\mrF_2(m^2) &=& \frac{1}{\epb} - \ln\frac{m^2}{\muRs} \spc
\nl
\mrF_n(m^2) &=& \frac{1}{(n-1)\,(n-2)}\,(m^2)^{2-n} \spc \qquad n \ge 3 \spp
\eqa

\item[{\bf{$\mrG$ family}}]

\bqa
\mrG_1(m^2) &= & \frac{1}{4}\,m^4\,(\frac{1}{\epb} - \ln\frac{m^2}{\muRs} + \frac{3}{2} ) \spc
\nl
\mrG_n(m^2) &=& \frac{1}{2\,(n-1)}\,\mrF_{n-1}(m^2) \spc \qquad n \ge 2 \spp
\eqa

\item[{\bf{$\mrH$ family}}]

\bqa
\mrH_1(m^2) &=& - \frac{1}{24}\,m^6\,(\frac{1}{\epb} - \ln\frac{m^2}{\muRs} + \frac{11}{6}) \spc
\nl
\mrH_n(m^2) &=& \frac{1}{2\,(n-1)}\,\mrG_{n-1}(m^2) \spc \qquad n \ge 2 \spp
\eqa

\end{itemize}

 \clearpage
\bibliographystyle{elsarticle-num}
\bibliography{XEFT}


\end{document}